\documentclass[a4paper,12pt]{article}
\pdfoutput=1
\usepackage{jheppub}
\usepackage[T1]{fontenc} 
\title{\boldmath Production of Heavy neutrino in next-to-leading order QCD at the LHC and beyond }
\preprint{\today}
\author[a,1]{Arindam Das,\note{Corresponding author.}}
\author[b]{Partha Konar,}
\author[c]{Swapan Majhi}
\affiliation[a]{Department of Physics and Astronomy, University of Alabama, Tuscaloosa, Alabama, 35487, USA}
\affiliation[b]{Theoretical Physics Group, Physical Research Laboratory, Ahmedabad-380009, India}
\affiliation[c]{Department of Physics, Achhruram Memorial College, Purulia, WB, 723202, India}
\emailAdd{adas8@ua.edu}
\emailAdd{konar@prl.res.in}
\emailAdd{majhi.majhi@gmail.com}
\newcommand{\beq}{\begin{equation}}
\newcommand{\eeq}{\end{equation}}
\newcommand{\bea}{\begin{eqnarray}}
\newcommand{\eea}{\end{eqnarray}}
\newcommand{\beas}{\begin{eqnarray*}}
\newcommand{\eeas}{\end{eqnarray*}}
\newcommand{\bi}{\begin{itemize}}
\newcommand{\ei}{\end{itemize}}

\def\tev{\,{\ifmmode\mathrm {TeV}\else TeV\fi}}
\def\gev{\,{\ifmmode\mathrm {GeV}\else GeV\fi}}
\def\to{\rightarrow}

\abstract{Majorana and pseudo-Dirac heavy neutrinos are introduced into the type-I and inverse seesaw models, respectively, in explaining the naturally small neutrino mass.
TeV scale heavy neutrinos can also be accommodated to have a sizable mixing with the Standard Model light neutrinos, through which they can be produced and detected at the high energy colliders. 
In this paper we consider the Next-to-Leading Order QCD corrections to the heavy neutrino production, and study the scale variation in cross-sections as well as the kinematic distributions
with different final states at $14$ TeV LHC and also in the context of $100$ TeV hadron collider. 
The repertoire of the Majorana neutrino is realized through the characteristic signature of the same-sign dilepton pair, whereas, due to a small lepton number violation, the
pseudo-Dirac heavy neutrino can manifest the trileptons associated with missing energy in the final state. 
Using the $\sqrt{s}=8$ TeV,  $20.3$ fb$^{-1}$ and $19.7$ fb$^{-1}$ data at the ATLAS and CMS respectively, we obtain prospective scale dependent upper bounds of the light-heavy neutrino mixing angles for the Majorana heavy neutrinos at the $14$ TeV LHC and $100$ TeV collider. 
Further exploiting a recent study on the anomalous multilepton search by CMS at $\sqrt{s}=8$ TeV with $19.5$ fb$^{-1}$ data, we also obtain the prospective scale dependent upper bounds on the mixing angles for the pseudo-Dirac neutrinos.  
We thus project a scale dependent prospective reach using the NLO processes at the $14$ TeV LHC.
}
\begin{document} 
\maketitle
\flushbottom
\section{Introduction}
\label{sec:intro}
The journey of Large Hadron Collider (LHC) in its 7 and 8 TeV run has been extremely successful in discovering, and further constraining the properties of long-awaited Higgs boson~\cite{:2012gu,:2012gk} of Standard Model (SM). However,  LHC is still lacking any clinching signature yet from the beyond the Standard Model (BSM) physics. With general wisdom, there exists a broad agreement in admitting the SM as, at most a very accurate description for low energy effective theory of particle physics. This notion is embolden from the fact, that the SM falls short to explain divers outstanding issues both in theory and in explaining some crucial  experimental observations.

The most recent observations on the neutrino oscillation phenomena \cite{Neut1,Neut2,Neut3,Neut4,Neut5,Neut6} have established that tiny neutrino mass and the flavor mixing of the SM neutrinos which is one of the divers mysteries in SM. The existence of such a tiny neutrino mass requires us to extend the SM. The seesaw mechanism \cite{seesaw1, typeI-seesaw1, typeI-seesaw2, typeI-seesaw3, typeI-seesaw4, typeI-seesaw5, seesaw2} is probably the simplest idea to extend the SM, which can explain the small neutrino mass naturally. The SM-singlet heavy right-handed Majorana neutrinos induce the dimension five operators \cite{Weinberg} leading to very small light Majorana neutrino masses. If such heavy neutrino mass lies in the electroweak scale, then the heavy neutrinos can be produced in the high energy colliders \cite{Keung:1983uu,Datta:1992qw, Datta:1993nm, AguilarSaavedra:2009ik, Chen:2011hc, Das:2012ze, Das:2014jxa, Das:2015toa, Dev:2015pga, Das:2016akd, Gluza:2016qqv, delAguila:2008hw, delAguila:2007qnc, AguilarSaavedra:2012gf, Nayak:2015zka, Nayak:2013dza, AguilarSaavedra:2012fu, LalAwasthi:2011aa, Fong:2011xh, Dias:2011sq, Ibarra:2011xn, ILC1, ILC2, type-I1, type-I2, Batell:2015aha, Leonardi:2015qna, p-photon, konar1, konar2, ruiz,  Dutta:1994wz, Haba:2009sd, Matsumoto:2010zg, Mondal:2012jv, Helo:2013esa, Hessler:2014ssa, Deppisch:2015qwa, Arganda:2015ija, Dib:2015oka, Degrande:2016aje} from various initial states. The heavy neutrinos are singlet under the SM gauge group therefore they can be coupled to the SM gauge bosons through the mixing with the light SM neutrinos vis Dirac Yukawa coupling. The Dirac Yukawa coupling can be sizable for the electroweak scale heavy neutrinos in general Casas-Ibarra parameterization \cite{Ibarra1}, while reproducing the neutrino oscillation data.

There is another kind of seesaw mechanism, commonly known as inverse seesaw \cite{inverse-seesaw1, inverse-seesaw2}, where the tiny Majorana mass is generated from the small lepton number violating parameters, rather than being suppressed by the heavy neutrino mass in conventional seesaw mechanism. In case of inverse seesaw the  heavy neutrinos are pseudo-Dirac and the Dirac Yukawa coupling could be of order one, satisfying the neutrino oscillation data. Thus at the high energy colliders the pseudo-Dirac heavy neutrinos can be produced through the sizable mixing with the SM neutrinos. In our analysis we choose the LHC at the center of mass energy $\sqrt{s}=14$ TeV and a proposed proton-proton collider at the center of mass energy $\sqrt{s}=100$ TeV \cite{Arkani-Hamed:2015vfh} which can enlighten the new physics era including the heavy neutrino physics more precisely with its higher fact finding ability.

Apart from these seesaw mechanisms there are different other simple ways which can also be tested in colliders. Type-II seesaw where the SM is extended by an SU(2) triplet scalar, see \cite{Magg, Cheng:1980qt, Lazarides:1980nt, Mohapatra, Nesti, Chun:2012zu, Chun:2012jw, Dev:2013ff, Chun:2013vma, Gu, Haba:2016zbu,Han:2015hba, Han:2015sca} for detailed studies. The other one is the type-III seesaw which is obtained by the extension of the SM with an SU(2) triplet fermion, see \cite{Foot, Franceschini:2008pz, Abada:2008ea, Aguilar-Saavedra:2013twa, Eboli:2011ia, Ahn:2011pq, Biggio:2011ja, France1} for detailed studies, (see \cite{Ruiz:2015zca} for the NLO analysis in type-III seesaw). Additional interesting possibility of generate naturally small neutrino mass is from higher-dimensional operators at the TeV scale and thus testable at the collider \cite{Babu:2009aq, Bambhaniya:2013yca, Gogoladze:2008wz}.

 The heavy neutrino can be produced at the high energy colliders from various initial states among them the leading contributions come from the processes generated from the quark-quark $(q\overline{q^{\prime}})$, quark-gluon $(qg)$ and gluon-gluon $(gg)$ initial states. Among these processes the $q\overline{q^{\prime}}$ initial state is the commonly studied leading order (LO) production channel for the heavy neutrinos, where as, the other channels can contribute in  its Next-to-Leading-Order (NLO) and Next-to-Next-to-Leading-Order (NNLO) QCD corrections together with the corresponding LO processes. In this paper we concentrate on the QCD NLO production processes including the virtual correction contributions and the real emission processes. For the LO processes we demonstrate the production of the heavy neutrino for different factorization $(\mu_{F})$ scales associated to the parton density functions (PDFs) considering the $14$ TeV LHC and in the context of proposed $100$ TeV hadron collider. On the other hand, NLO processes are studied with different choices of  factorization $(\mu_{F})$ as well as renormalization $(\mu_{R})$ scales juxtaposing together with LO contributions.

The paper is organized in the following way. In Sec.~\ref{sec:neutrino mass} we introduce the type-I seesaw and inverse seesaw models. These are the primary models we concentrate in our present analysis. In Sec.~\ref{sec:calc} we calculate the production cross-section of the heavy neutrino at the high energy colliders. We discuss the methodology followed with different choice of parameters in estimating in leading order and next-to-leading order production of heavy neutrinos. We also opened up discussion on the scale variation related to these production cross-sections. In Sec.~\ref{sec:result} we study the scale dependent kinematic distributions of different kinematic measurable quantities in the heavy neutrino production from the trilepton plus missing energy final state. In Sec.~\ref{sec:Mix}  we utilize the current Large Hadron Collider (LHC) data from ATLAS and CMS to put scale dependent upper bounds on the mixing angles between the light- heavy neutrinos. Sec.~\ref{sec:conc} is dedicated to the conclusion.

\section{Neutrino Mass Mechanism}
\label{sec:neutrino mass}
In type-I seesaw \cite{seesaw1, typeI-seesaw1, typeI-seesaw2, typeI-seesaw3, typeI-seesaw4, typeI-seesaw5, seesaw2}, we introduce
  SM gauge-singlet right handed Majorana neutrinos $N_R^{\beta}$,    
  where $\beta$ is the flavor index. $N_R^{\beta}$ couple with SM lepton doublets 
  $\ell_{L}^{\alpha}$ and the SM Higgs doublet $H$.
The relevant part of the Lagrangian is
\bea
\mathcal{L} \supset -Y_D^{\alpha\beta} \overline{\ell_L^{\alpha}}H N_R^{\beta} 
                   -\frac{1}{2} m_N^{\alpha \beta} \overline{N_R^{\alpha C}} N_R^{\beta}  + H. c. .
\label{typeI}
\eea
After the spontaneous electroweak symmetry breaking
   by the vacuum expectation value (VEV), 
   $ H =\begin{pmatrix} \frac{v}{\sqrt{2}} \\  0 \end{pmatrix}$, 
    we obtain the Dirac mass matrix as $M_{D}= \frac{Y_D v}{\sqrt{2}}$.
Using the Dirac and Majorana mass matrices 
  we can write the neutrino mass matrix as 
\bea
M_{\nu}=\begin{pmatrix}
0&&M_{D}\\
M_{D}^{T}&&m_{N}
\end{pmatrix}.
\label{typeInu}
\eea
Diagonalizing this matrix we obtain the seesaw formula
 for the light Majorana neutrinos as 
\bea
m_{\nu} \simeq - M_{D} m_{N}^{-1} M_{D}^{T}.
\label{seesawI}
\eea
For $m_{N}\sim 100$ GeV, one may find that the extremely minuscule $Y_{D} \sim 10^{-6}$ is needed to construct some light neutrino mass of order $m_{\nu}\sim 0.1$ eV. However, using the general parameterization based on Casas-Ibarra \cite{Ibarra1}, one gets the Yukawa coupling expressed in terms of a orthogonal matrix which remains completely  arbitrary and hence can be large.
Following this mechanism $Y_{D}$ can be phenomenologically viable, as large as order one, and this is the case we consider in our present work.

There is another seesaw mechanism, so-called inverse seesaw \cite{inverse-seesaw1, inverse-seesaw2},
   where the light Majorana neutrino mass is generated through tiny 
   lepton number violation.
The relevant part of the Lagrangian is given by
\bea
\mathcal{L} \supset - Y_D^{\alpha\beta} \overline{\ell_L^{\alpha}} H N_R^{\beta}- m_N^{\alpha \beta} \overline{S_L^{\alpha}} N_R^{\beta} -\frac{1}{2} \mu_{\alpha \beta} \overline{S_L^{\alpha}}S_L^{\beta^{C}} + H. c. ,
\label{InvYuk}
\eea 
where  $N_R^{\alpha}$ and $S_L^{\beta}$ are two SM-singlet heavy neutrinos
   with the same lepton numbers, $m_N$ is the Dirac mass matrix, and
   $\mu$ is a small Majorana mass matrix violating the lepton numbers.
After the electroweak symmetry breaking we obtain the neutrino mass matrix as 
\bea
M_{\nu}=\begin{pmatrix}
0&&M_{D}&&0\\
M_{D}^{T}&&0&&m_{N}^{T}\\
0&&m_{N}&&\mu
\end{pmatrix}.
\label{InvMat}
\eea
Diagonalizing this mass matrix we obtain the light neutrino mass matrix 
\bea
M_{\nu} \simeq M_{D} m_{N}^{-1} \, \mu \,  m_{N}^{-1^{T}} M_{D}^{T}.
\label{numass}
\eea
Note that the smallness of the light neutrino mass originates 
  from the small lepton number violating term $\mu$. 
The smallness of $\mu$ allows the $m_{D}m_{N}^{-1}$ parameter
  to be order one even for an electroweak scale heavy neutrino.
Since the scale of $\mu$ is much smaller than the scale of $m_{N}$,
  the heavy neutrinos become the pseudo-Dirac particles. 
This is the main difference between the type-I and the inverse seesaws. 
  
\begin{figure*}
\begin{center}
\includegraphics[scale=0.5]{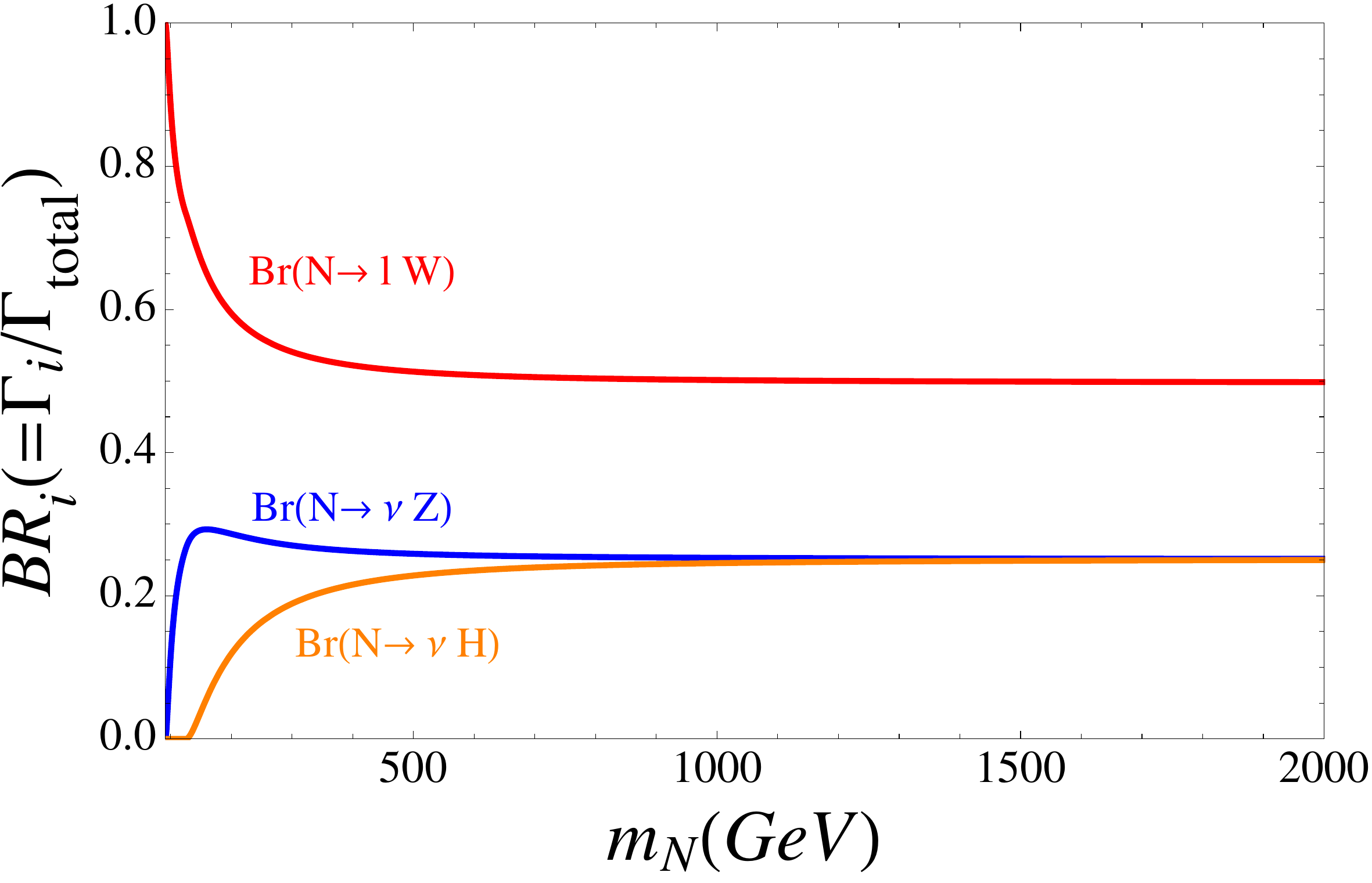}
\caption{Heavy neutrino branching ratios ($BR_i$) for different decay modes are shown with respect to the heavy neutrino mass $\left(m_{N}\right)$.}
\label{fig:BR}
\end{center}
\end{figure*}

Through the seesaw mechanism, a flavor eigenstate ($\nu$) of 
  the SM neutrino is expressed in terms of the mass eigenstates 
  of the light ($\nu_m$) and heavy ($N_m$) Majorana neutrinos such as 
\bea 
  \nu \simeq  \nu_m  + V_{\ell N} N_m,  
\eea 
where $V_{\ell N}$  is the mixing between the SM neutrino and the SM-singlet heavy neutrino, 
  and we have assumed a small mixing, $|V_{\ell N}| \ll 1$.  
Using the mass eigenstates, the charged current interaction for the heavy neutrino 
  is given by 
\bea 
\mathcal{L}_{CC} \supset 
 -\frac{g}{\sqrt{2}} W_{\mu}
  \bar{\ell} \gamma^{\mu} P_L   V_{\ell N} N_m  + h.c., 
\label{CC}
\eea
where $\ell$ denotes the three generations of the charged leptons in the vector form, and 
  $P_L =\frac{1}{2} (1- \gamma_5)$ is the projection operator. 
Similarly, the neutral current interaction is given by 

\bea 
\mathcal{L}_{NC} \supset 
 -\frac{g}{2 c_w}  Z_{\mu} 
\left[ 
  \overline{N_m} \gamma^{\mu} P_L  |V_{\ell N}|^2 N_m 
+ \left\{ 
  \overline{\nu_m} \gamma^{\mu} P_L V_{\ell N}  N_m 
  + h.c. \right\} 
\right] , 
\label{NC}
\eea
 where $c_w=\cos \theta_w$ with $\theta_w$ being the weak mixing angle. 
The main decay modes of the heavy neutrino are 
 $N \to \ell W$, $\nu_{\ell} Z$, $\nu_{\ell} h$. 
The corresponding partial decay widths are respectively given by
\bea
\Gamma(N \rightarrow \ell W) 
 &=& \frac{g^2 |V_{\ell N}|^{2}}{64 \pi} 
 \frac{ (m_{N}^2 - m_W^2)^2 (m_{N}^2+2 m_W^2)}{m_{N}^3 m_W^2} ,
\nonumber \\
\Gamma(N \rightarrow \nu_\ell Z) 
 &=& \frac{g^2 |V_{\ell N}|^{2}}{128 \pi c_w^2} 
 \frac{ (m_{N}^2 - m_Z^2)^2 (m_{N}^2+2 m_Z^2)}{m_{N}^3 m_Z^2} ,
\nonumber \\
\Gamma(N \rightarrow \nu_\ell h) 
 &=& \frac{ |V_{\ell N}|^2 (m_{N}^2-m_h^2)^2}{32 \pi m_{N}} 
 \left( \frac{1}{v }\right)^2.
\label{widths}
\eea 
The decay width of heavy neutrino into charged gauge bosons being twice as large as neutral one owing to the two degrees of freedom $(W^{\pm})$.
We plot the branching ratios $BR_i \left(= {\Gamma_{i}}/{\Gamma_{\rm total}}\right)$ of the respective decay modes $\left(\Gamma_{i}\right)$ with respect to the total decay decay width $\left(\Gamma_{\rm total}\right)$ of the heavy neutrino into $W$, $Z$ and Higgs bosons in Fig.~\ref{fig:BR}
as a function of the heavy neutrino mass $\left(m_{N}\right)$.
Note that for larger values of $m_{N}$, the branching ratios can be obtained as 
\bea
BR\left(N\rightarrow \ell W\right) : BR\left(N\rightarrow \nu Z\right) : BR\left(N\rightarrow \nu H\right) \simeq 2: 1: 1.
\eea 

\section{Heavy neutrino production at the high energy colliders}
\label{sec:calc}
We implement our model in the event generator  {\tt MadGraph5-aMC@NLO} \cite{MG, MG5, aMC} and calculate the production cross-section of the heavy neutrino at the LO and NLO respectively. 
The full automation of NLO computation is based on two main steps. 
The code for the evaluation of one loop is made through {\tt MADLOOP}\cite{madloop} and 
the born and real-emission amplitudes have been computed through {\tt MadFKS}\cite{madfks} together with the integration and matching scheme of {\tt MC@NLO}. 
MadLoop evaluate one loop amplitude by using Ossola-Papadopoulos-Pittau {\tt OPP}\cite{opp} integrand-reduction 
technique which is implemented in {\tt CutTools}\cite{CutTools}. 
In {\tt MadFKS}, subtraction method have been used by {\tt FKS}\cite{FKS} formalism. 
The showering and hadronization of the events were performed with {\tt PYTHIA6.4} for LO and {\tt PYTHIA6Q} for the NLO processes~\cite{Pyth} bundled in {\tt MadGraph} with {\tt anti-$k_{T}$} algorithm. The hadronic jets are clustered with {\tt anti-kT} formalism using {\tt FastJet}\cite{FJ}~\footnote{In our final drafting phase Ref. [56] appeared with the NLO prediction of the heavy neutrino production using MadGraphMC@NLO which overlaps with a part our result consistently depending upon the choices of scale and selection cuts. It is important to mention that we have used our independently developed code for the type-I and Inverse seesaw mechanisms in MadGraphMC@NLO, fixing bugs with the active support from the MadGraph team \cite{BugReport1} and finally with private communications. The implementation of scale variations were suggested in \cite{Questions}, although we finally
used the straightforward method bypassing the SyaCalc.}.

\begin{figure*}
\begin{center}
\includegraphics[scale=0.88]{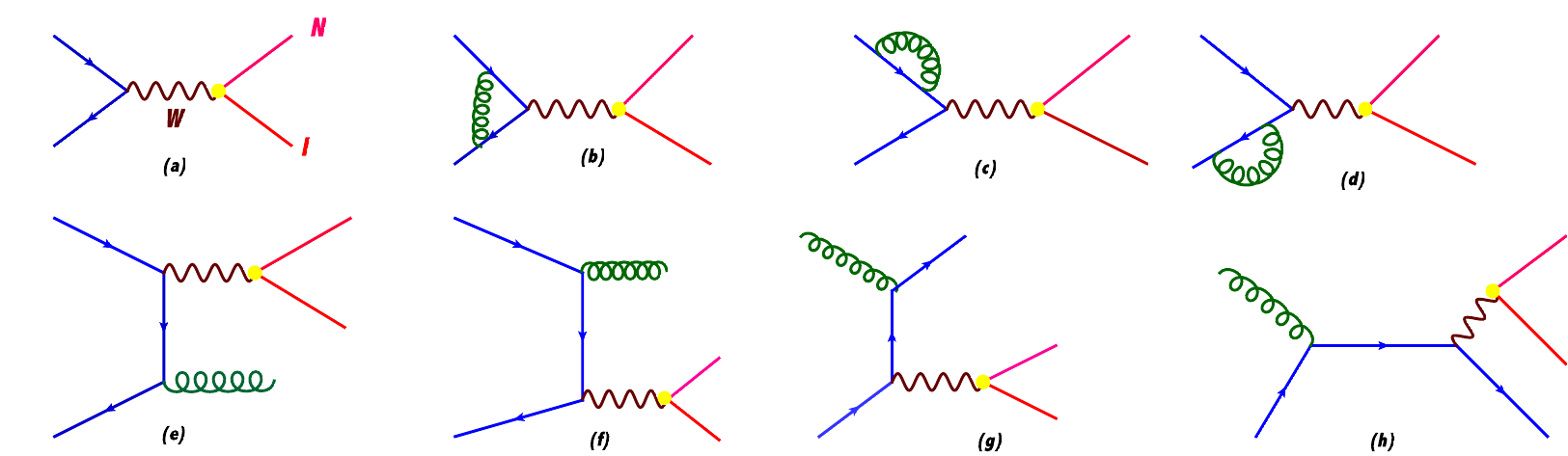}
\caption{Representative Feynman diagrams for the leading order $N\ell$ production from the $q\overline{q^{'}}$ process  at hadron collider at the Born process or LO$(a)$. Corresponding NLO diagrams including Virtual Corrections$(b-d)$ and Real Emissions$(e-h)$ contributing from different initial states are shown in rest of the diagrams.} 
\label{fig:Nl}
\end{center}
\end{figure*}

The hadronic cross-sections have been calculated by convoluting LO (NLO) 
parton distribution functions (PDF), namely, {\tt CTEQ6L1 (CTEQ6M)}  
with LO (NLO) partonic cross-section which has been done through {\tt MadGraph5-aMC@NLO}.
We choose $\alpha_s(m_Z) = 0.130$ in {\tt CTEQ6L1} for LO and  $\alpha_s(M_Z) = 0.1180$ in {\tt CTEQ6M} for NLO according to \cite{Madevent}, $m_Z = 91.188$ GeV, $m_W = 80.423$  GeV and
  $G_F = 1.166 \times 10^{-5}$  GeV$^{-2}$ as electroweak input parameters.  
Thereof, $\alpha_{QED} = 1/132.54$ and $\sin^2{\theta_W} = 0.22217$ are
 computed via LO electroweak relations. In this analysis we have considered 
two types of finals states. One of them is $N\ell$ from the $W$ boson 
mediated process. In Fig.~\ref{fig:Nl} the leading Feynman diagrams including 
the Born level in Fig.~\ref{fig:Nl}($a$), virtual diagrams in Fig.~\ref{fig:Nl}$(b-d)$ and 
real emissions diagrams in Fig.~\ref{fig:Nl}$(e-h)$ initiated from quark-antiquark or 
quark-gluon are demonstrated. The other final state we considered is the 
$N\nu$ from the $Z$ boson mediated process which we can extract easily 
from Fig.~\ref{fig:Nl}.
%
\begin{figure*}[t]
\begin{center}
\includegraphics[scale=0.28]{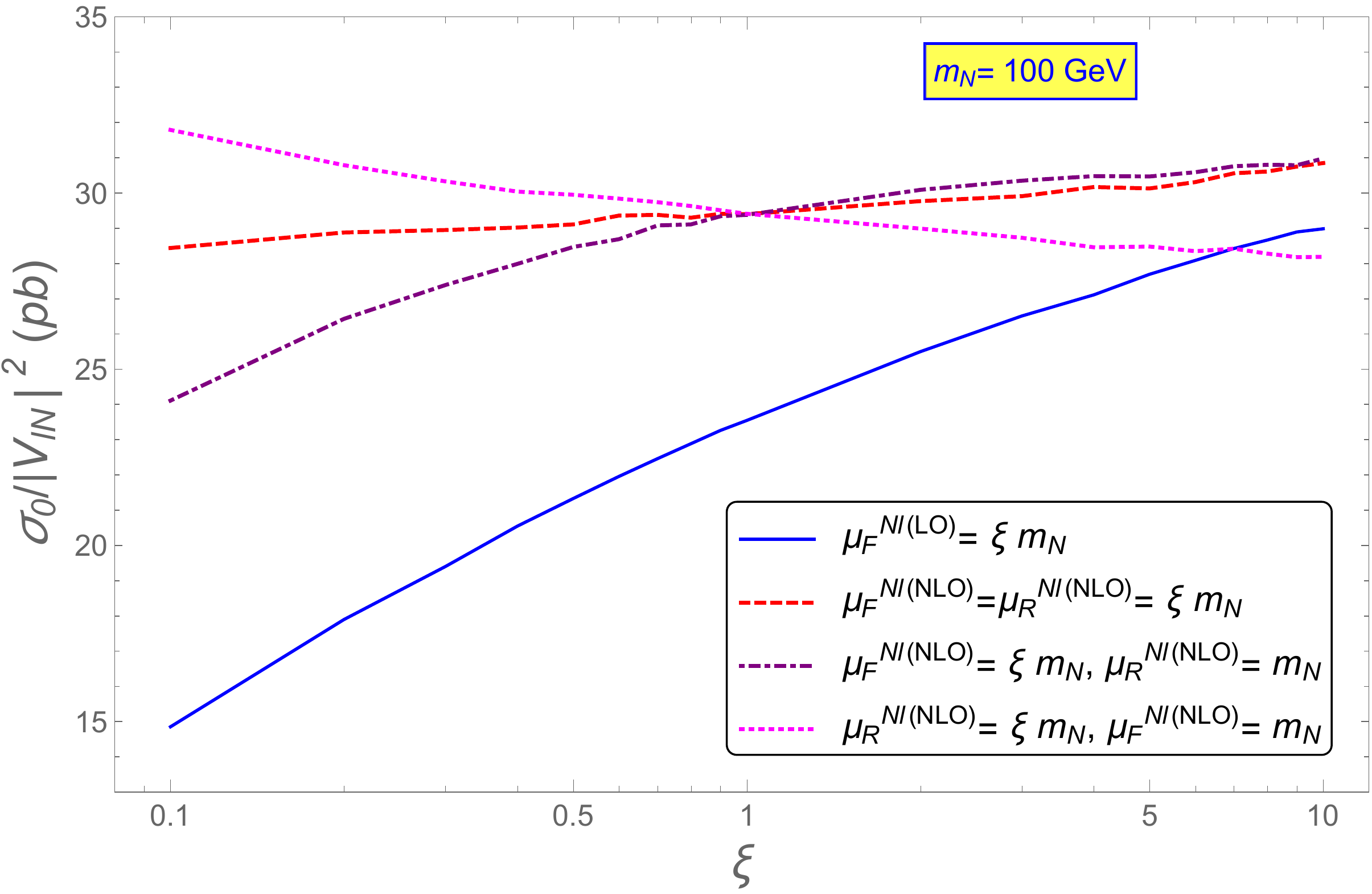}
\hspace{0.5 cm}
\includegraphics[scale=0.28]{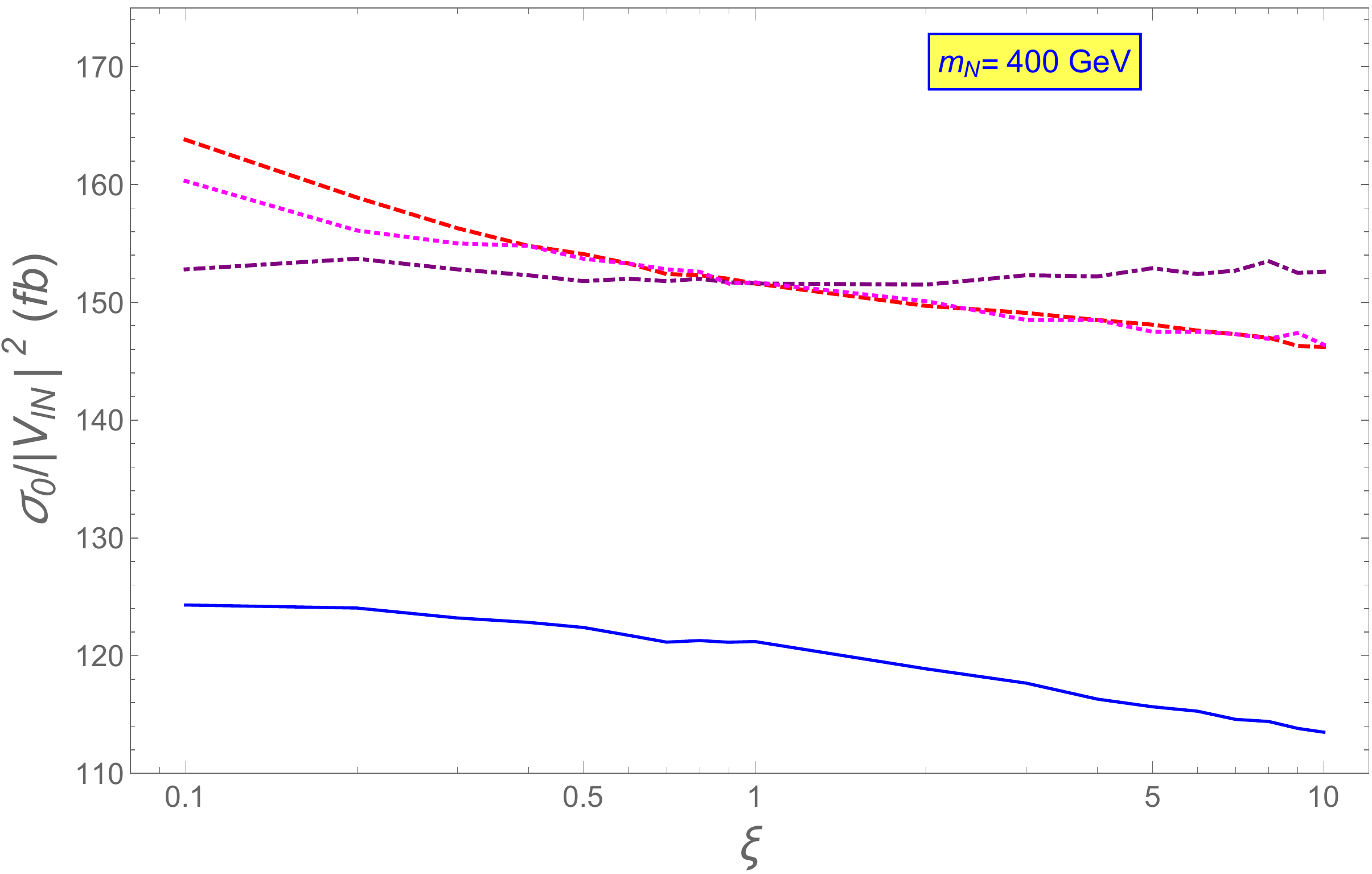}\\
\vspace{0.5 cm}
\includegraphics[scale=0.28]{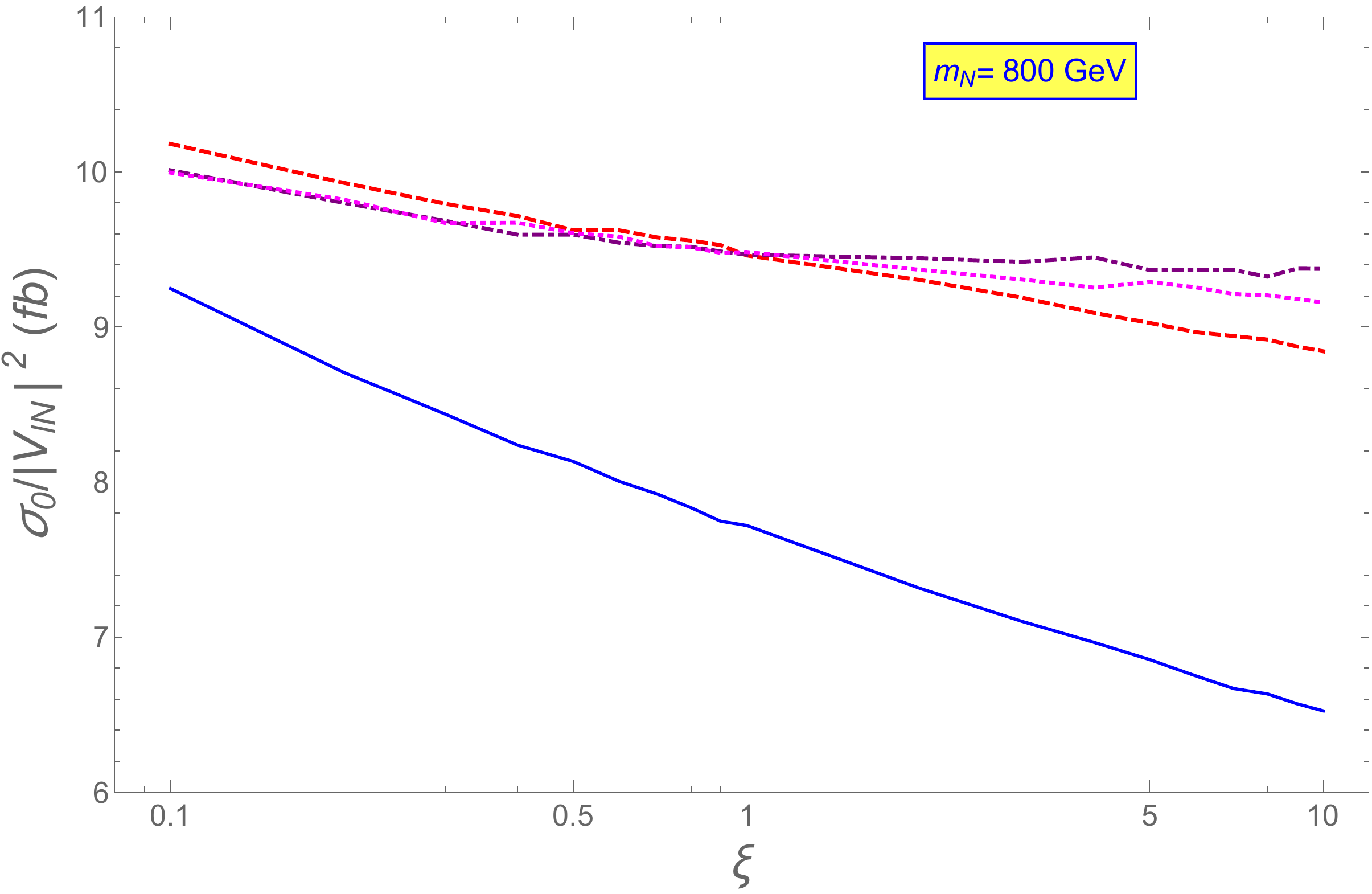}
\hspace{0.5 cm}
\includegraphics[scale=0.28]{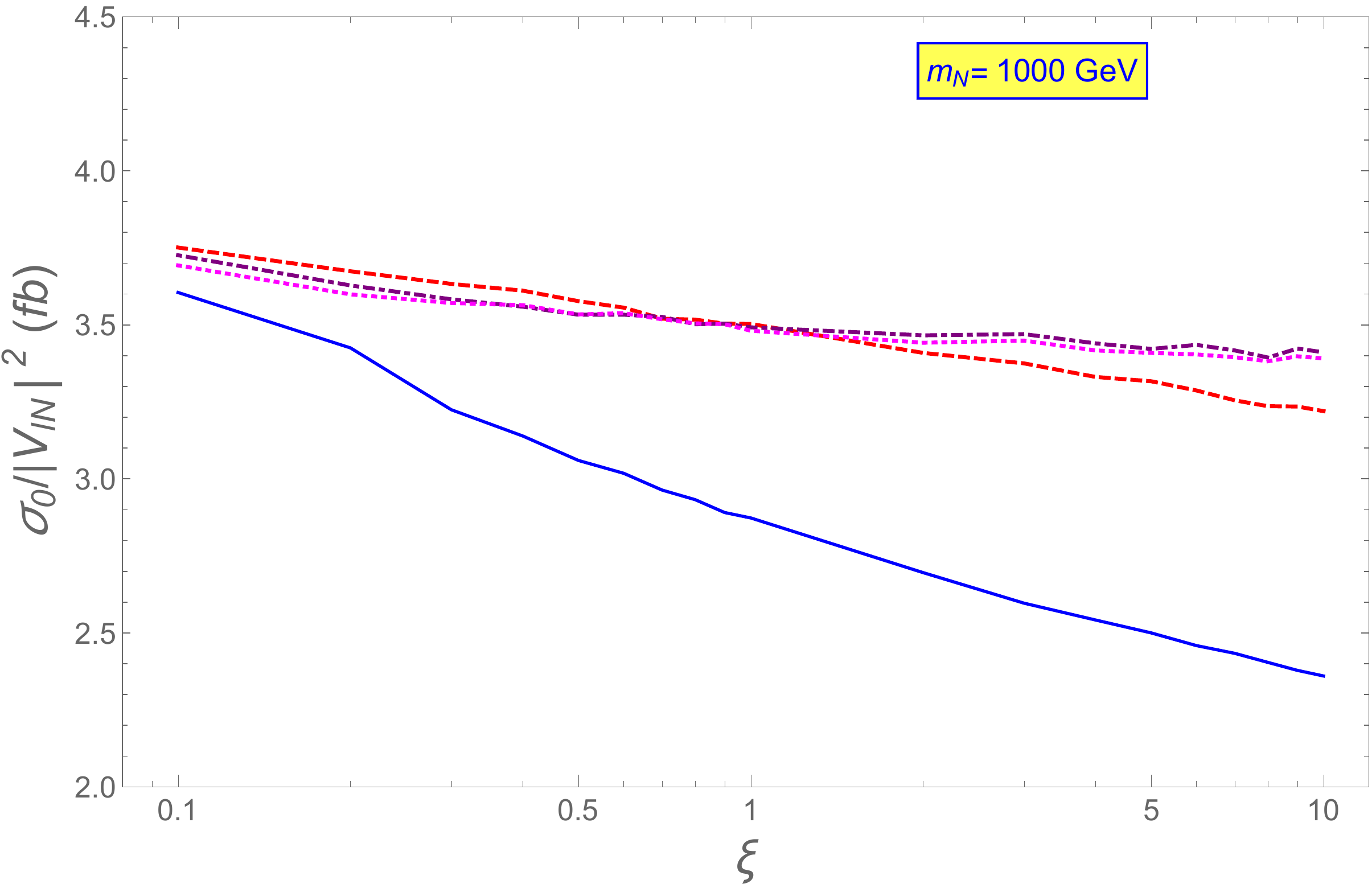}
\caption{Scale variation of the heavy neutrino production process $ p p \rightarrow \ell N$ at the $14$ TeV LHC comparing the LO with the NLO estimates at different scale choices. Plots are shown for four different heavy neutrino mass $m_N$. Cross-sections are shown as normalized with the value of $|V_{\ell N}|^2$.}
\label{fig:scale_dep_Nl_14}
\end{center}
\end{figure*}
%
\begin{figure*}[t]
\begin{center}
\includegraphics[scale=0.28]{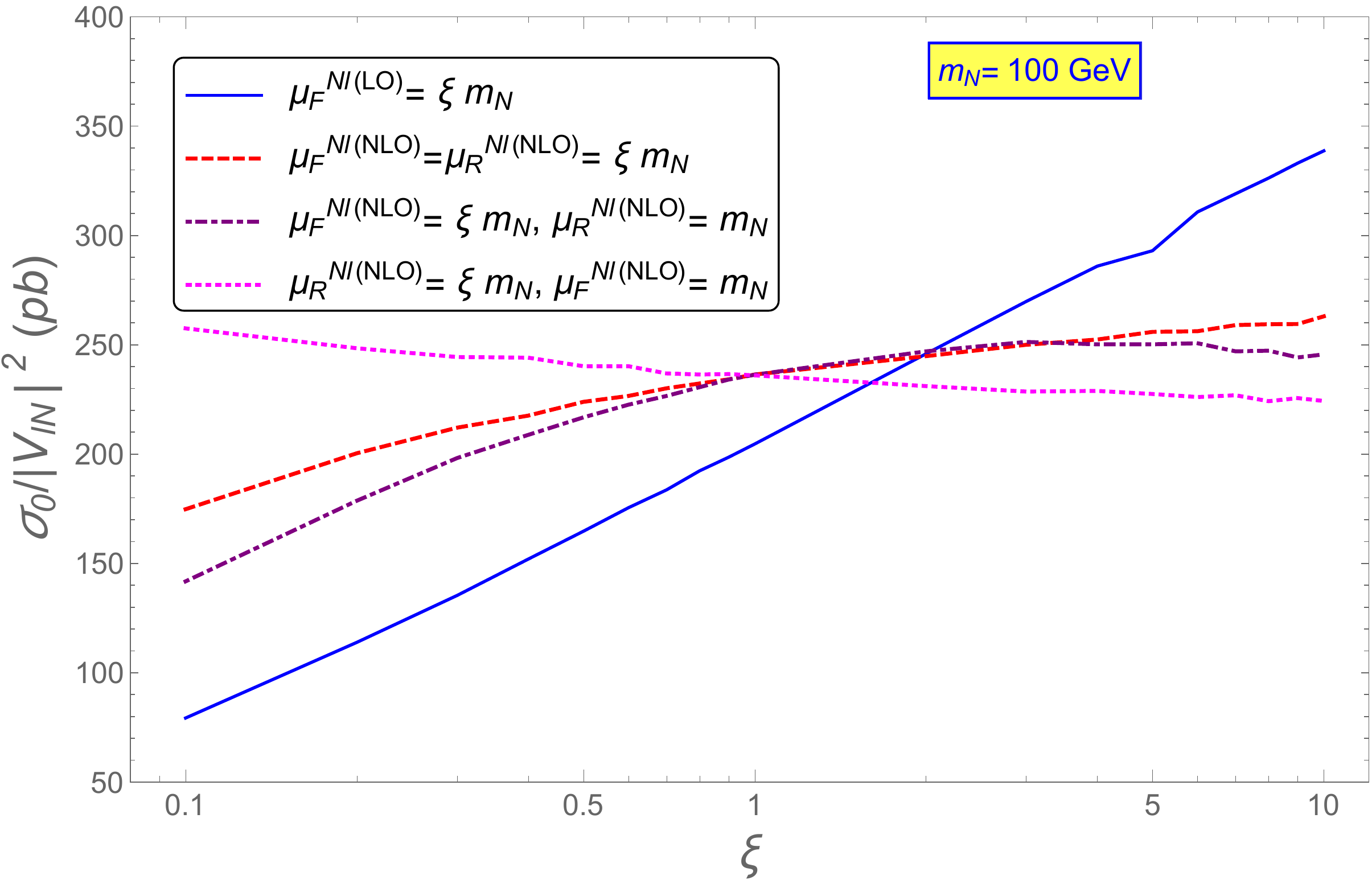}
\hspace{0.5 cm}
\includegraphics[scale=0.28]{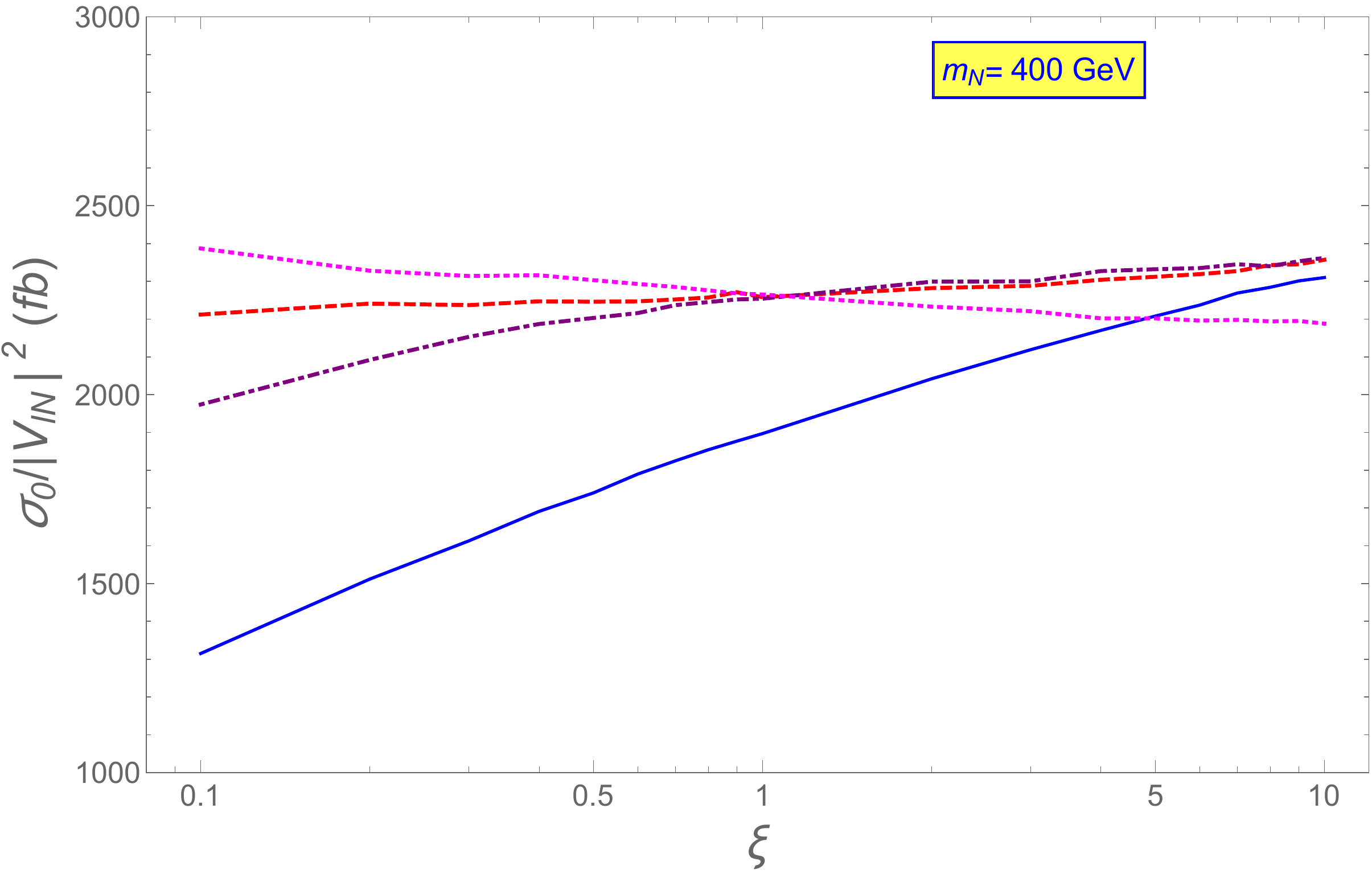}\\
\vspace{0.5 cm}
\includegraphics[scale=0.28]{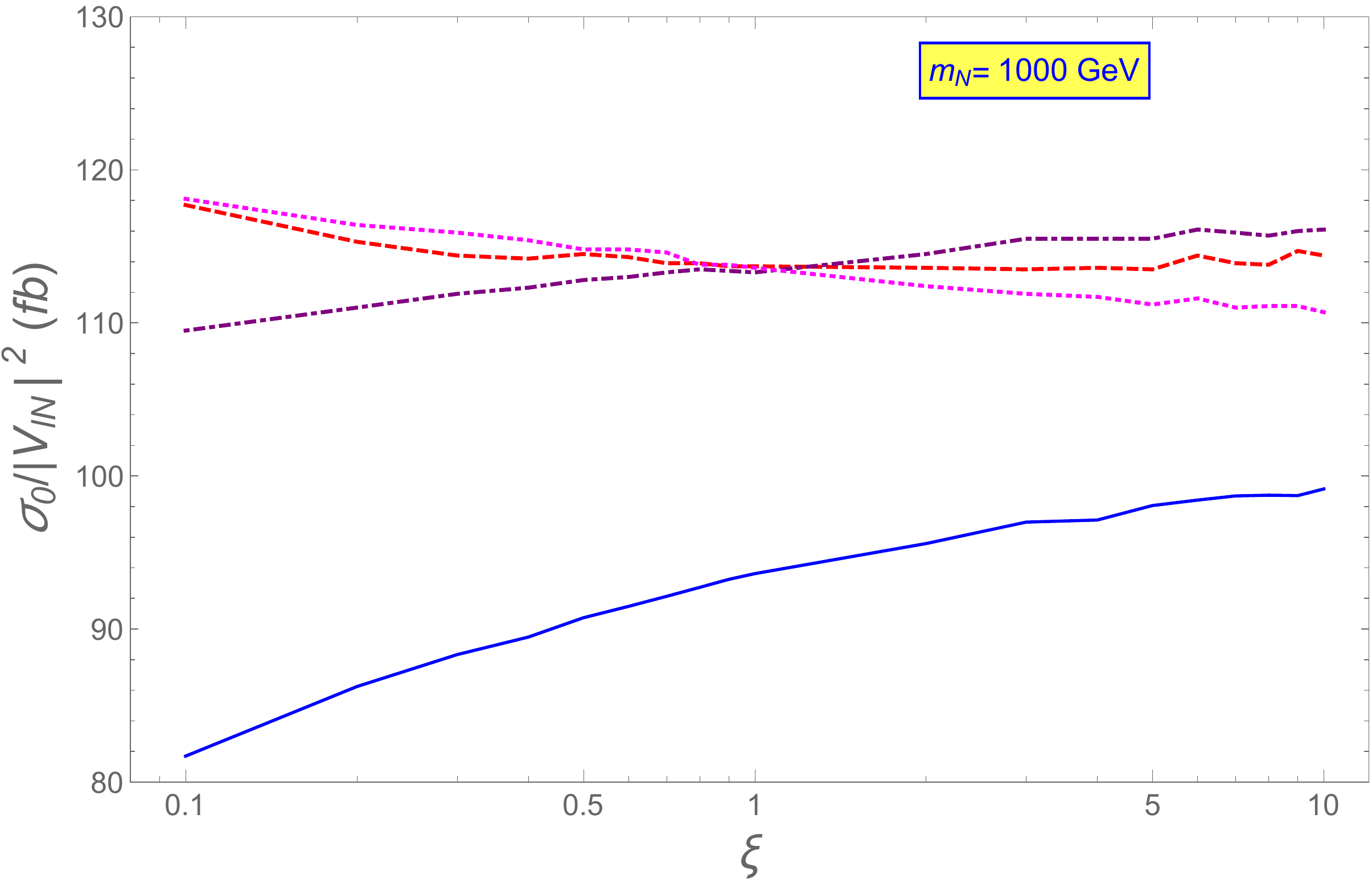}
\hspace{0.5 cm}
\includegraphics[scale=0.28]{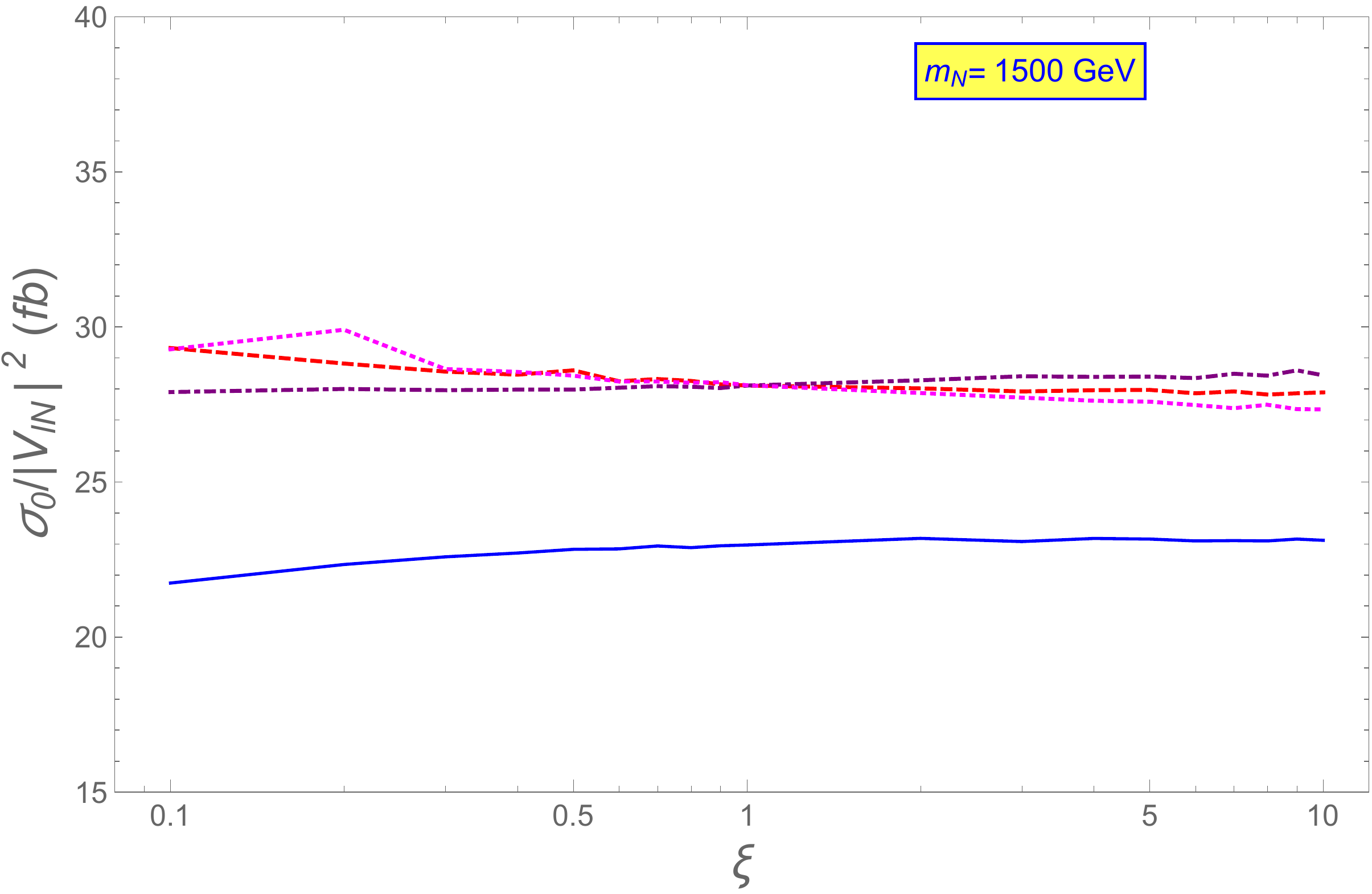}
\caption{Scale variation of the heavy neutrino production process $ p p \rightarrow \ell N$ at the $100$ TeV HC comparing the LO with the NLO estimates at different scale choices. Plots are shown for four different heavy neutrino mass $m_N$. Cross-sections are shown as normalized with the value of $|V_{\ell N}|^2$.}
\label{fig:scale_dep_Nl_100}
\end{center}
\end{figure*}
We have computed the LO cross-sections for the fixed mass with the variation of
 factorisation scale ($\mu_F$). Since the LO cross-section depends only on the 
$\mu_{F}$ through LO PDFs and we varied as 
\bea
\mu_{F} =\xi m_{N}
\label{muF}
\eea
where  $\xi$ is the scale factor varying between $0.1$ to $10$. 
Whereas the NLO cross-section depends on both the scale, namely,
 the factorization scale ($\mu_{F}$) through PDFs and the renormalisation scale
($\mu_{R}$) through NLO partonic cross-section (mainly due to the couplings 
renormalisation). For simplicity, throughout the present analysis we have 
considered to vary both these scales as,
\bea
\mu_{F} = \mu_{R}= \xi m_{N}    \, \, \, \,  \, \, \, \, \text{with} \, \,\, \,  0.1\le \xi \le 10.
\label{muR}
\eea

We have produced the scale dependent cross-sections normalized by the square of
 the mixing angle $|V_{\ell N}|^{2}$ for a fixed choice of heavy neutrino mass $m_{N}$ 
at $100$ GeV, $400$ GeV, $800$ GeV and $1$ TeV at the $14$ TeV LHC for the LO 
and NLO processes with varying $\xi$ between $0.1$ to $10$. 
This scale dependence are shown in Fig.~\ref{fig:scale_dep_Nl_14}.
In the same plot, we also display the theoretical scale uncertainty in NLO calculation due to $\mu_{F}$ 
alone by fixing the renormalisation scale at the corresponding heavy neutrino mass ($\mu_{R} = m_N$). 
Since later dependence only enters at the the NLO level in the form of $\alpha_s(\mu_{R})$, one expects 
the $\mu_{F}$ scale dependence which actually soften in NLO calculation.
The other scenario by fixing the factorization scale at the corresponding heavy neutrino mass ($\mu_{F} = m_N$) 
is also shown by changing only the $\mu_{R}$ scale in the same plot.
For $m_{N}= 100$ GeV, the leading order cross-section $\sigma_{LO}$ varies 
sharply and increasing almost monotonically  
by a factor of two approximately with 
increase in scale factor ($\xi$), which indicates a substantial amount of
 theoretical uncertainty present in the LO result.  
 This is because of only the LO quark-antiquark flux with varying scale factor $\xi^{2}$. 
 Whereas in NLO, it is three fold - the scale dependent logarithmic terms present 
in partonic cross-sections, the NLO PDF fluxes (namely, quark-antiquark, quark-gluon and antiquark-gluon) as well as strong coupling 
constant and hence the strong scale dependent part cancels among themselves. 
 Therefore this strong scale dependence has been
soften by including the NLO calculation which varies slowly with the 
scale.
At the larger choices of  heavy neutrino mass, both the LO and the NLO cross 
sections decrease with $\xi$ variation, although the basic feature of 
softening of NLO variations are evident in all such examples.

The scale dependent results for $100$ TeV collider considering 
$m_{N}$ at $100$ GeV, $400$ GeV, $1$ TeV and $1.5$ TeV are also shown 
in Fig.~\ref{fig:scale_dep_Nl_100}. 
Here the scale variation of the leading order cross-section varies rather 
sharply especially for lower value of  $m_{N}$, which provides the 
 LO cross-section dominating over the NLO prediction  
for $\xi > 2$. This steep rising of LO cross-section is mainly due to the
LO PDF (CTEQ6L1) sets. 
On the other hand the NLO cross-sections 
reduce this PDF scale uncertainty significantly so that the NLO cross-section 
remains almost flat with respect to $\xi$ for all $m_{N}$ at the $100$ TeV hadron collider.

\begin{figure*}[t]
\centering
\includegraphics[scale=0.4]{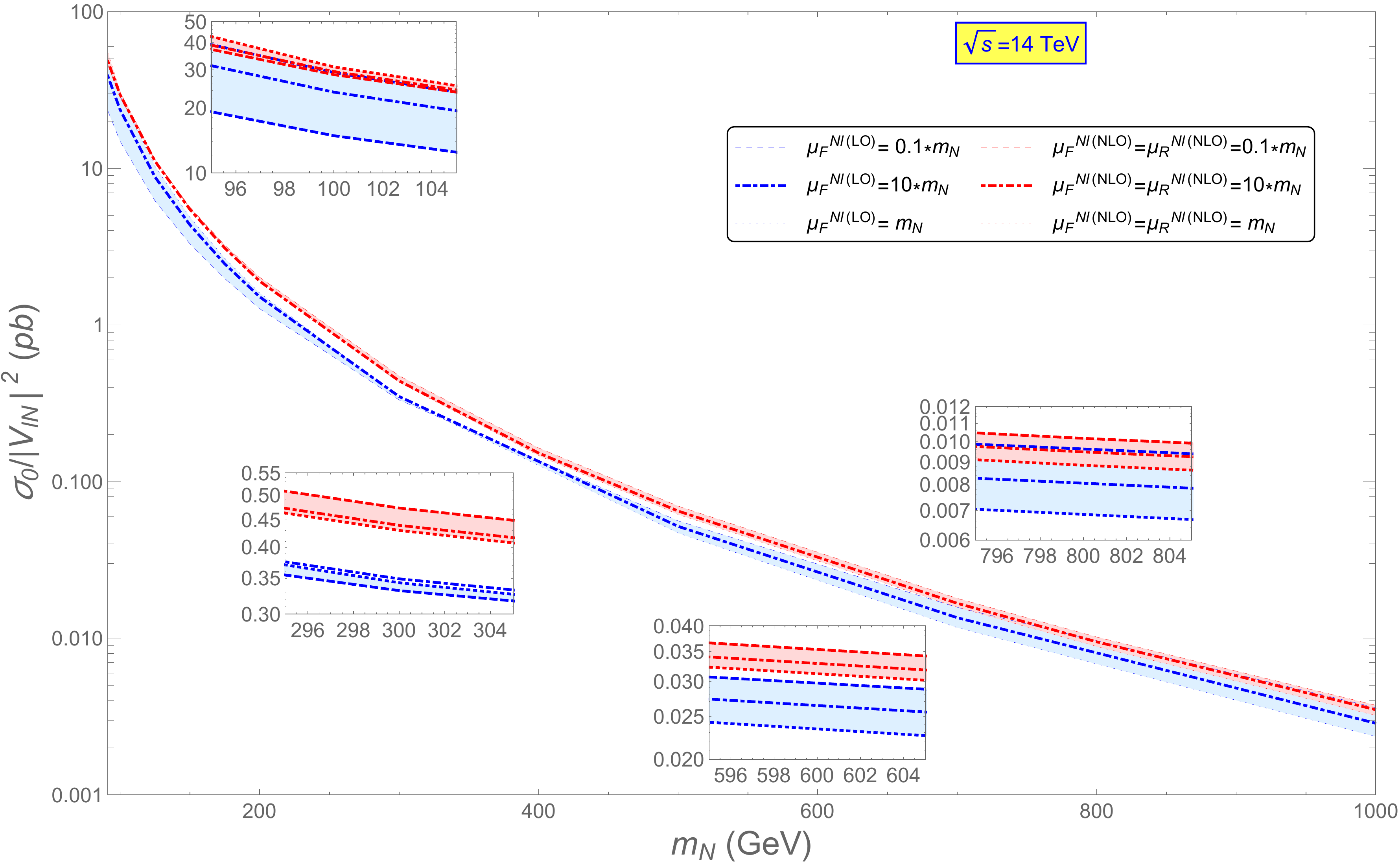}

\caption{Heavy neutrino production cross-section from $ p p \rightarrow \ell N$ at the $14$ TeV LHC as a function of $m_N$. Both  LO and NLO predictions are shown with the scale variation effect as a band. The cross-sections are normalized by the square of the mixing angles. Inset plots showing with zoomed bands at different masses.}
\label{fig:x-sec_Nl_14}
\end{figure*}

\begin{figure*}[t]
\centering
\includegraphics[scale=0.4]{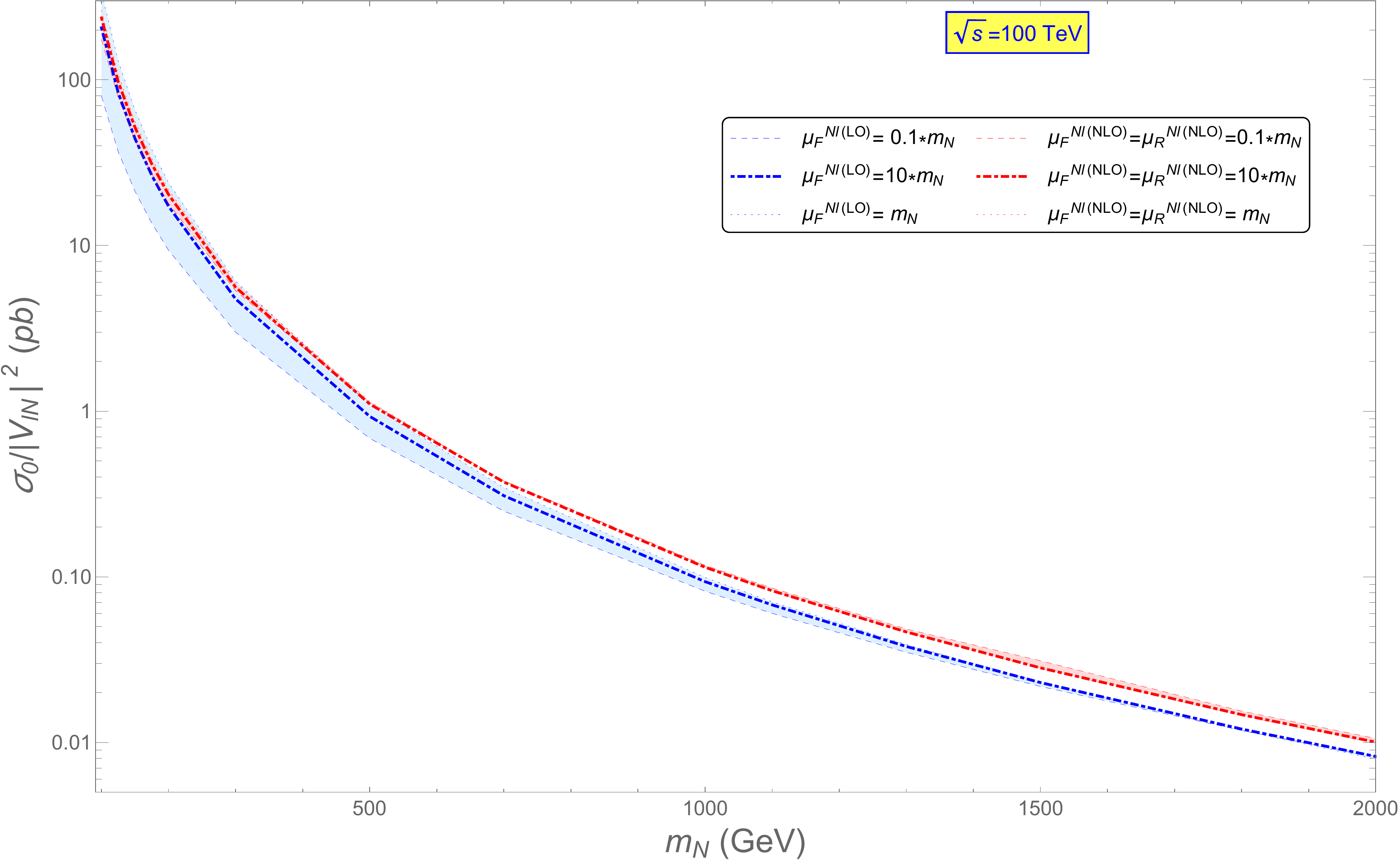}
\caption{Heavy neutrino production cross-section from $ p p \rightarrow \ell N$ at the $100$ TeV hadron collider as a function of $m_N$. Both  LO and NLO predictions are shown with the scale variation effect as a band.}
\label{fig:x-sec_Nl_100}
\end{figure*}

The scale variations of the heavy neutrino production 
cross-sections, normalized by the square of the mixing matrix, is further 
demonstrated as a function of $m_{N}$ at $14$ TeV LHC and $100$ TeV hadron 
collider in Fig.~\ref{fig:x-sec_Nl_14} and Fig.~\ref{fig:x-sec_Nl_100}. 
In these figures, the blue (red) bands shows the scale dependence of the 
$\sigma_{LO}$ ($\sigma_{NLO}$). 
It is clear from these figures that the strong scale dependence reduces 
significantly at NLO cross-section as compared to LO one.


\begin{figure*}[t]
\begin{center}
\includegraphics[scale=0.28]{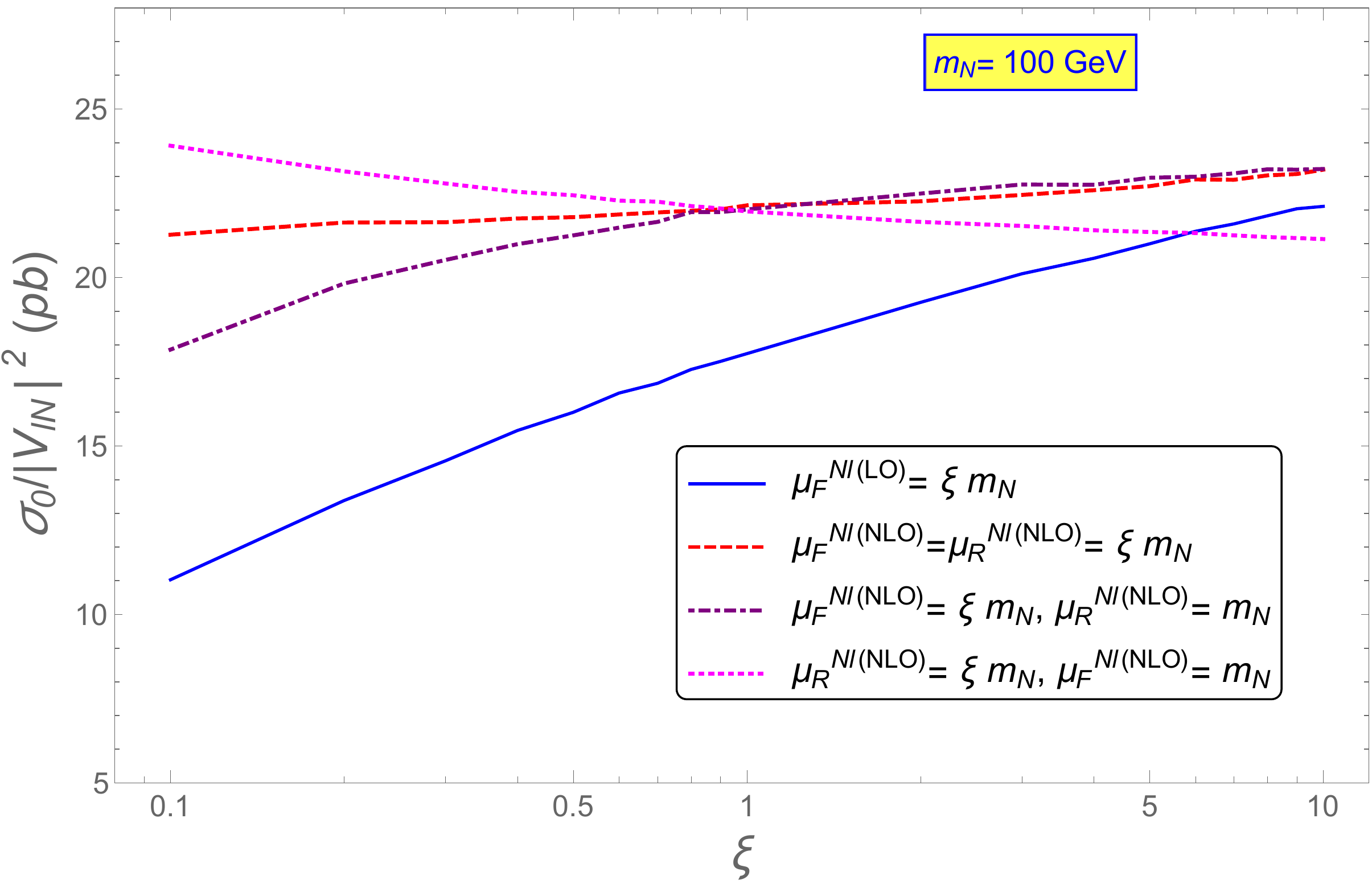}
\hspace{0.5 cm}
\includegraphics[scale=0.28]{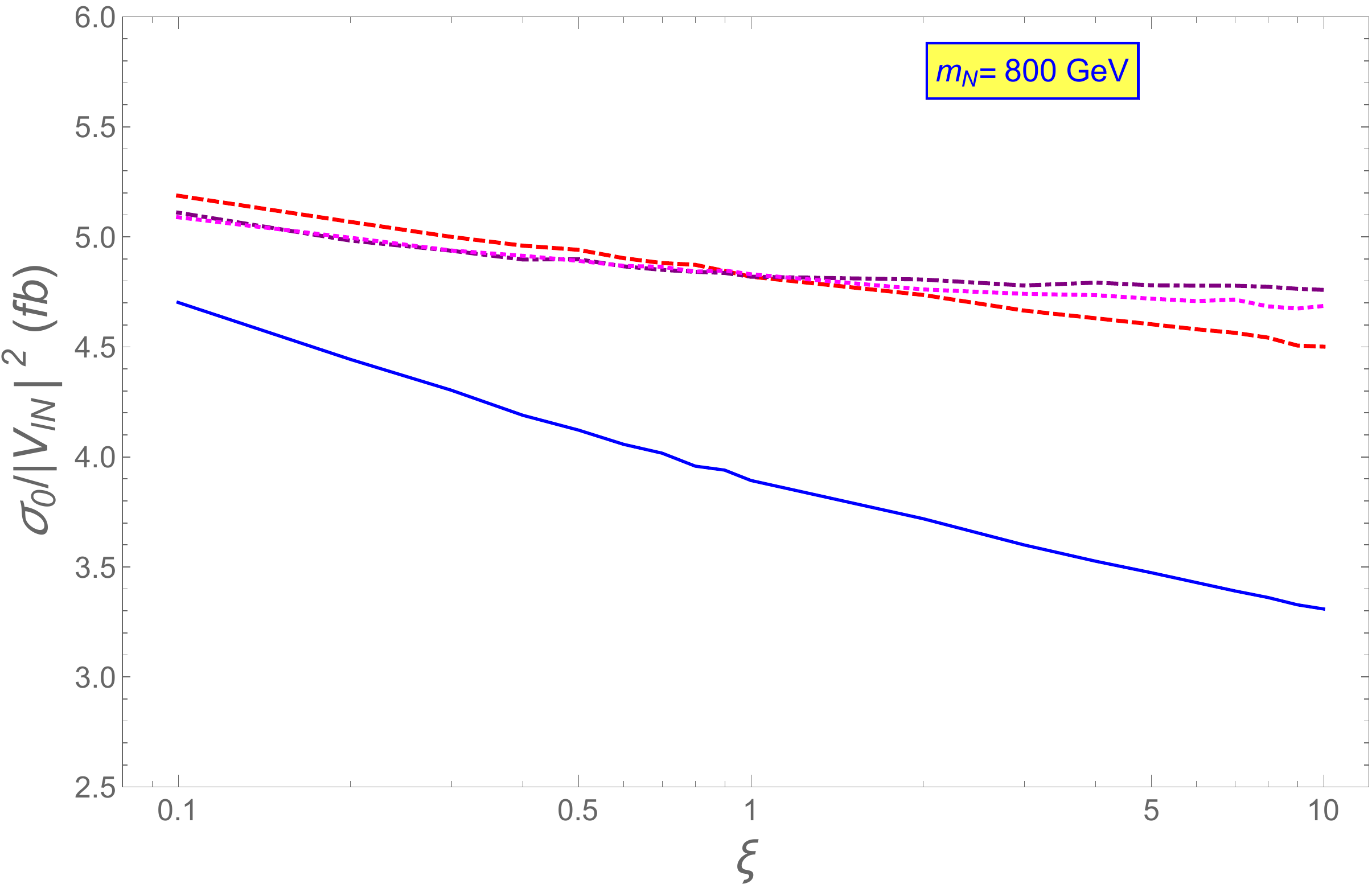}\\
\vspace{0.5 cm}
\includegraphics[scale=0.28]{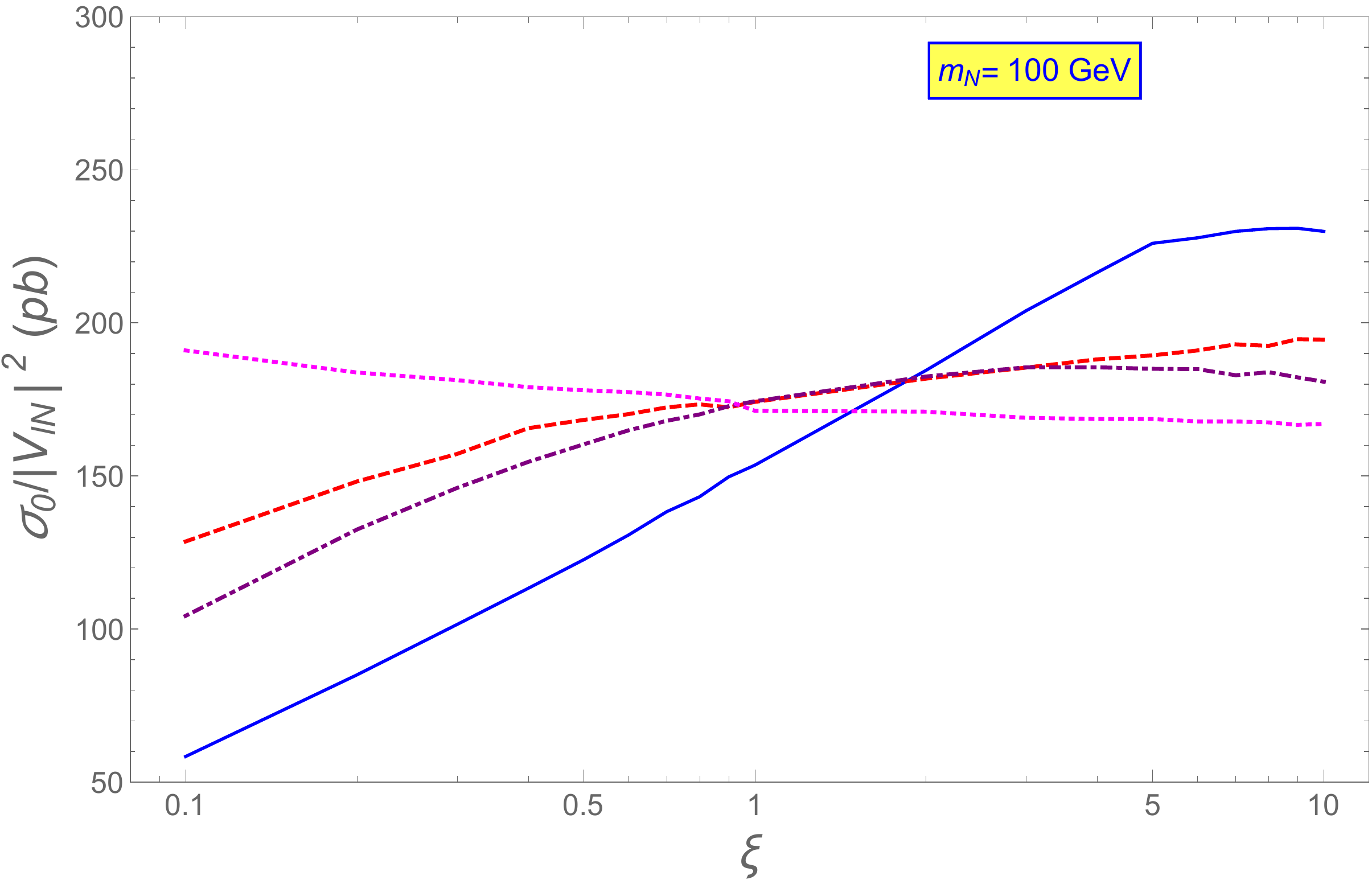}
\hspace{0.5 cm}
\includegraphics[scale=0.28]{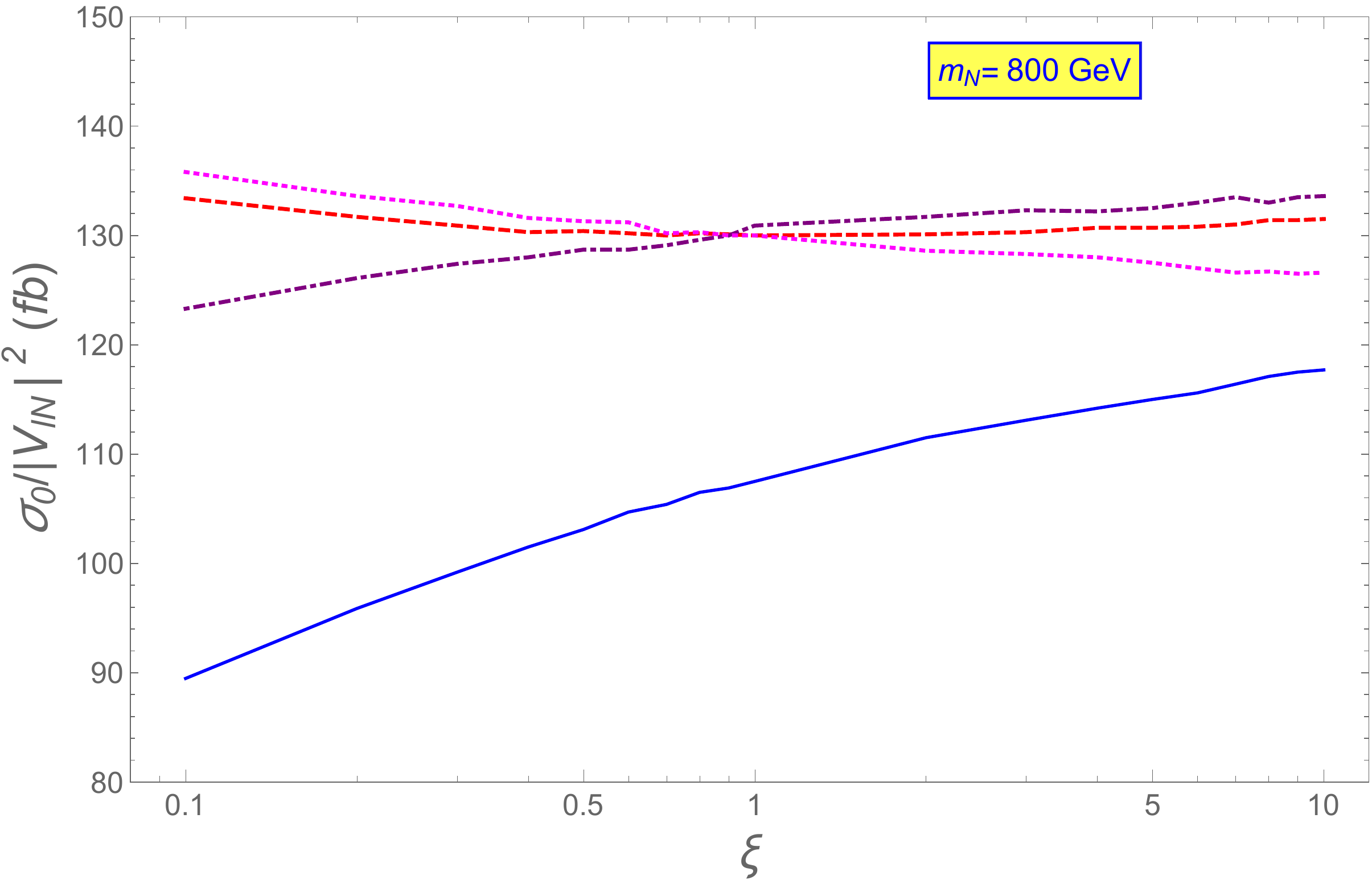}
\caption{Scale variation of the heavy neutrino production process $ p p \rightarrow \nu N$ at the (upper) 14 TeV LHC (lower) 100 TeV hadron collider comparing the LO with the NLO estimates at different scale choices. Plots are shown for two different heavy neutrino mass $m_N$. Cross-sections are shown as normalized with the value of $|V_{\ell N}|^2$.}
\label{fig:scale_dep_Nnu}
\end{center}
\end{figure*}

\begin{figure*}[t]
\centering
\includegraphics[scale=0.2]{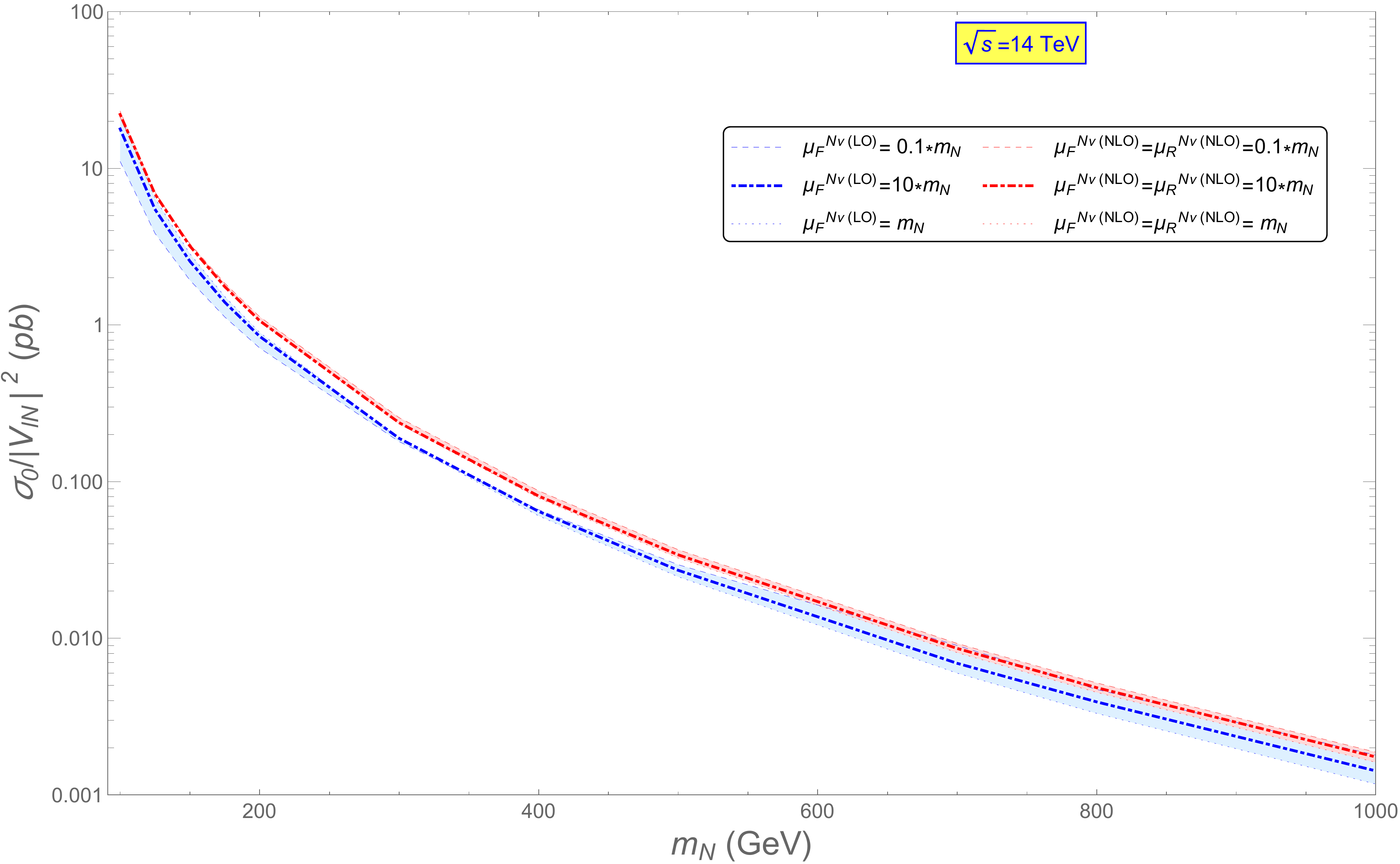}
\hspace{0.3 cm}
\includegraphics[scale=0.2]{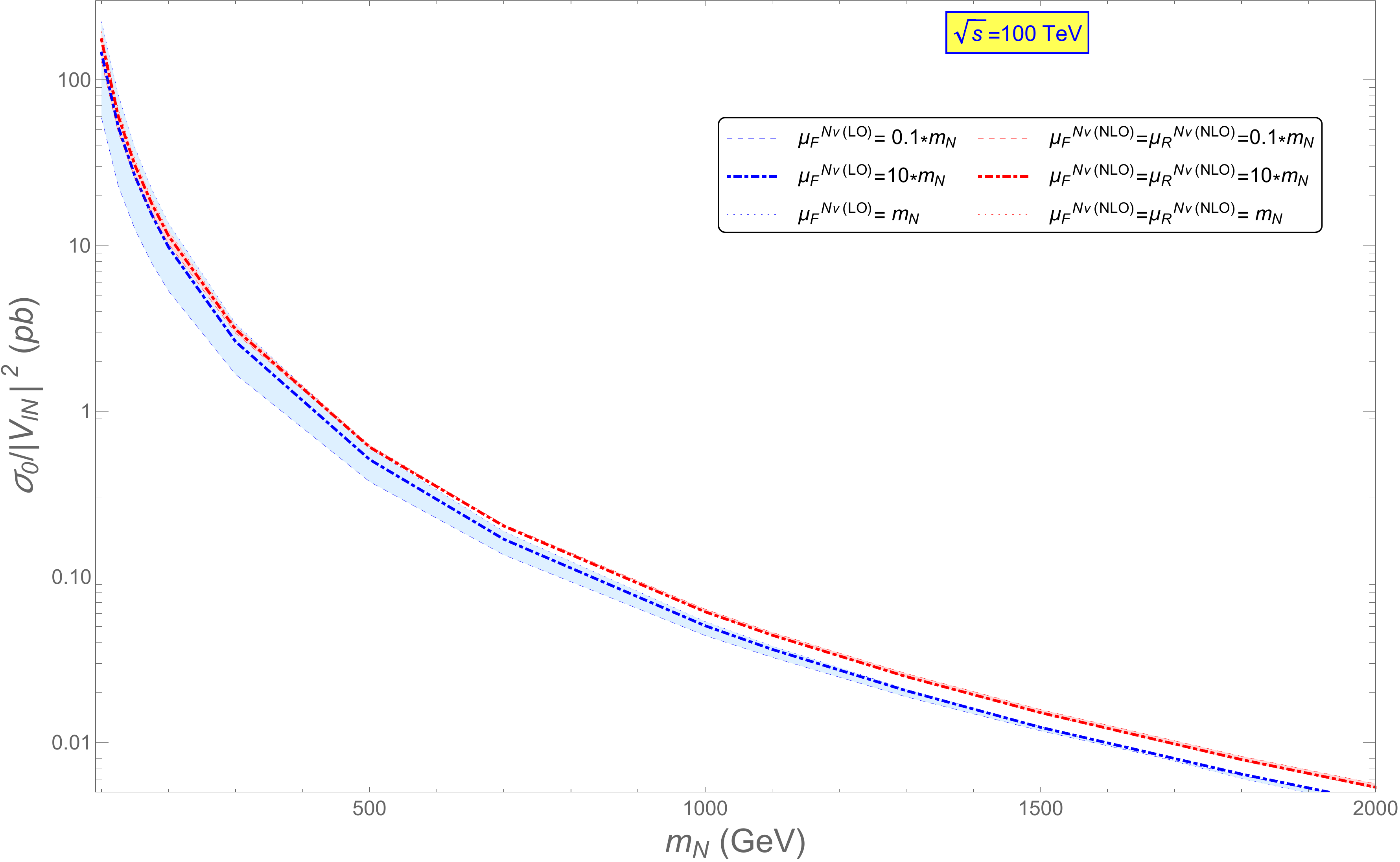}
\caption{Heavy neutrino production cross-section normalized by the square of the mixing angle from $ p p \rightarrow \nu N$ at the 14 TeV LHC as a function of $m_N$ at the $14$ TeV(Left panel) and 100 TeV(right panel). Both  LO and NLO predictions are shown with the scale variation effect as a band.}
\label{fig:x-sec_Nnu_14_100}
\end{figure*}

We also study the next-to-leading order predictions for $N\nu$ final state mediated by the $Z$ boson. The scale dependent cross-sections at $14$ TeV LHC and $100$ TeV hadron collider are given in Fig.~\ref{fig:scale_dep_Nnu}.  We consider the heavy neutrino mass at $100$ GeV and $800$ GeV.  We used the same scale dependence for the $N\nu$ final state as we did in $N\ell$ final state  for LO and NLO respectively.  At 14 TeV the LO cross-section at $m_{N}=100$ GeV  increases with $\xi$ at a faster rate than the NLO process whereas for $m_{N}=800$ GeV the LO cross-section decreases with the increase in $\xi$  with a faster rate than that in NLO. On the other hand at $100$ TeV collider the LO process at $m_{N}=100$ GeV takes over the NLO for $\xi > 1.9$. For $m_{N}=800$ GeV  the LO cross-section increases with the increase in $\xi$ but the NLO cross-section remains more or less same for $\xi > 0.5$. 
Following our earlier demonstration on $N \ell$, we have also shown in Fig.~\ref{fig:x-sec_Nnu_14_100}
the cross-sections for the $N\nu$ final state as a function of $m_{N}$ at $14$ TeV LHC and $100$ TeV collider along with the corresponding scale dependence bands.


\section{Scale dependent kinematic distributions in trilepton channel}
\label{sec:result}

%
\begin{figure*}[t]
\begin{center}

\includegraphics[scale=0.17]{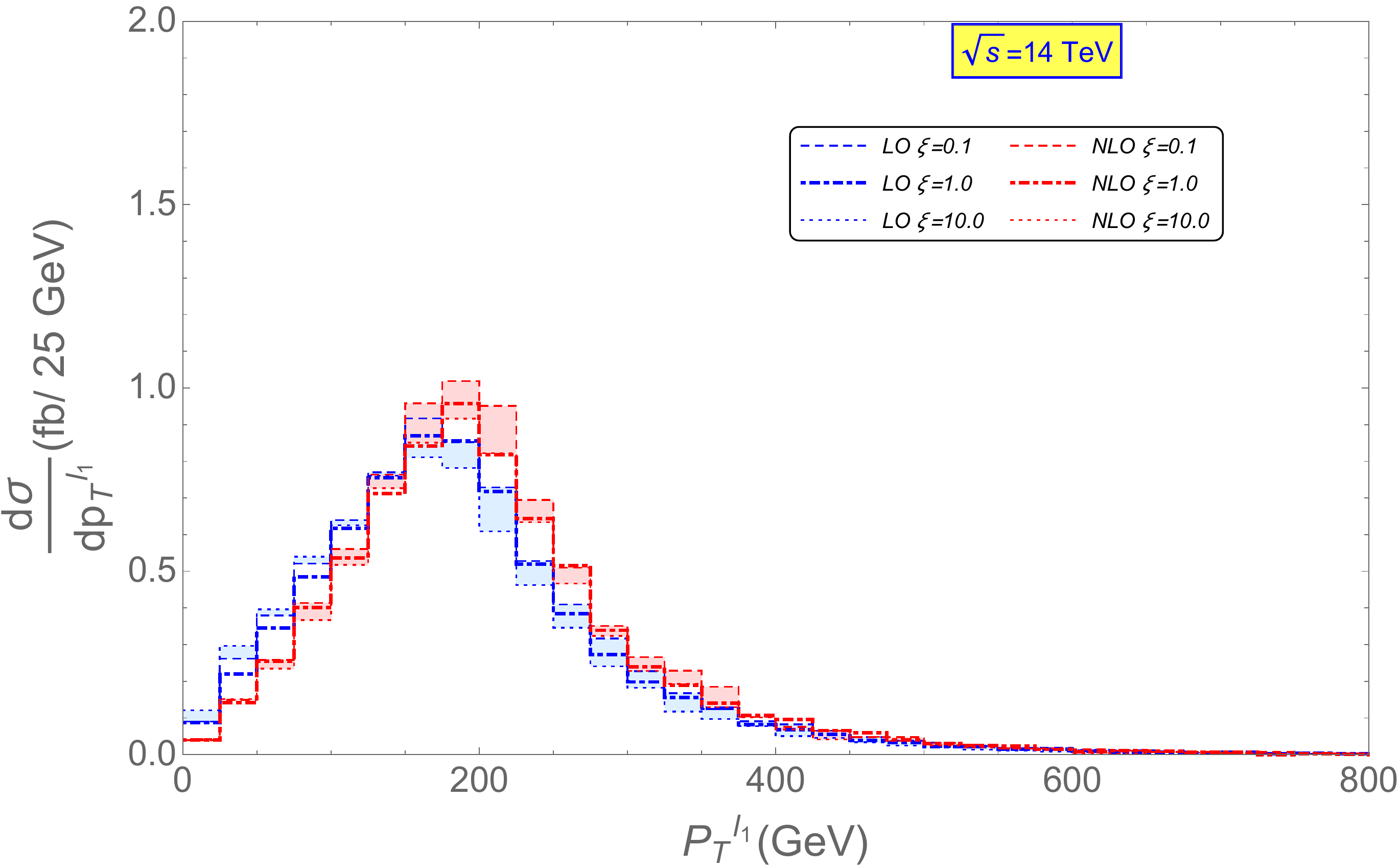}
\includegraphics[scale=0.17]{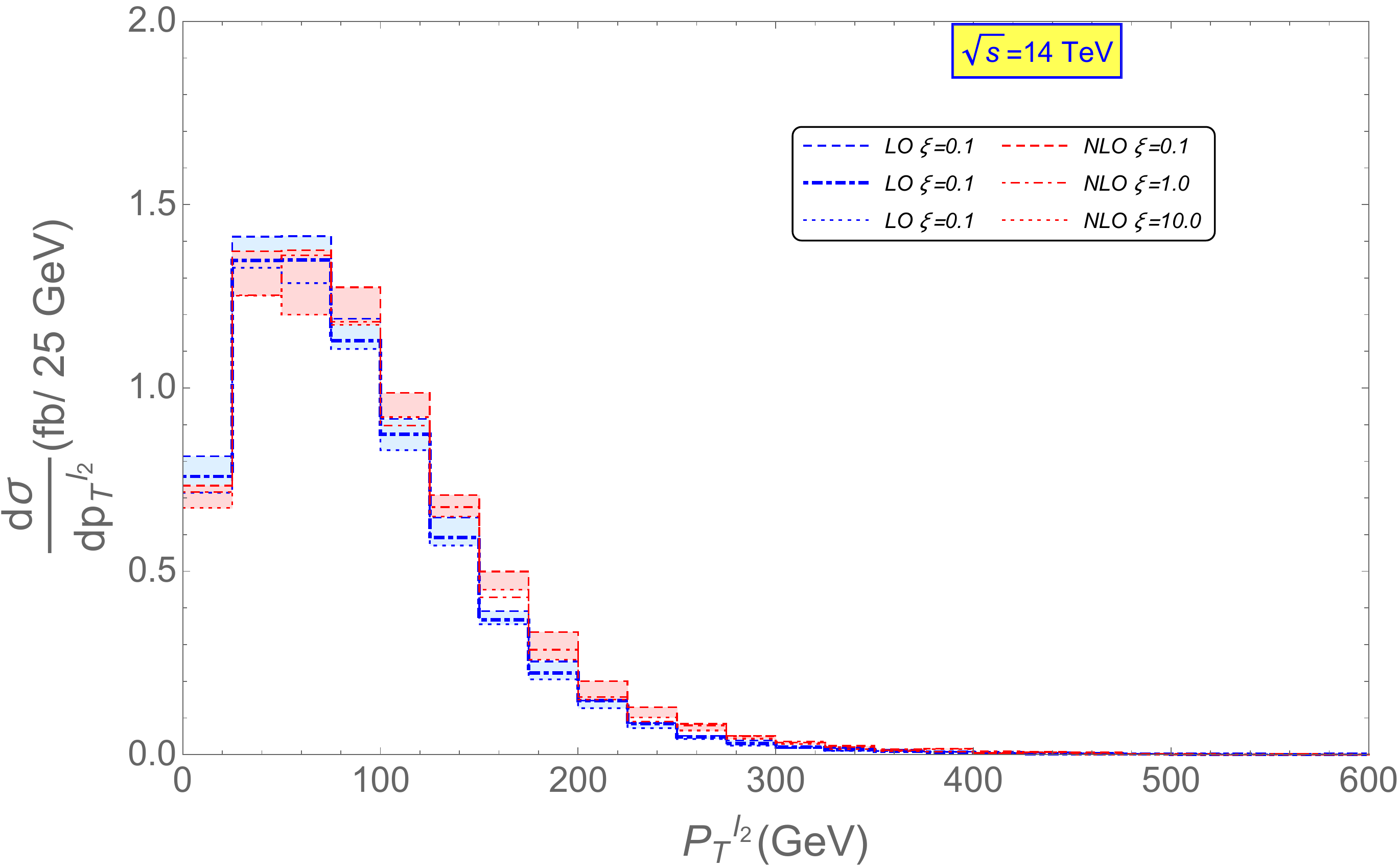}
\includegraphics[scale=0.17]{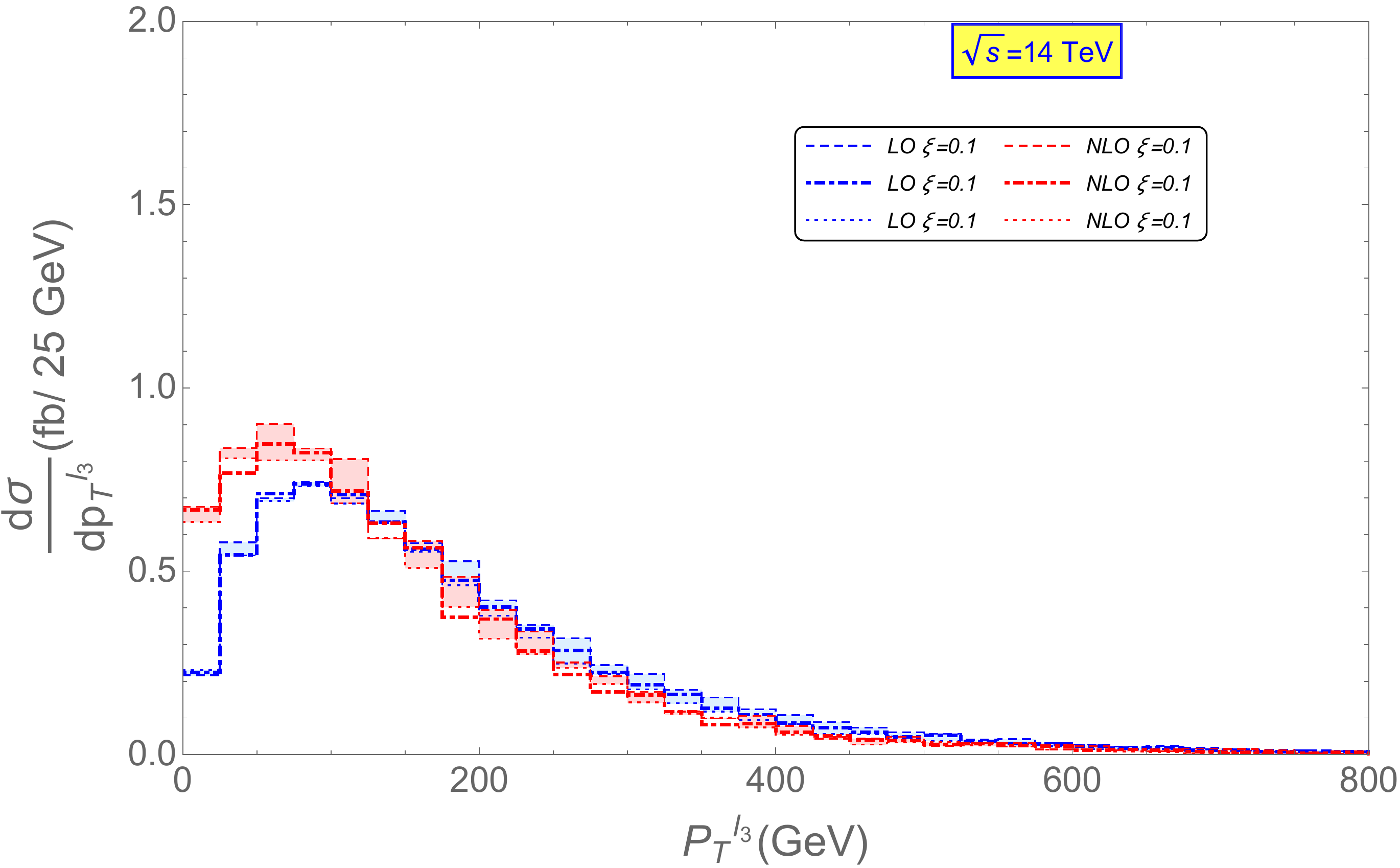}\\
\includegraphics[scale=0.17]{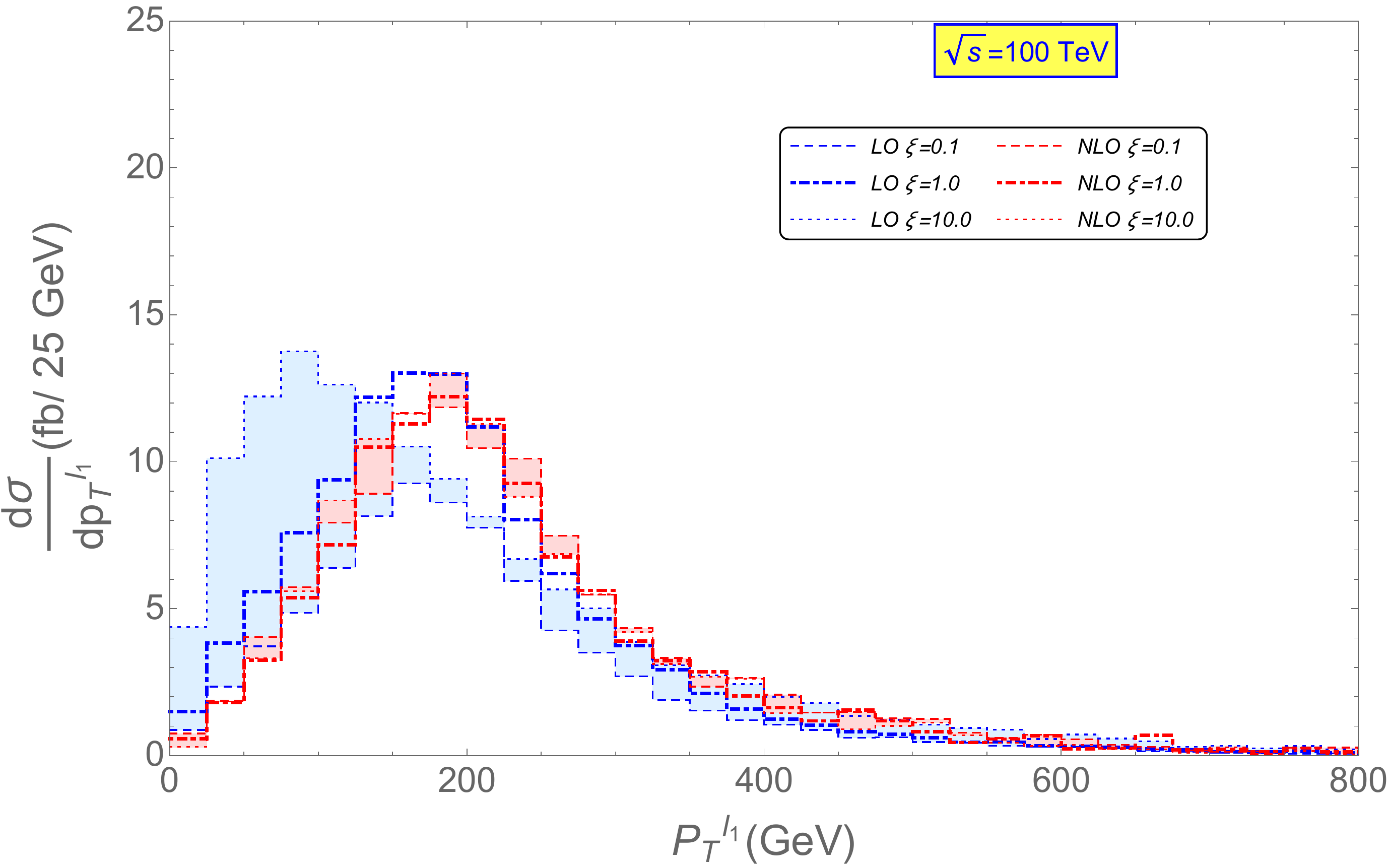}
\includegraphics[scale=0.17]{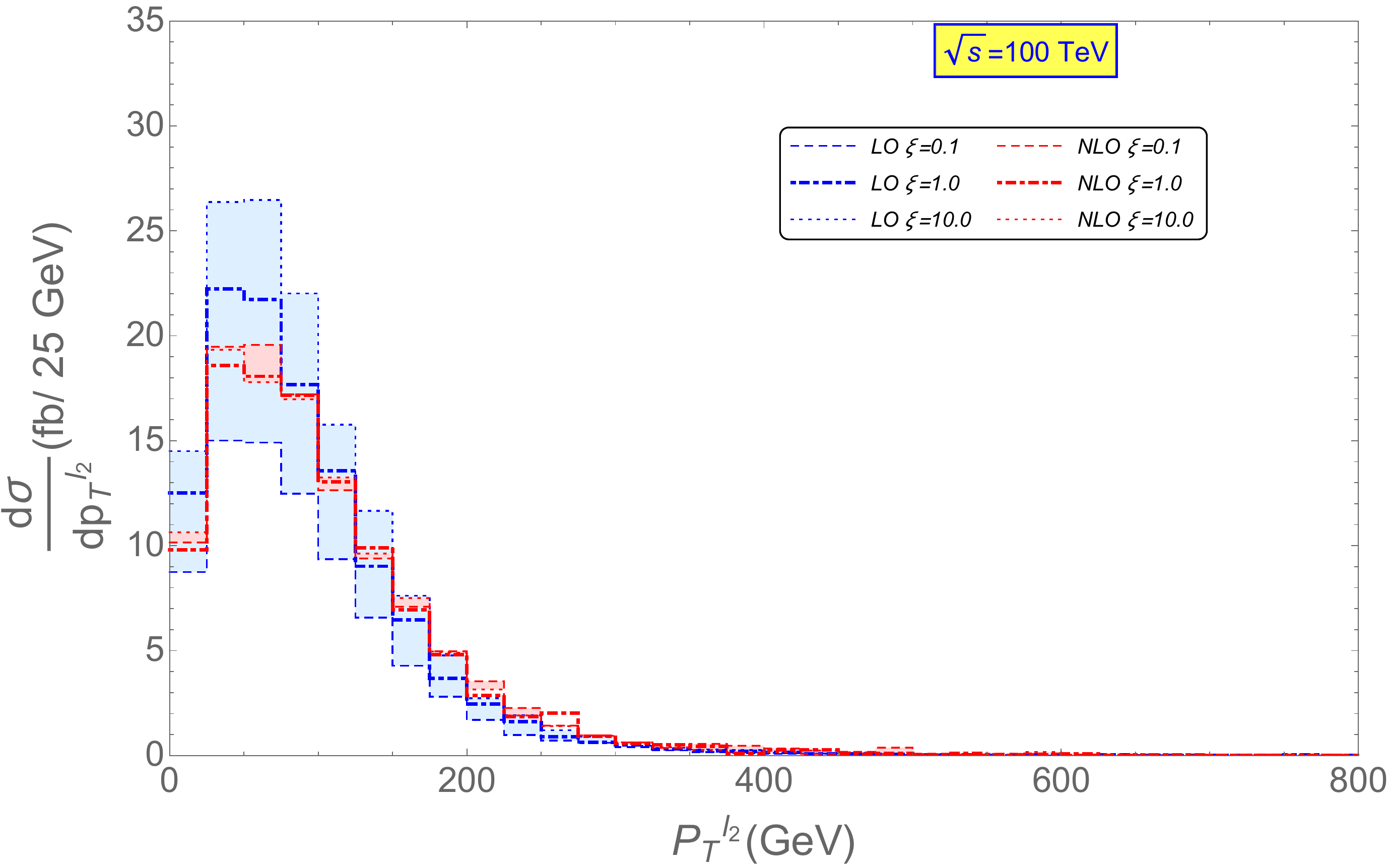}
\includegraphics[scale=0.17]{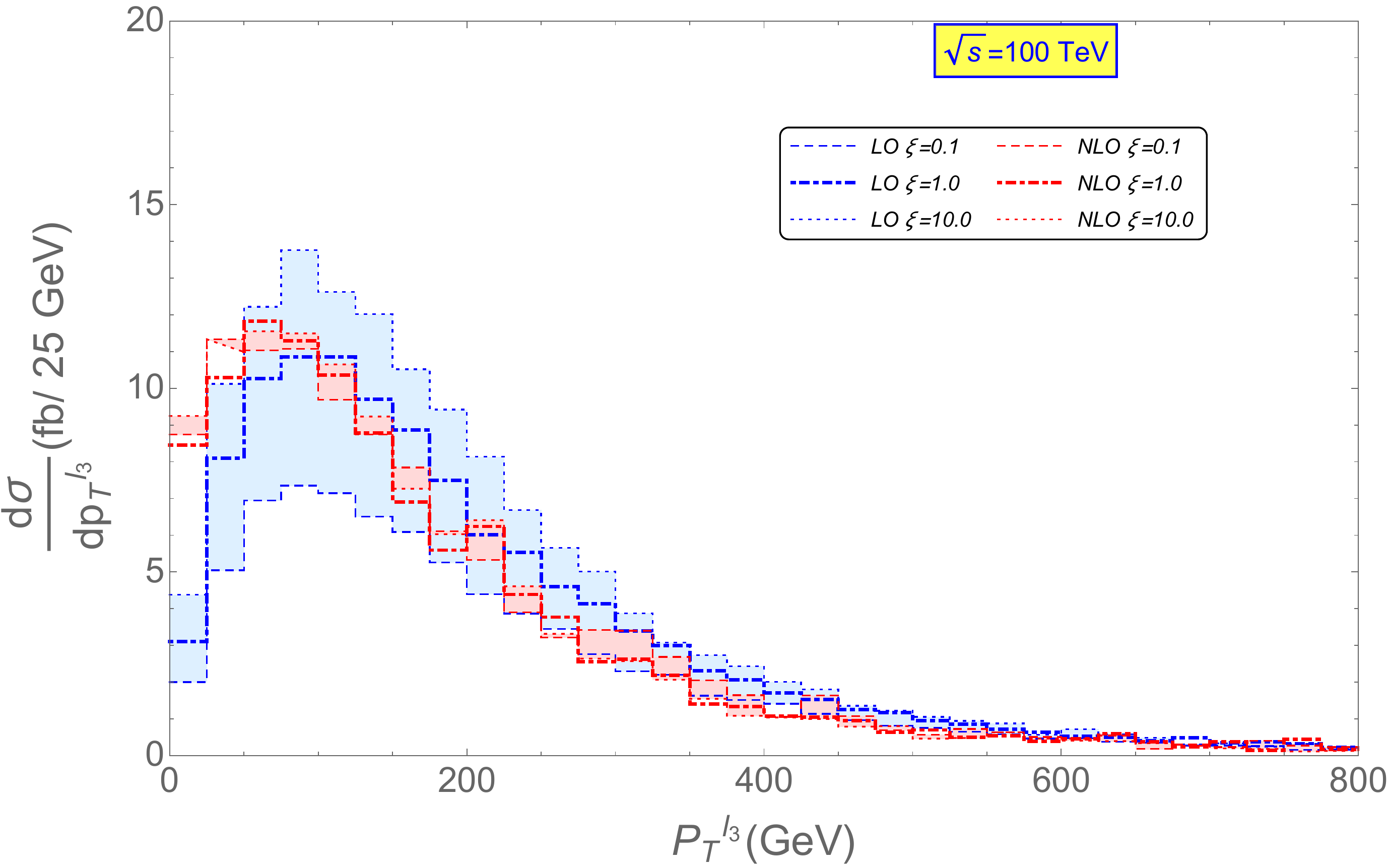}\\
\includegraphics[scale=0.17]{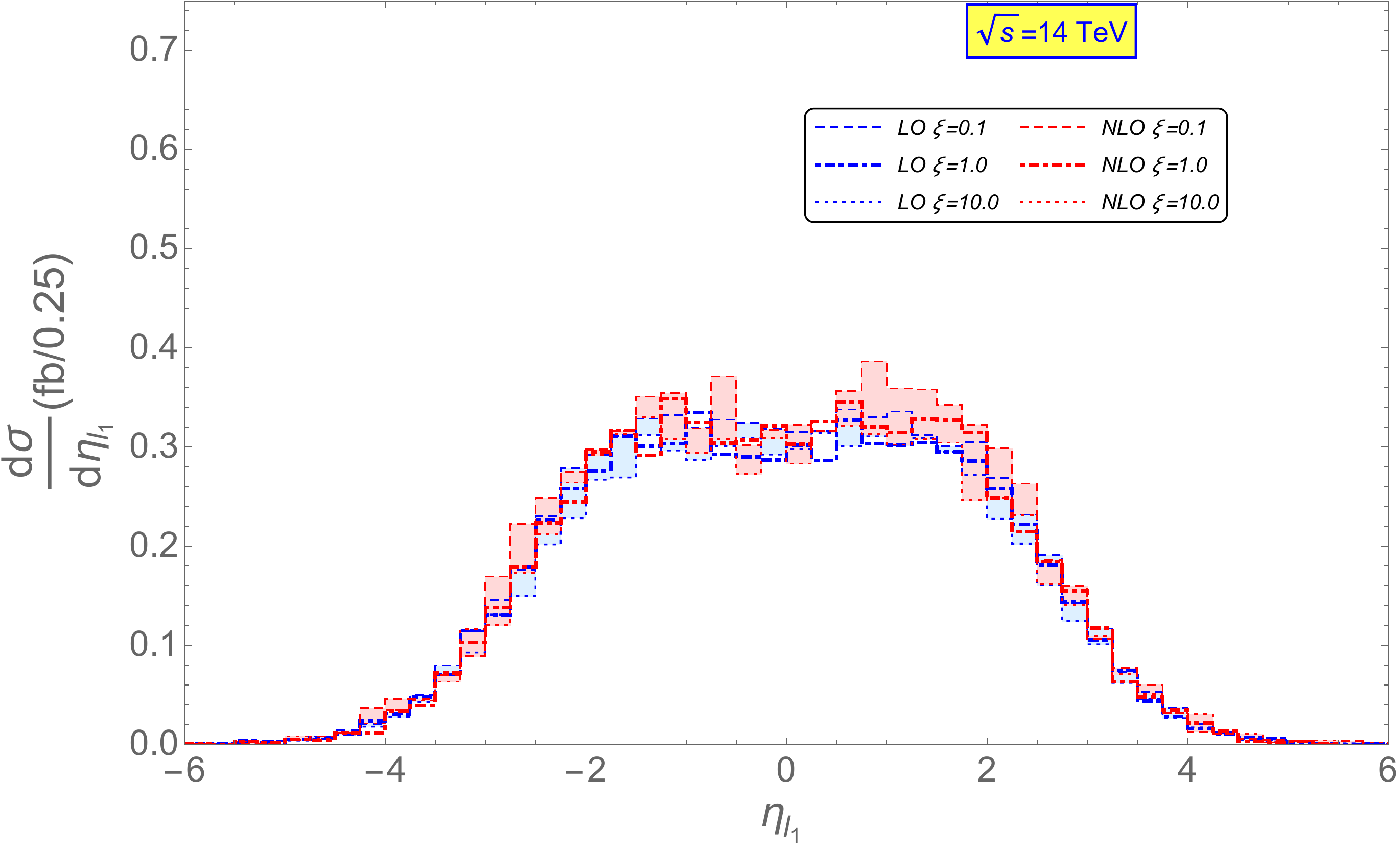}
\includegraphics[scale=0.17]{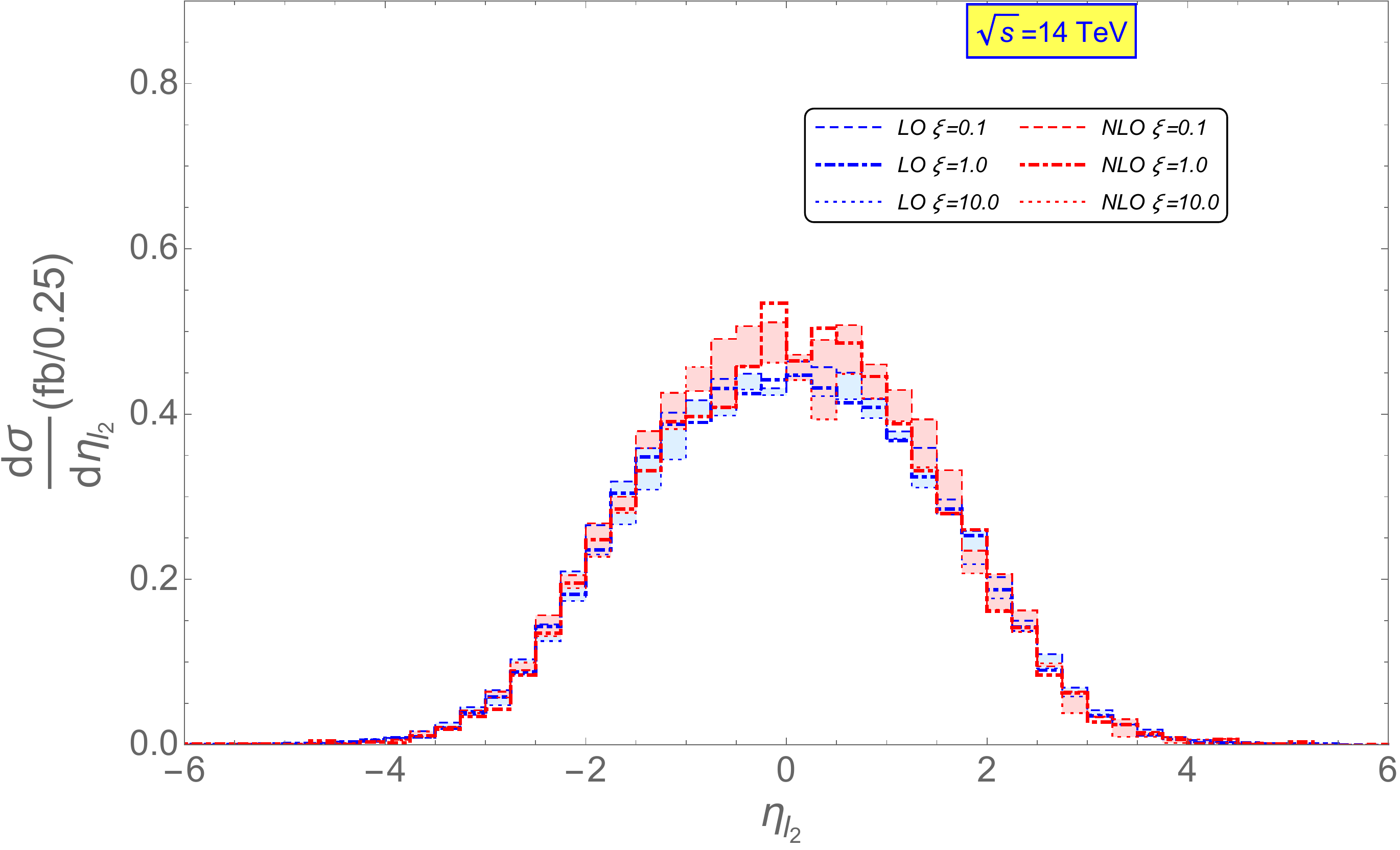}
\includegraphics[scale=0.17]{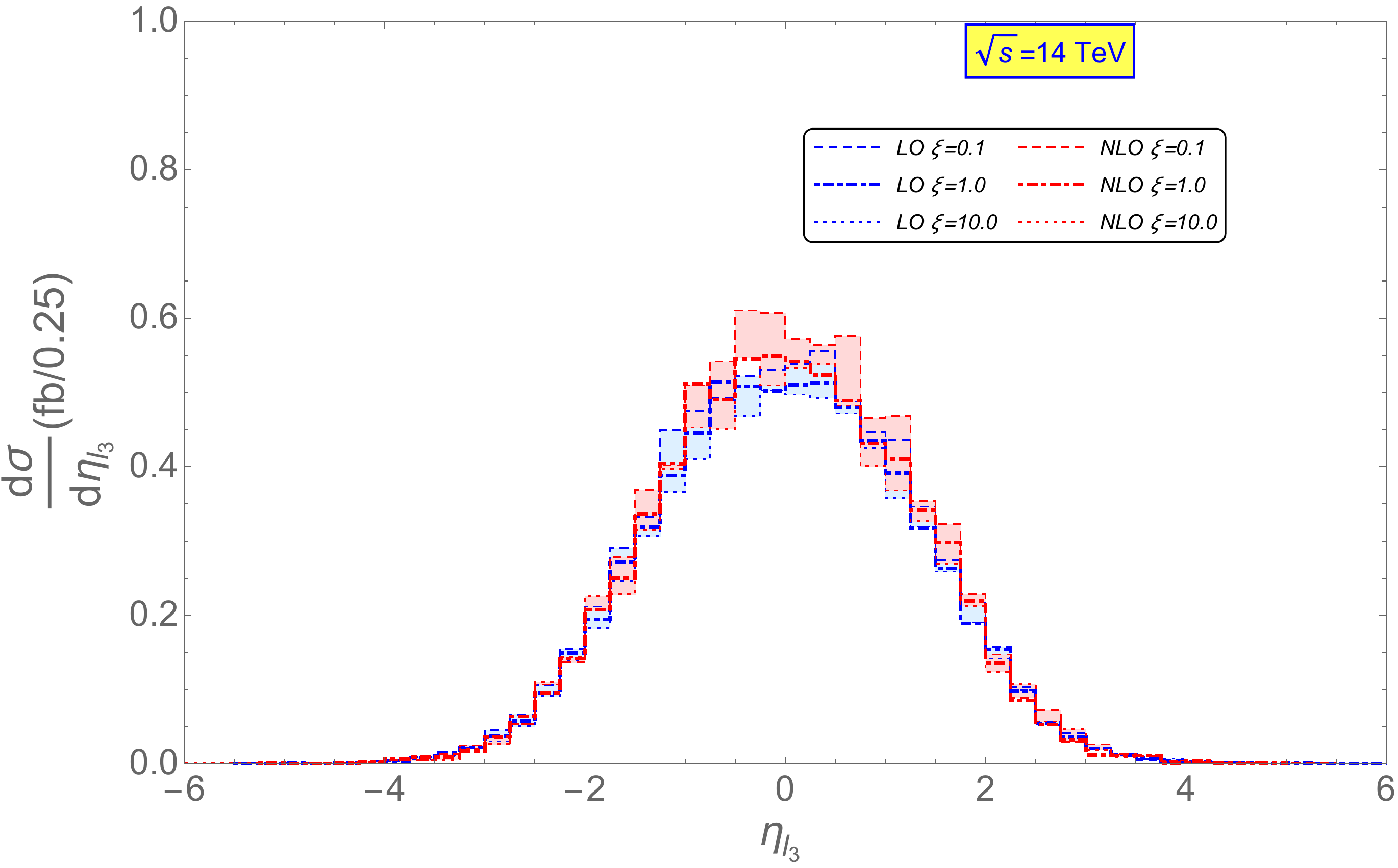}\\
\includegraphics[scale=0.17]{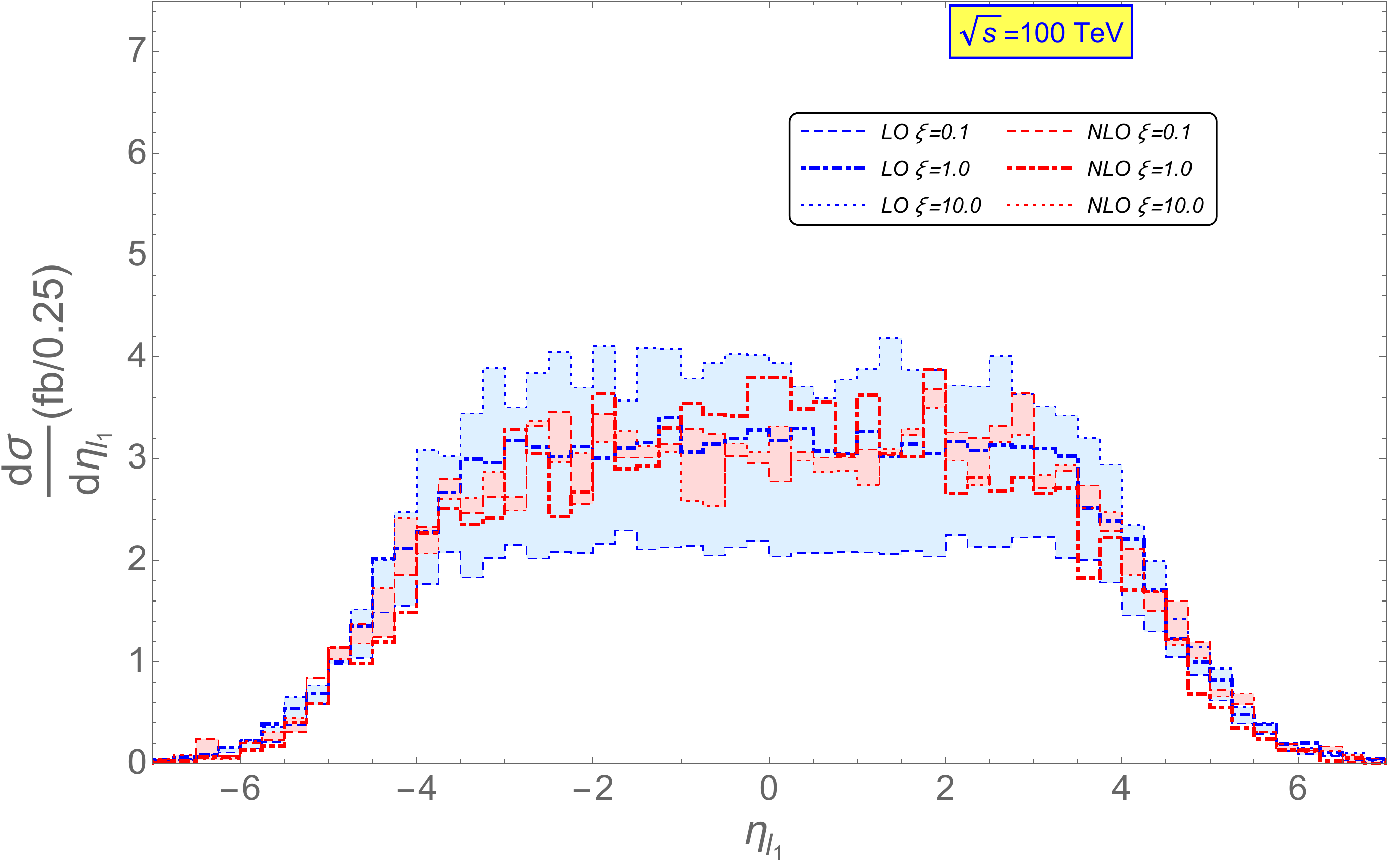}
\includegraphics[scale=0.17]{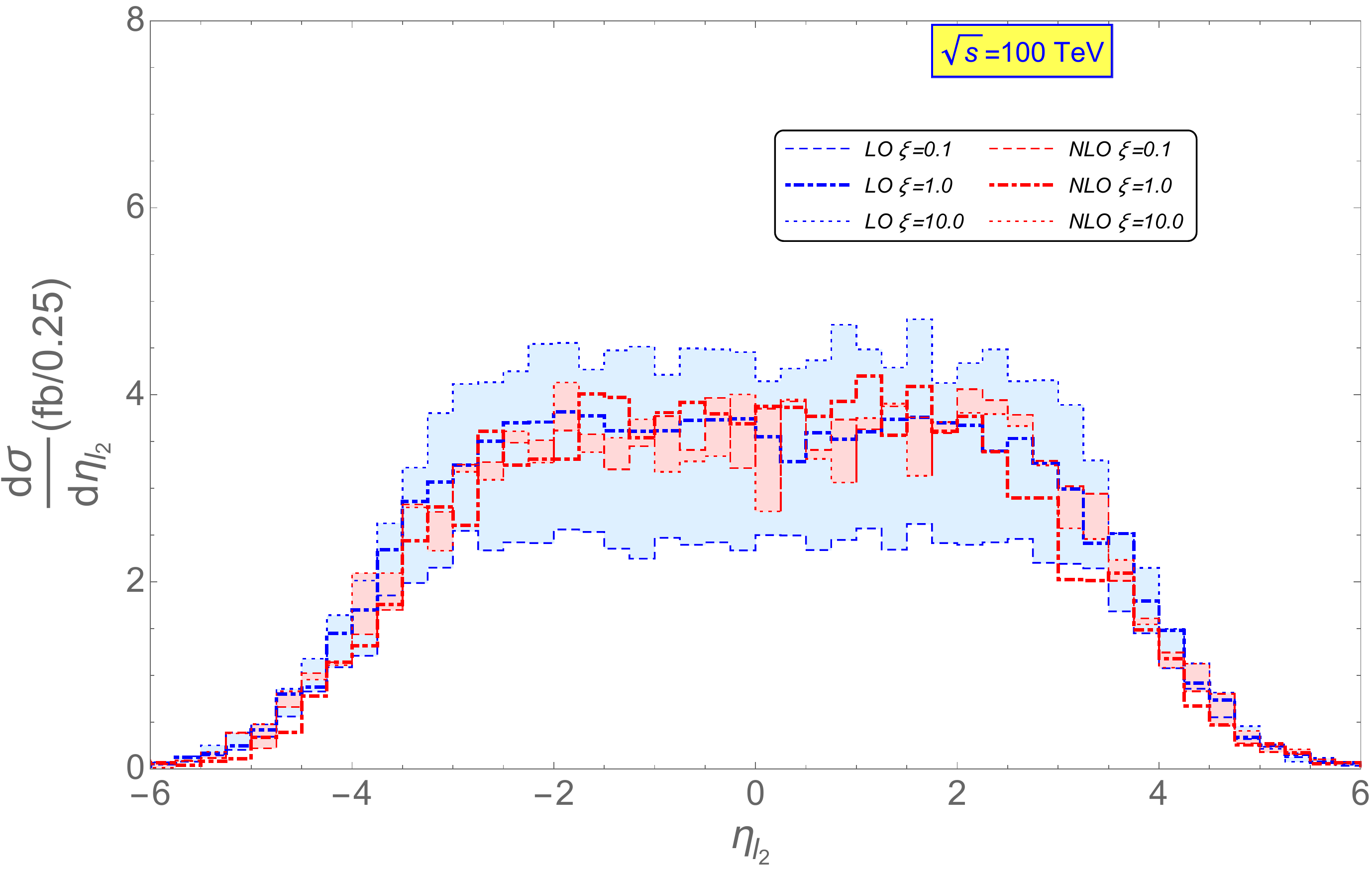}
\includegraphics[scale=0.17]{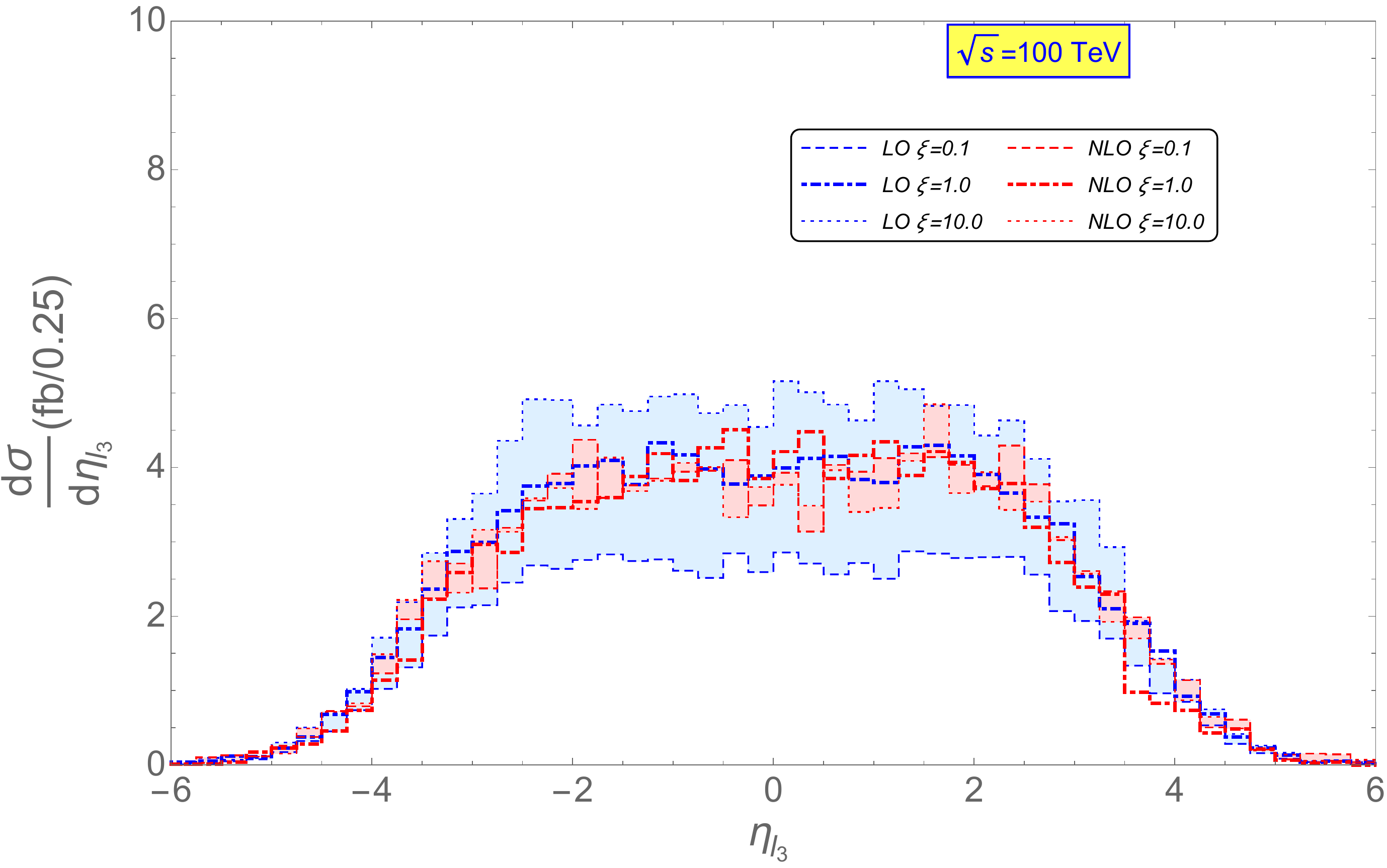}

\caption{Scale variation of the differential scattering cross-section as a function of the transverse momentum of leptons($p_{T}^{\ell_{i}}$, $i=1, 2, 3$)  and pseudorapidity of leptons($\eta^{\ell_{i}}$, $i=1, 2, 3$) in case of trilepton production channel for $m_{N}=400$ GeV. The first row corresponds to the $p_{T}^{\ell_{i}}$ distributions at $\sqrt{s}=14$ TeV LHC whereas the second row represents the same at $\sqrt{s}=100$ TeV collider. The third row corresponds to the $\eta^{\ell^{i}}$ distributions at $\sqrt{s}=14$ TeV LHC whereas the fourth row represents the same at the $\sqrt{s}=100$ TeV collider. The differential scattering cross-section distributions are normalized by $|V_{\ell N}|^2$.}
\label{fig:pTl_14_100}
\end{center}
\end{figure*}

\begin{figure*}[t]
\begin{center}
\includegraphics[scale=0.17]{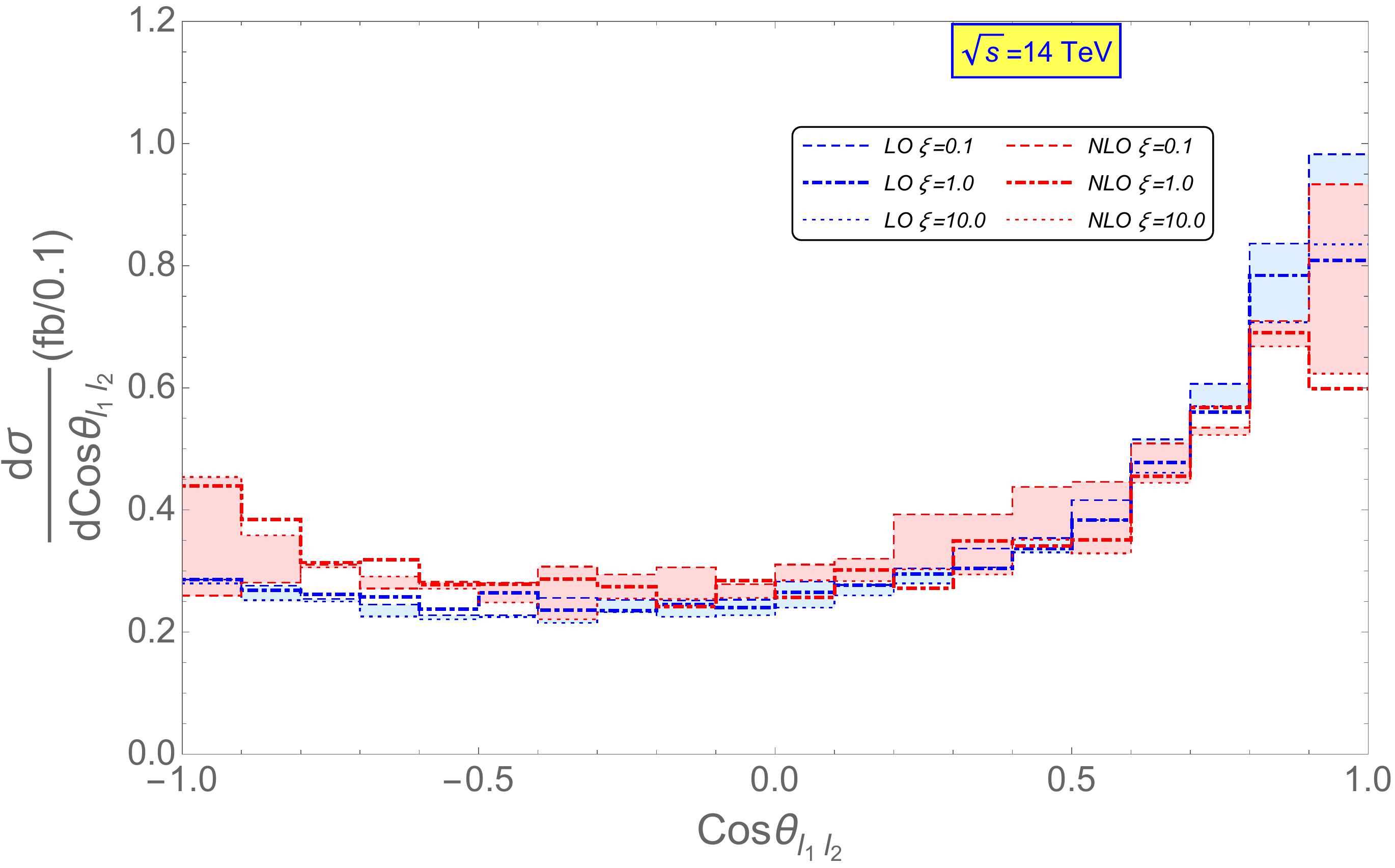}
\includegraphics[scale=0.17]{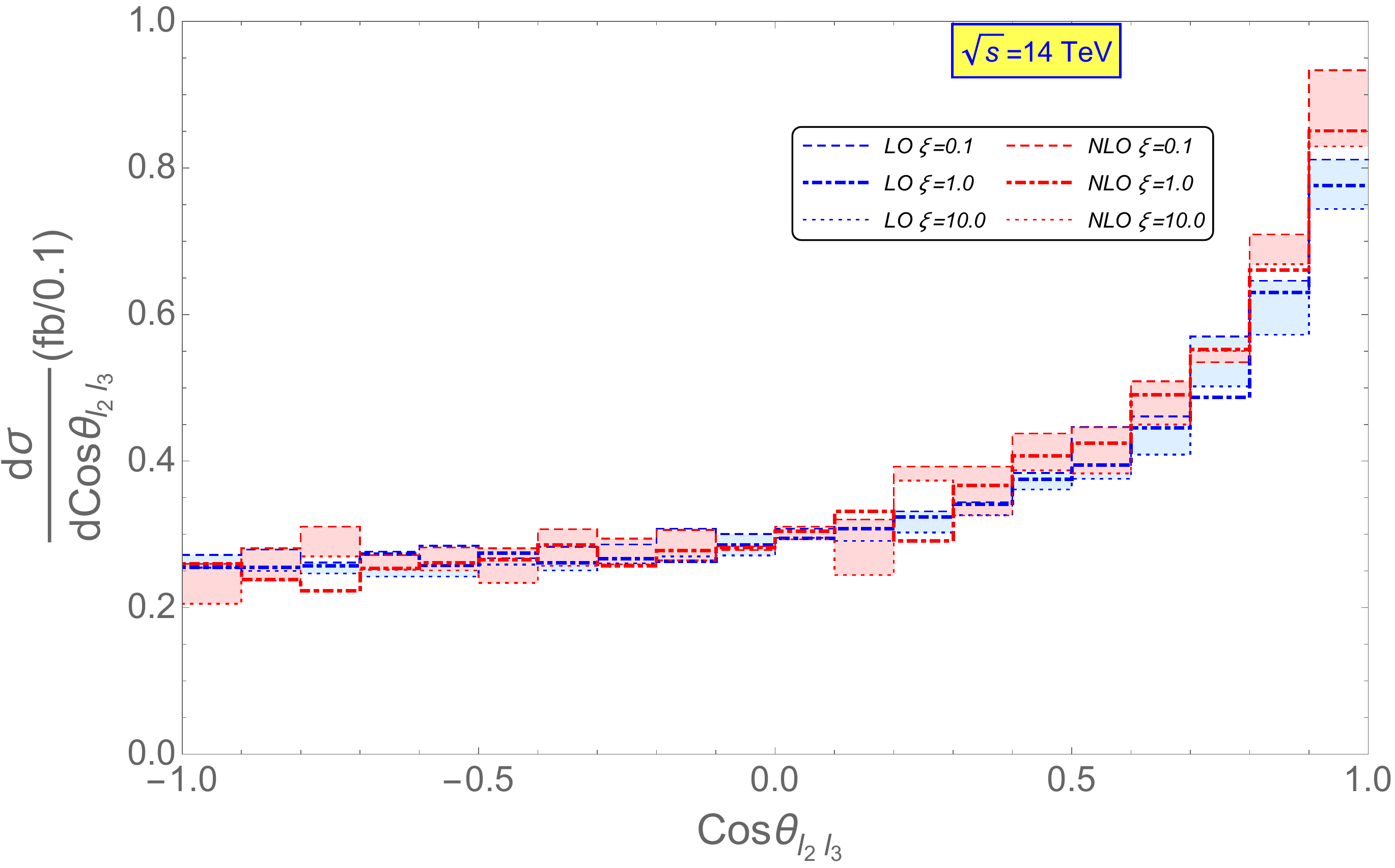}
\includegraphics[scale=0.17]{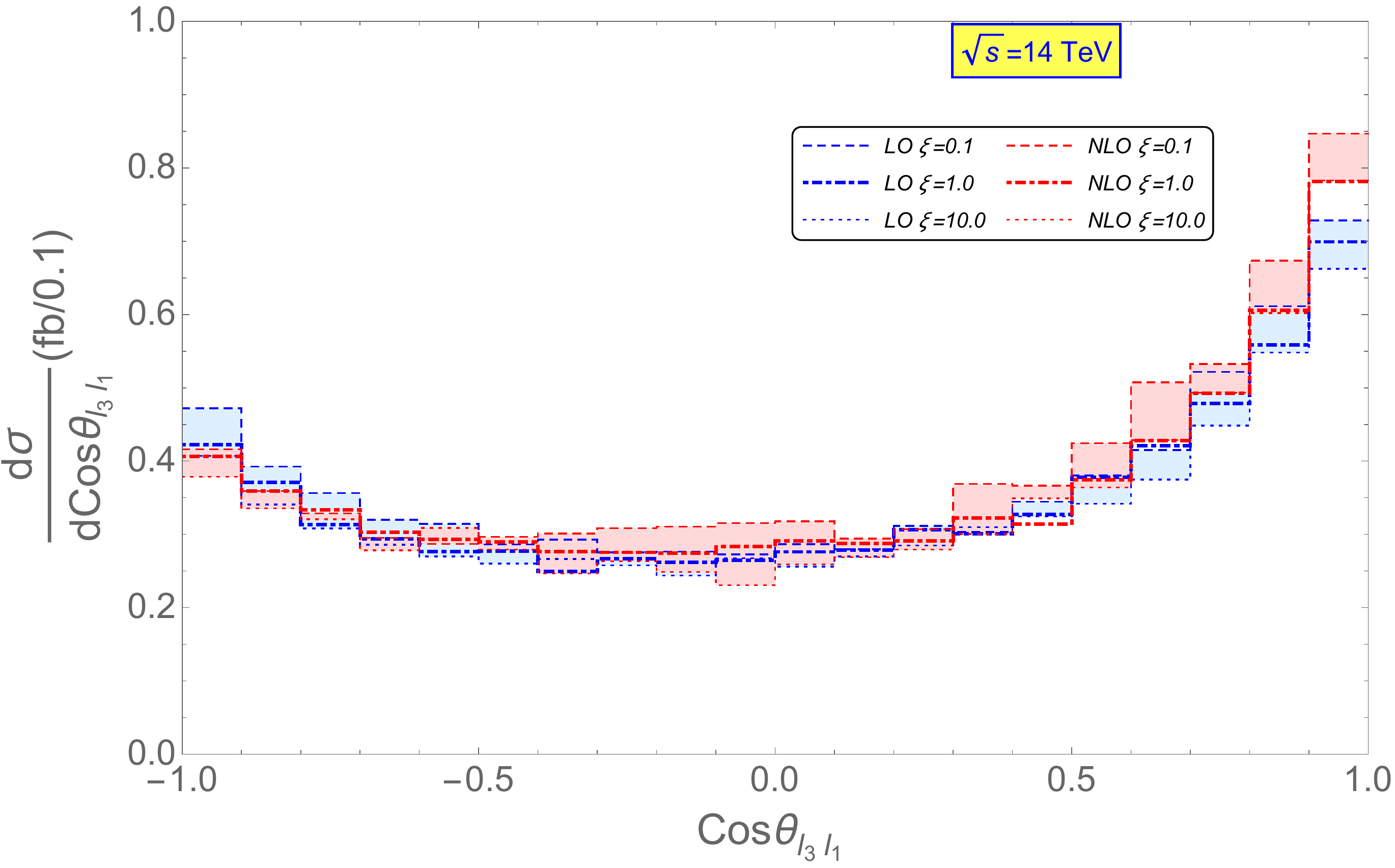}\\
\includegraphics[scale=0.17]{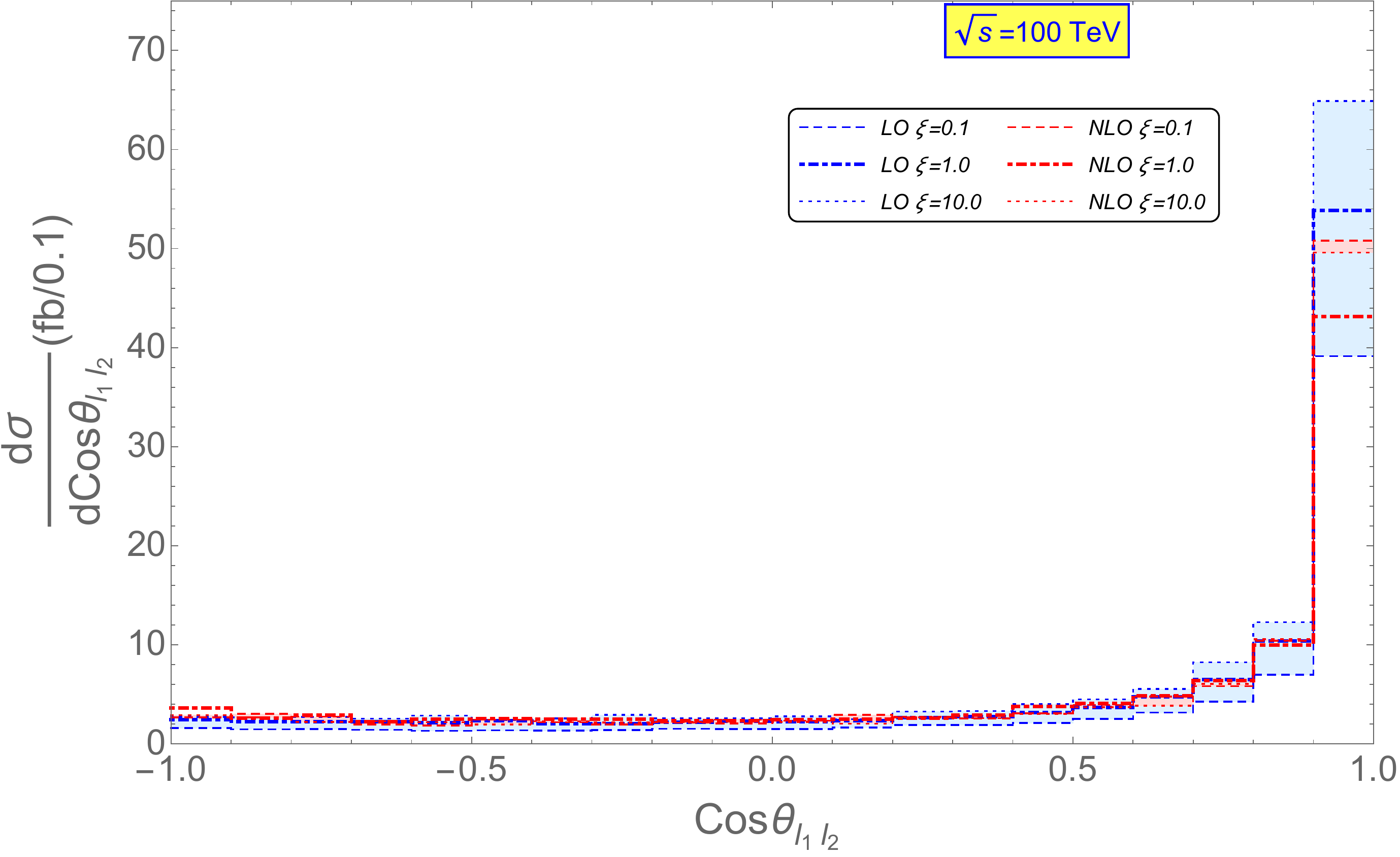}
\includegraphics[scale=0.17]{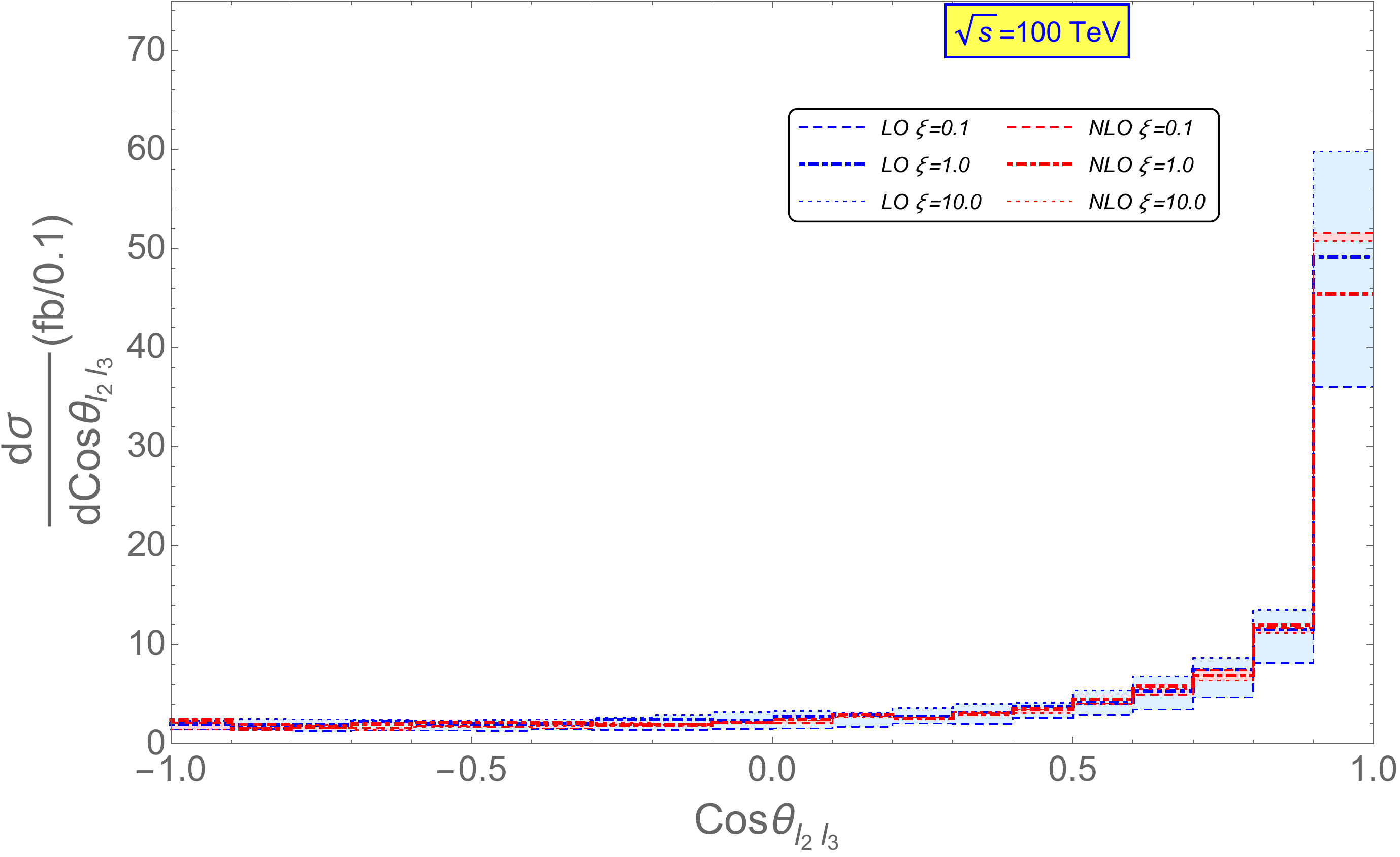}
\includegraphics[scale=0.17]{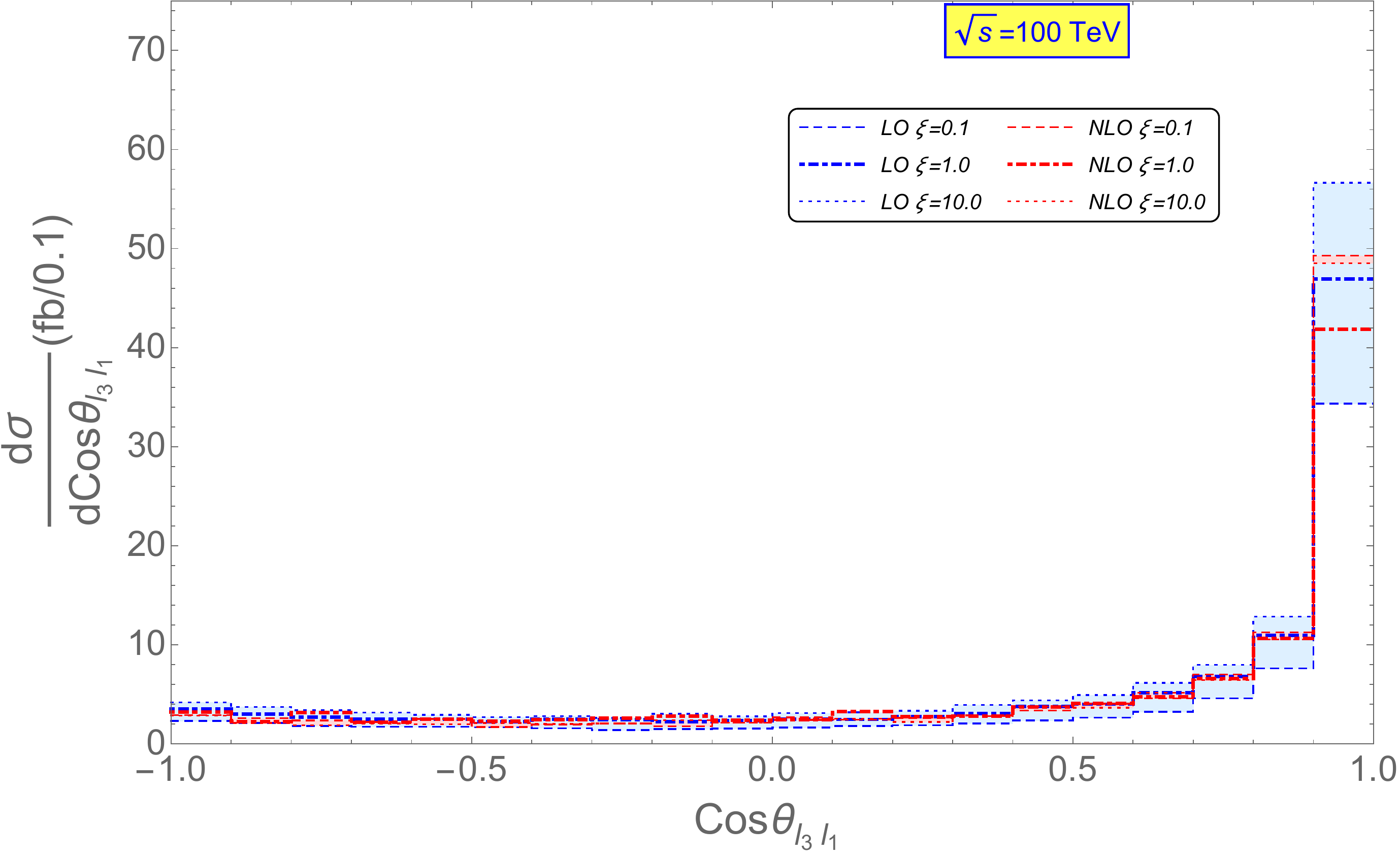}\\
\includegraphics[scale=0.17]{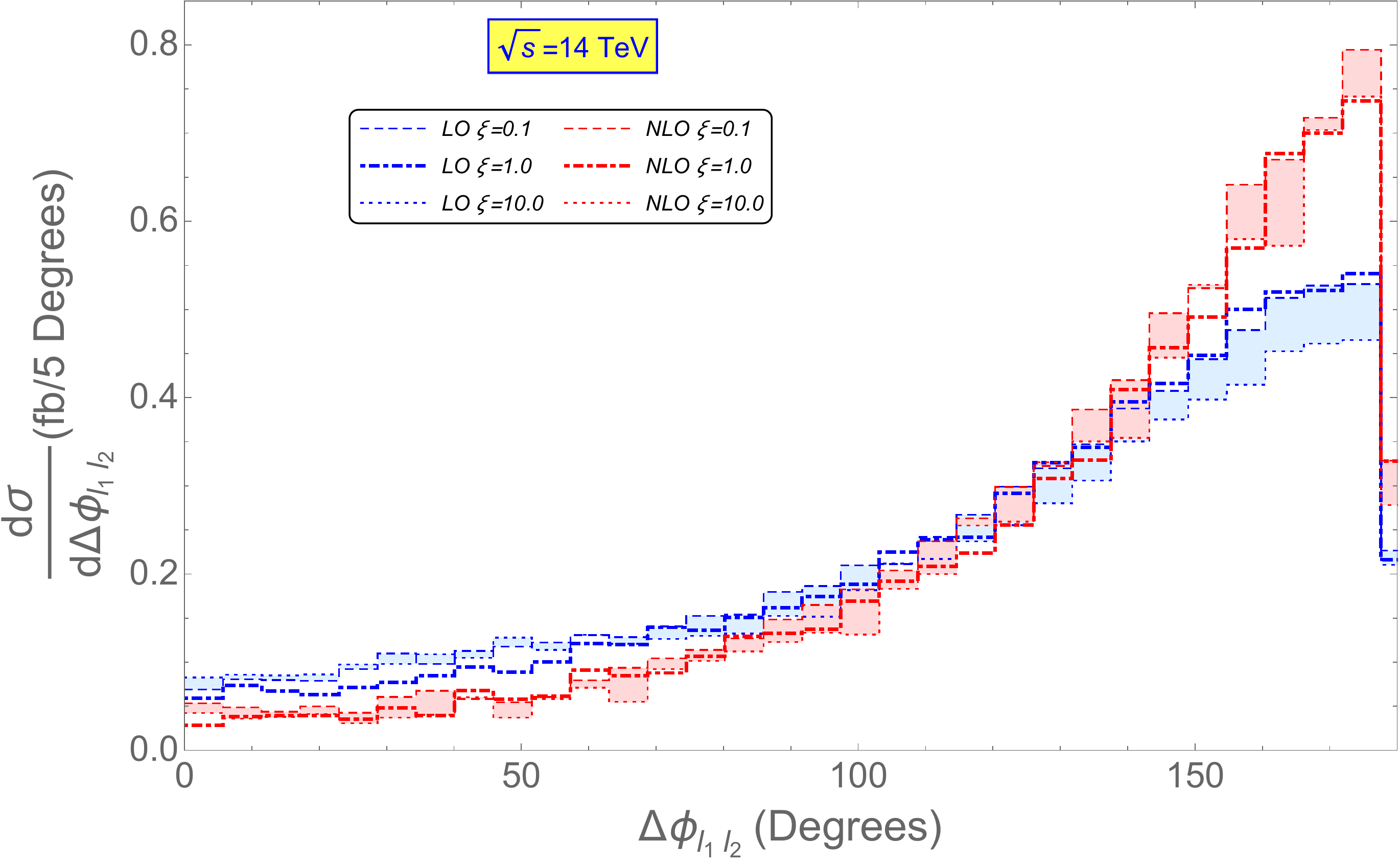}
\includegraphics[scale=0.17]{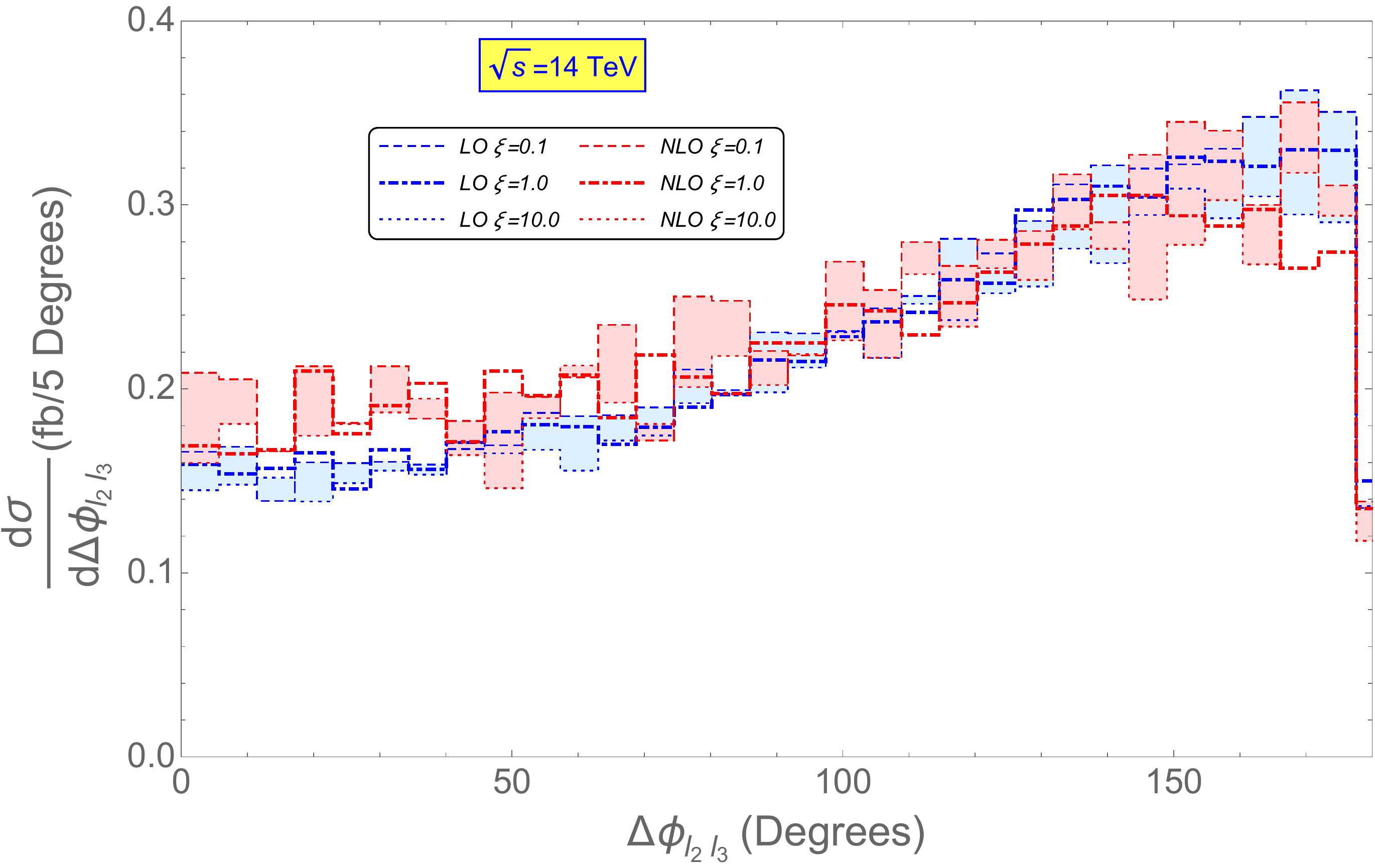}
\includegraphics[scale=0.17]{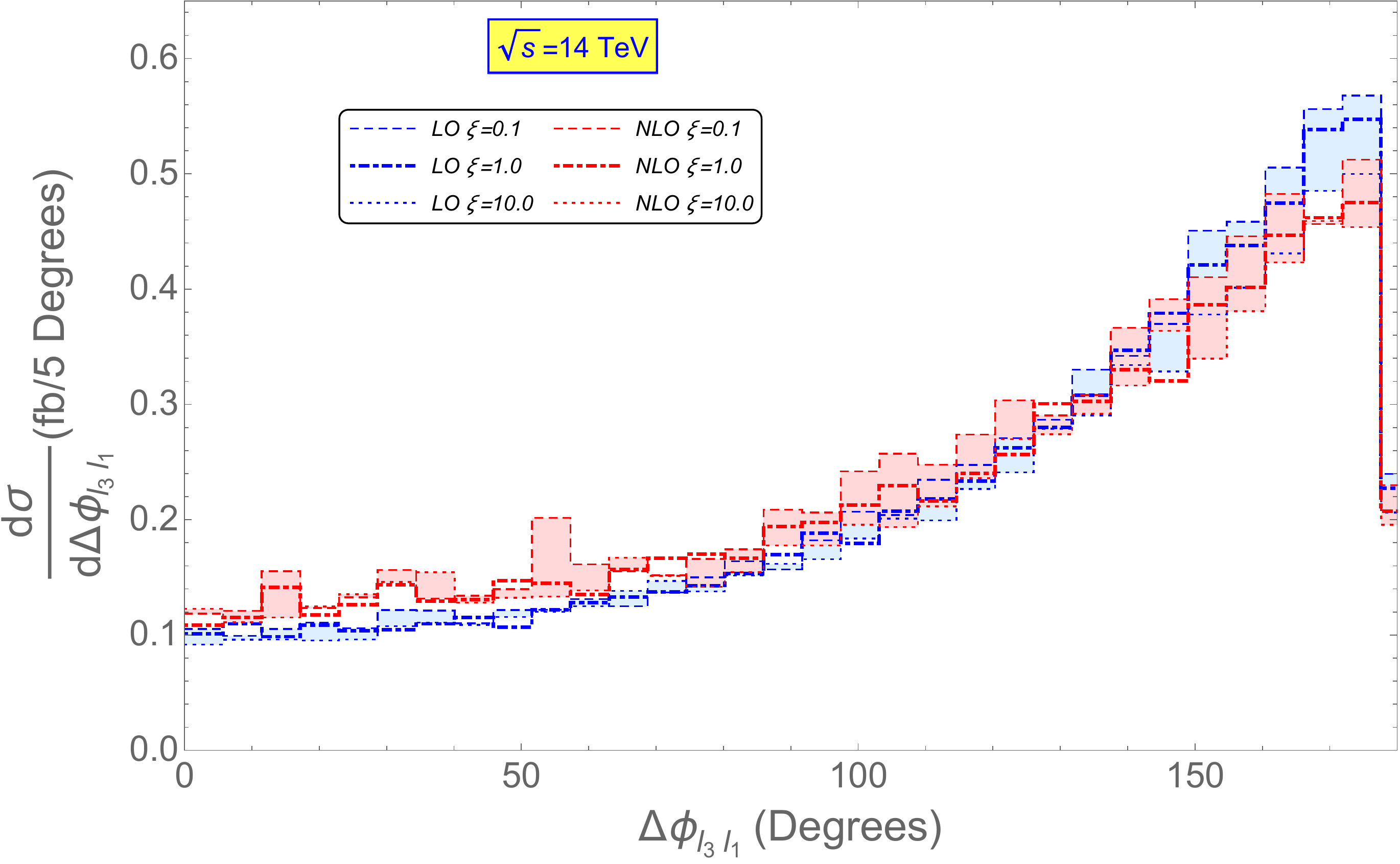}\\
\includegraphics[scale=0.17]{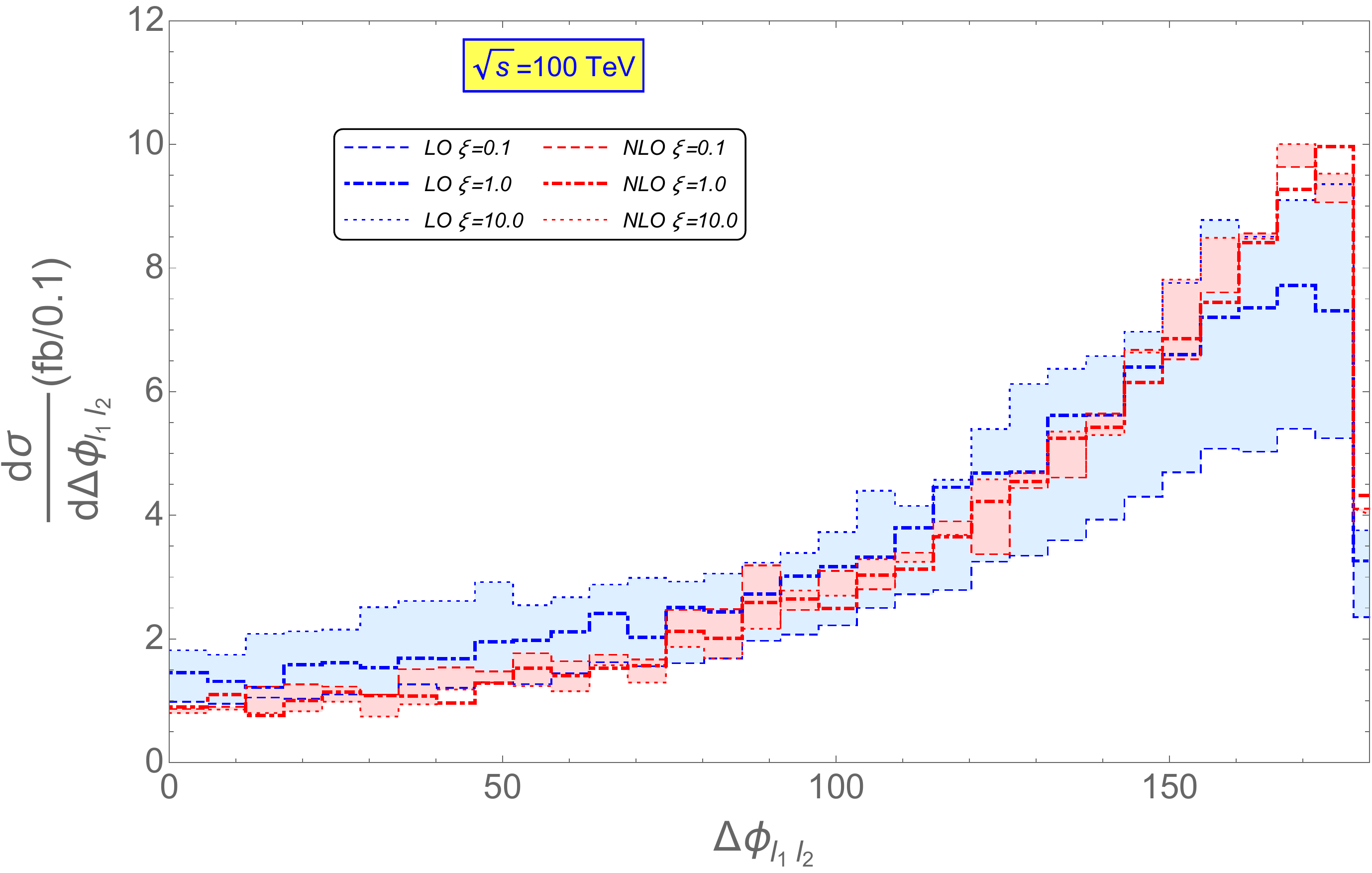}
\includegraphics[scale=0.17]{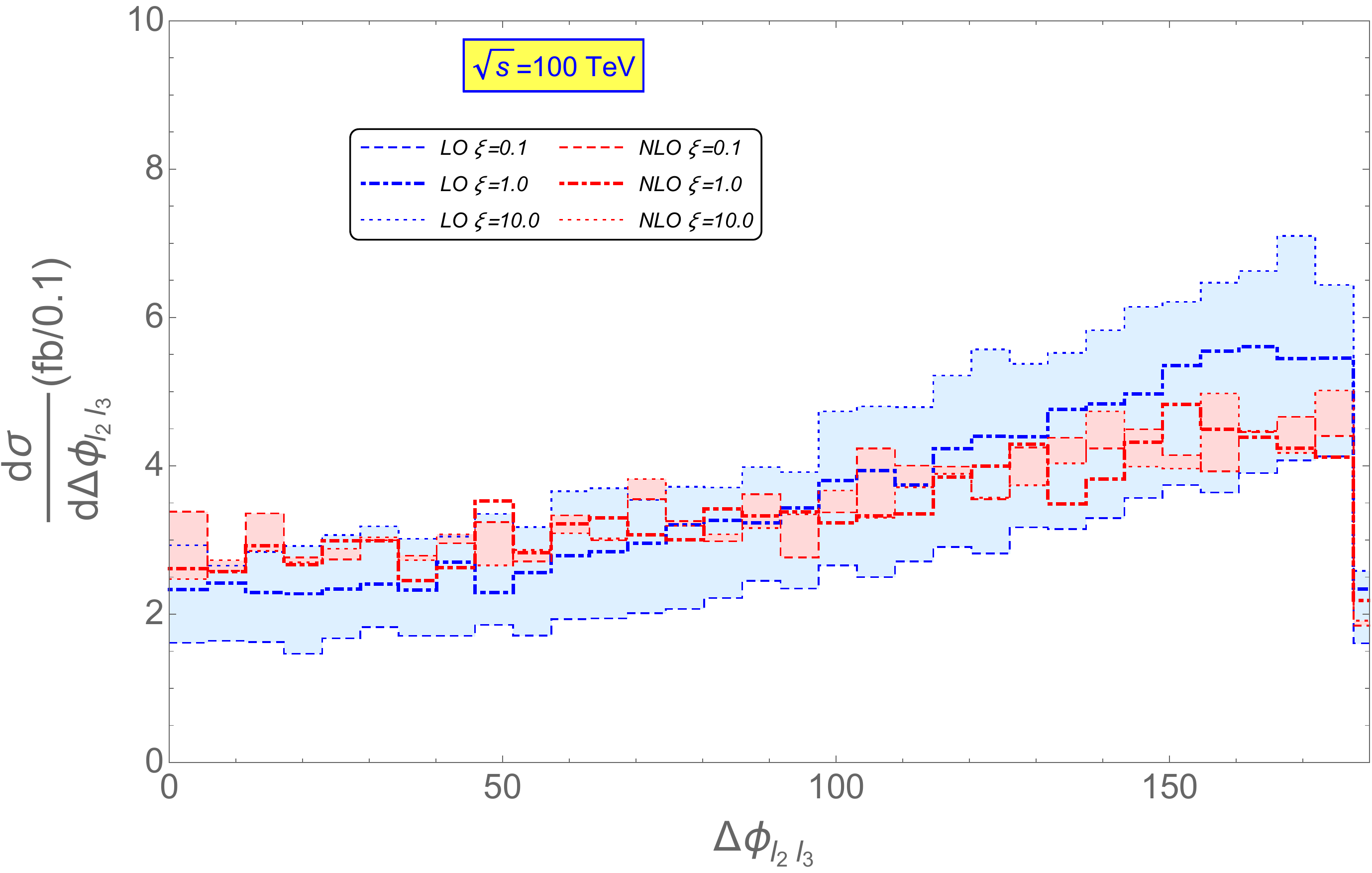}
\includegraphics[scale=0.17]{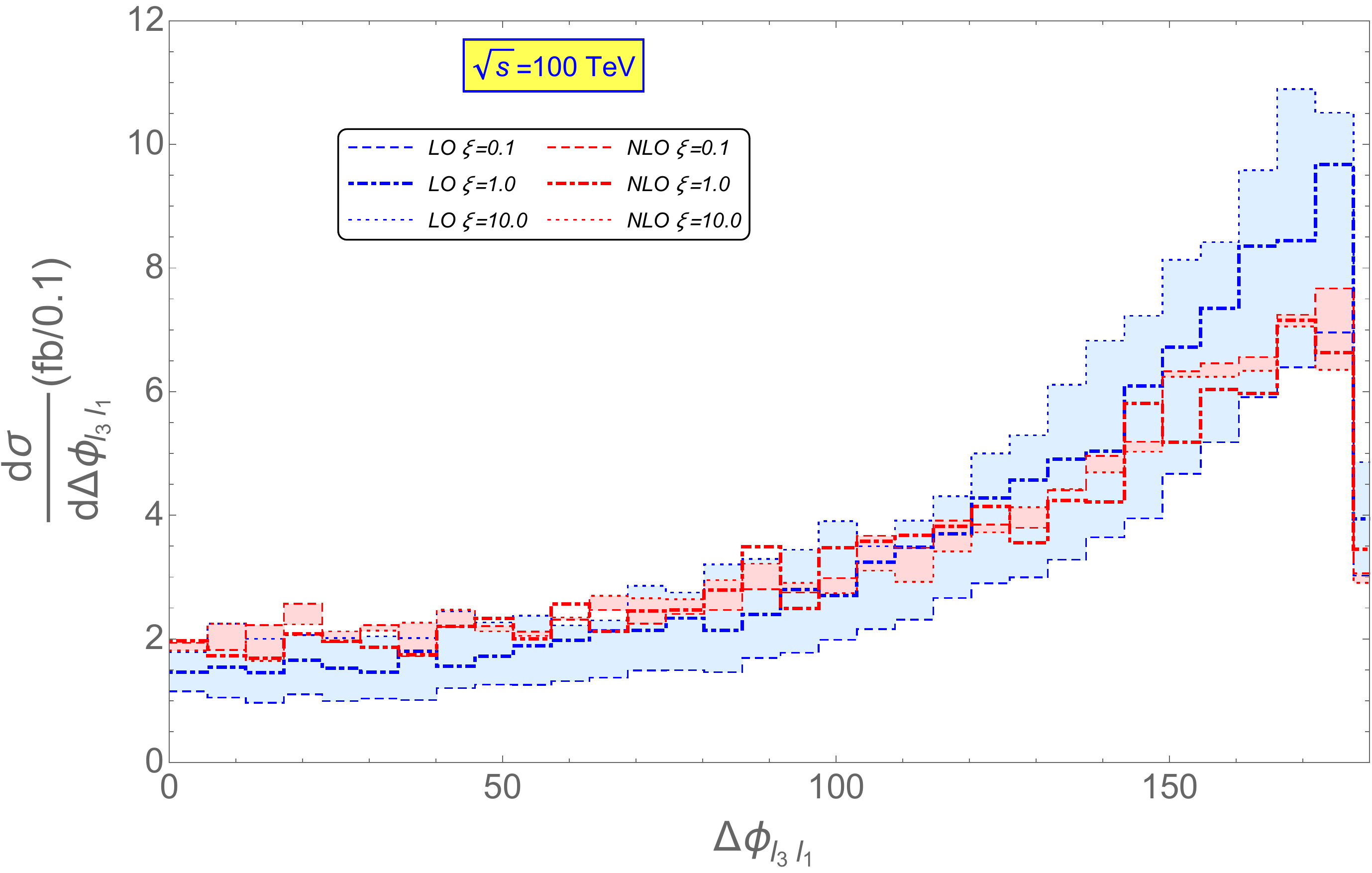}
\caption{Scale variation of the differential scattering cross-section distribution with respect to $\cos\theta_{\ell_{i}\ell_{j}}$ of the leptons with $i=1$, $2$, $3$, $j=1$, $2$, $3$ and $i\neq j$ and $\Delta\phi_{\ell_{i}\ell_{j}}$ of the leptons with $i=1$, $2$, $3$, $j=1$, $2$, $3$ and $i\neq j$ for the trilepton production channel for $m_{N}=400$ GeV. The first row corresponds to $\cos\theta_{\ell_{i}\ell_{j}}$ at $\sqrt{s}=14$ TeV LHC whereas the second row represents the $\sqrt{s}=100$ TeV collider. The third row corresponds to $\Delta\phi_{\ell_{i}\ell_{j}}$ at $\sqrt{s}=14$ TeV LHC whereas the fourth row represents the same at $\sqrt{s}=100$ TeV collider. The differential cross-section distributions are normalized by $|V_{\ell N}|^2$.}
\label{fig:costhetall_14_100}
\end{center}
\end{figure*}

In this section we consider the pseudo-Dirac heavy neutrino production at the hadron collider and study its decay process for the $14$ TeV LHC and $100$ TeV collider. For further demonstration we choose the heavy neutrino with mass $m_{N}=400$  GeV. Different kinematic distributions are constructed both for the LO and NLO calculations by choosing both the factorization scale as well as renormalisation scale varying simultaneously (as in Eq.\ref{muR}) with a scale factor $\xi$ between $0.1$ and $10$.
The pseudo-Dirac heavy neutrino is involved in the inverse seesaw mechanism to generate the neutrino mass. The collider phenomenology of inverse seesaw mechanism has been studied in \cite{Das:2012ze, Das:2014jxa, Das:2015toa, Das:2016akd} for the LO process with trilepton final state
for a fixed scale. In case of inverse seesaw mechanism the Yukawa coupling could be high enough to enhance the heavy neutrino production. However, due to the small lepton number violating parameter\footnote{See, \cite{Das:2012ze} for the detailed discussion about the
 smallness of the lepton number violating parameter in case of inverse seesaw mechanism.} the heavy neutrino becomes pseudo-Dirac.  In this analysis we consider the Single Flavor (SF) scenario, where only one heavy neutrino is light and accessible to the high energy colliders. 
  It couples with one generation of the lepton flavor. For simplicity, we consider that the heavy neutrino is coupled with the second generation of the lepton flavor\footnote{Another possibility could be possible where we can introduce two generations of the degenerate heavy neutrinos 
 and each generation couples with the single, corresponding lepton flavor which we can name as Flavor Diagonal(FD) case. See, \cite{Das:2014jxa, Das:2015toa} for the study on FD case.}. As a result the golden channel for the final state signal is the trilepton plus missing energy which is given by
\bea
pp && \rightarrow \ell_{1}^{+} N, N \rightarrow \ell_{2}^{-} W^{+}, W^{+} \rightarrow \ell_{3}^{+} \nu_{\ell_{3}} \nonumber \\
pp  && \rightarrow \ell_{1}^{-} \overline{N}, \overline{N} \rightarrow \ell_{2}^{+} W^{-}, W^{-} \rightarrow \ell_{3}^{-} \overline{\nu_{\ell_{3}}}. 
\label{trilep}
\eea
Within the same set of the model parameters as described in the last section,
we show the differential distributions of scattering cross-section as a function of the transverse momenta for these three leptons separately,
 $p_{T}^{\ell_{i}}$ for $i=1$, $ 2$, $ 3$ in Fig.~\ref{fig:pTl_14_100} for the $14$ TeV LHC and $100$ TeV collider. 
In case of $14$ TeV, the NLO distributions dominate over the LO distributions in the high transverse momentum region for $\ell_{1}$ and $\ell_{2}$. 
Whereas for $\ell_{3}$, the NLO distributions dominate 
over the LO 
distributions quite impressively in the low transverse momentum region.
This behavior is due to inclusion of extra radiation at the NLO process as well as showering effect.
In the $100$ TeV case, similar situations are demonstrated, however leading order scale uncertainties are exceptionally large, again due to LO PDF sets as mentioned earlier.

We exhibit the pseudo-rapidity distributions in Fig.~\ref{fig:pTl_14_100}.  The 
$\eta^{\ell_{2, 3}}$ distributions 
at the $14$ TeV are sharper than that of $\eta^{\ell_{1}}$, 
resulting the production of these two leptons in the central region. 
The scale variations for the LO and NLO cases are not very high in the 
$14$ TeV LHC. In comparison to that, the scale variations at the LO is very 
prominent in the $100$ TeV case. The scale variation for the NLO calculation 
soften strikingly as expected.

To study the different dilepton correlating observables in terms of LO and NLO
 calculation with their corresponding scale uncertainties, 
 in Fig.~\ref{fig:costhetall_14_100} we display the differential distributions 
for scattering cross-section with respect to the angles between the leptons, 
$\cos\theta_{\ell_{i}\ell_{j}}$ for $i=1$, $2$, $3$, $j=1$, $2$, $3$ and 
$i\neq j$.  
The production of the leptons with large polar angular separation is abundant, although
tend to choose smaller one since main production channel involves the contributions from 
both the valence and sea quarks. 
This effect is quite more evident at $100$ TeV machine, where the leptons are 
mostly produced at small polar angle. 
The difference between the azimuthal angle between the two leptons, namely, 
$\Delta\phi_{\ell_{i}\ell_{j}}$ of the leptons with $i=1$, $2$, $3$, $j=1$, 
$2$, $3$ and $i\neq j$ are also shown in Fig.~\ref{fig:costhetall_14_100}. 
One notice that both the $\Delta\phi_{\ell_{1}\ell_{2}}$ and $\Delta\phi_{\ell_{1}\ell_{3}}$ distributions show 
some enhancement when leptons are produced back to back, supported by the leptons produced 
in boosted heavy neutrino decay. This peak is further enhanced in the NLO calculation. 
However, $\Delta\phi_{\ell_{1}\ell_{2}}$ remains flat in both calculations. Scale uncertainty is also shown
to be substantially controlled in NLO estimates.

\begin{figure*}[t]
\begin{center}
\includegraphics[scale=0.25]{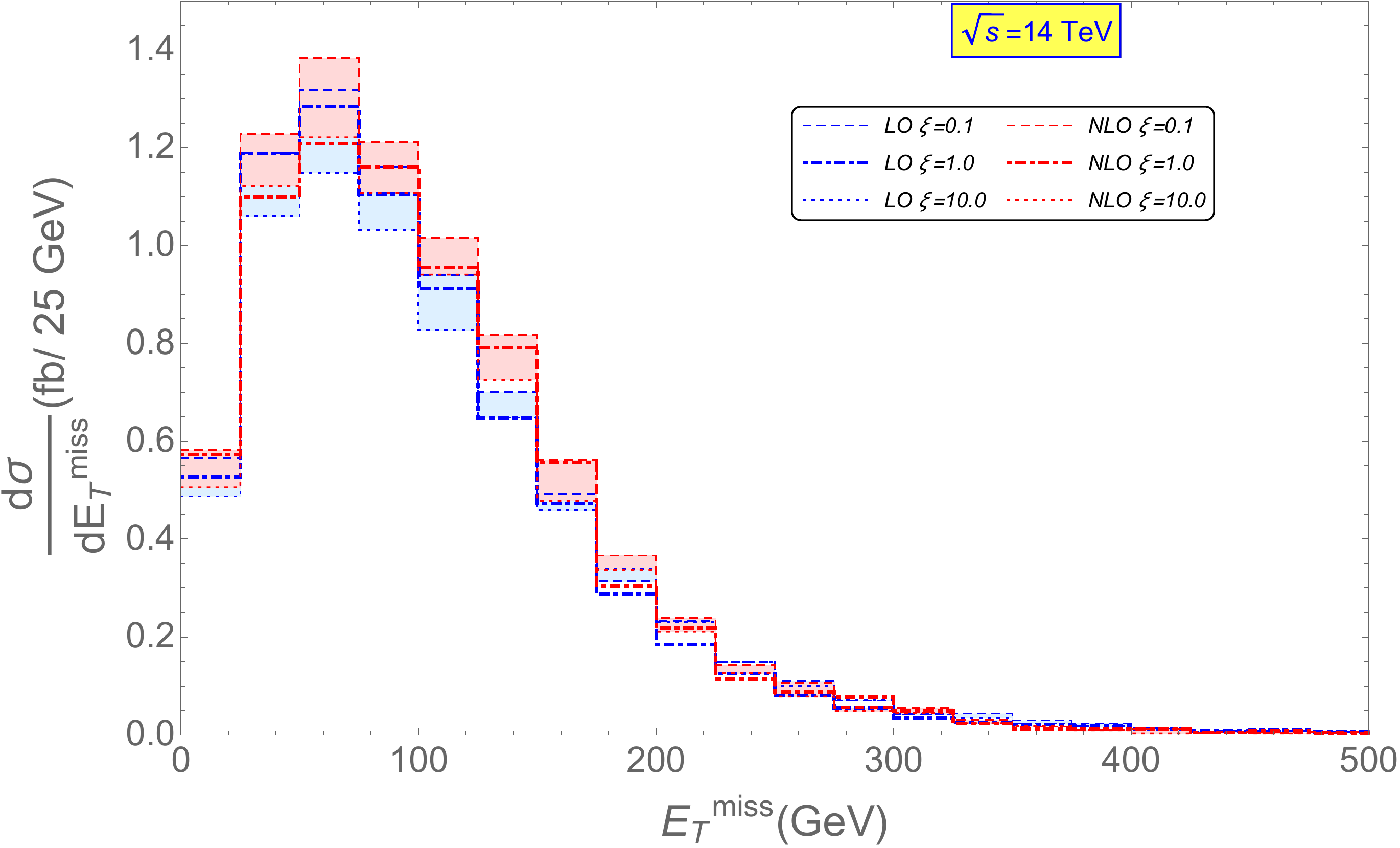}
\includegraphics[scale=0.25]{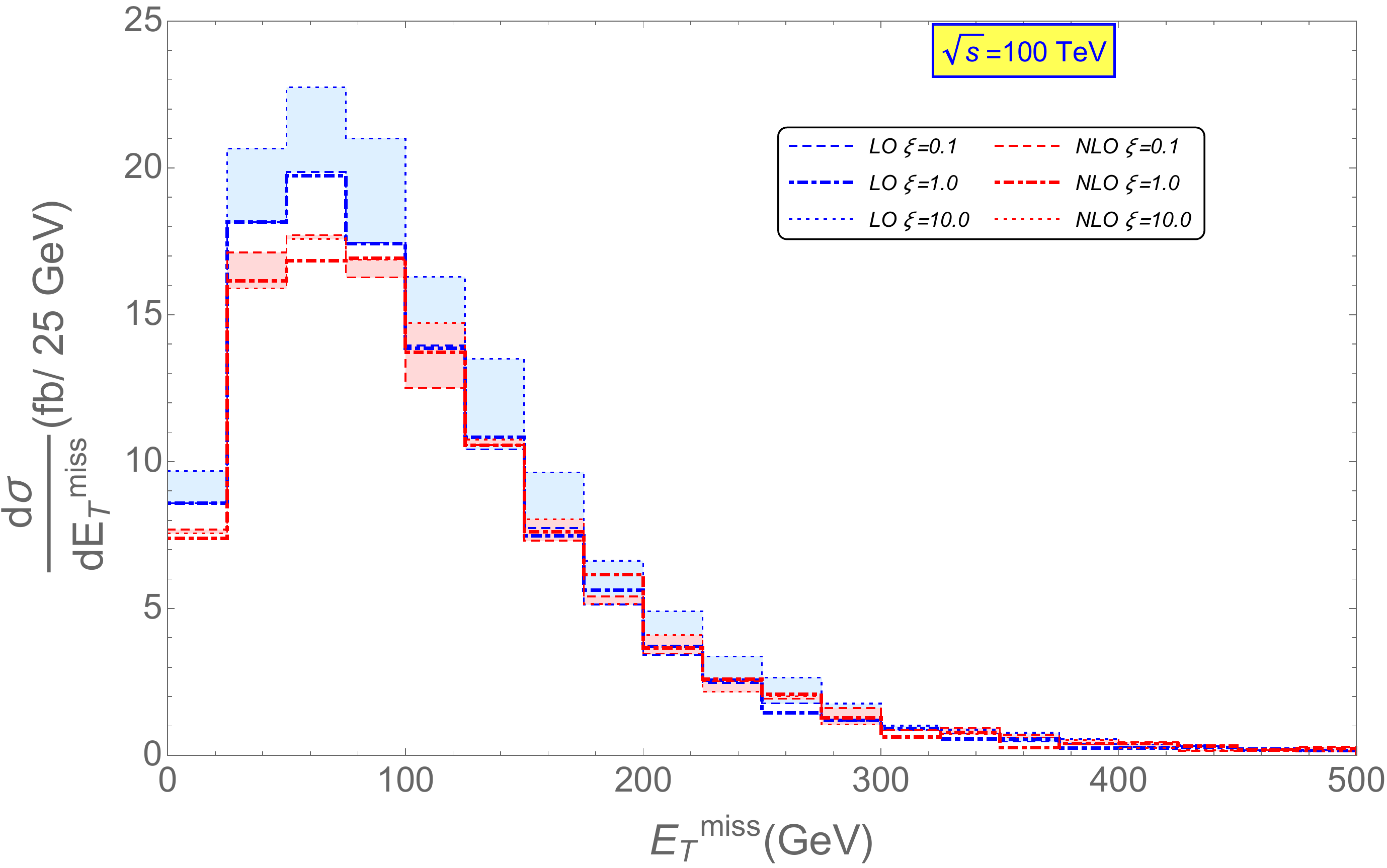}
\caption{Scale variation of the differential scattering cross-section as a function of the missing transverse momentum ($\eta^{\ell_{i}}$, $i=1, 2, 3$) in case of trilepton production channel for $m_{N}=400$ GeV. The left panel corresponds to case at $\sqrt{s}=14$ TeV LHC whereas the right panel represents the $\sqrt{s}=100$ TeV collider. The differential cross-section distributions are normalized by $|V_{\ell N}|^2$.}
\label{fig:ETmiss_14_100}
\end{center}
\end{figure*}

The scale dependent differential scattering cross-section as a function of 
missing transverse energy, $E_{T}^{miss}$, are given 
Fig.~\ref{fig:ETmiss_14_100} where the LO and NLO variations have good 
agreements at  $14$ TeV LHC. NLO distribution enhances at larger 
$E_{T}^{miss}$. We can make the same observation at the NLO and LO cases 
for the $100$ TeV collider.

\section{Scale dependent prospective upper bound on the mixing angles}
\label{sec:Mix}
In this section, we study the prospective upper bounds on the mixing angles between the heavy neutrino and the SM
light neutrinos. We consider two different scenario such as type-I seesaw which involves a heavy Majorana neutrino and inverse seesaw
which introduces a pseudo-Dirac heavy neutrino. Due to the large lepton number violation, the type-I seesaw scenario is 
observed through the characteristic same sign dilepton plus dijet final state. On the other hand due to a very small lepton number violating parameter,
optimal observable being the trilepton plus missing energy signal for inverse seesaw case. 
Using these mechanisms and existing searches done by LHC at $\sqrt{s}=8$ TeV(Run$-1$) we obtain 
prospective search reaches at the LHC at $\sqrt{s}=14$ TeV(Run$-2$) and proposed proton-proton collider at $\sqrt{s}=100$ TeV.

\subsection{Same sign dilepton plus two jet production signal}

For simplicity we consider the case that only one generation of the Majorana heavy neutrino
  is lighter enough and accessible to the LHC which couples to only the second generation of the lepton flavor. 
To generate the events in the {\tt MadGraph} we use the {\tt CTEQ6L1} for the LO and 
{\tt CTEQ6M} for the NLO$(\mu_{F}= \mu_{R})$ cases respectively. We study the scale dependent same sign dilepton plus dijet signal
cross-section as a function of the heavy neutrino mass $\left(m_{N}\right)$. The signal cross-section at the level of LO and NLO are calculated as $\sigma(\xi)_{LO}$ and $\sigma(\xi)_{NLO}$ respectively for the same sign dimuon production,
\bea
pp\to N \mu^{\pm} \to\mu^{\pm} \mu^{\pm} j j.
\eea
Thus corresponding values are expressed the $14$ TeV LHC $\left(\sigma(\xi)^{14}_{LO}, \sigma(\xi)^{14}_{NLO} \right)$ and $100$ TeV collider  $\left(\sigma(\xi)^{100}_{LO}, \sigma(\xi)^{100}_{NLO} \right)$.
Comparing our generated events with the recent ATLAS results \cite{Aad:2015xaa} at the $8$ TeV LHC with the luminosity $20.3$ fb$^{-1}$,
  we obtain an upper limit on the mixing angles between the Majorana type 
  heavy neutrino and the SM leptons as a function of $m_{N}$ for $\xi= 0.1$, $1.0$, $10.0$.
In the ATLAS analysis the upper bound of the production cross-section ($\sigma^{ATLAS}$) is obtained for the final state with
  the same sign di-muon plus dijet as a function of $m_{N}$.
Using these 14 TeV leading order (LO) scale dependent cross-sections, we obtain the prospective upper bounds on the mixing angles for different values of $\xi = \xi^{\prime}$ which is chosen to be either of $=\{0.1, 1, 10\}$, 
\bea
|V_{\ell N}|^{2}(\xi^{\prime})_{LO}^{14} \lesssim \frac{\sigma^{ATLAS}}{\sigma(\xi^{\prime})_{LO}^{14}},
\label{mixLO14xi}
\eea
whereas those for the NLO case at the $14$ TeV are given as
\bea
|V_{\ell N}|^{2}(\xi^{\prime})_{NLO}^{14} \lesssim \frac{\sigma^{ATLAS}}{\sigma(\xi^{\prime})_{NLO}^{14}},
\label{mixNLO14xi}
\eea

\begin{figure*}[t]
\begin{center}
\includegraphics[scale=0.185]{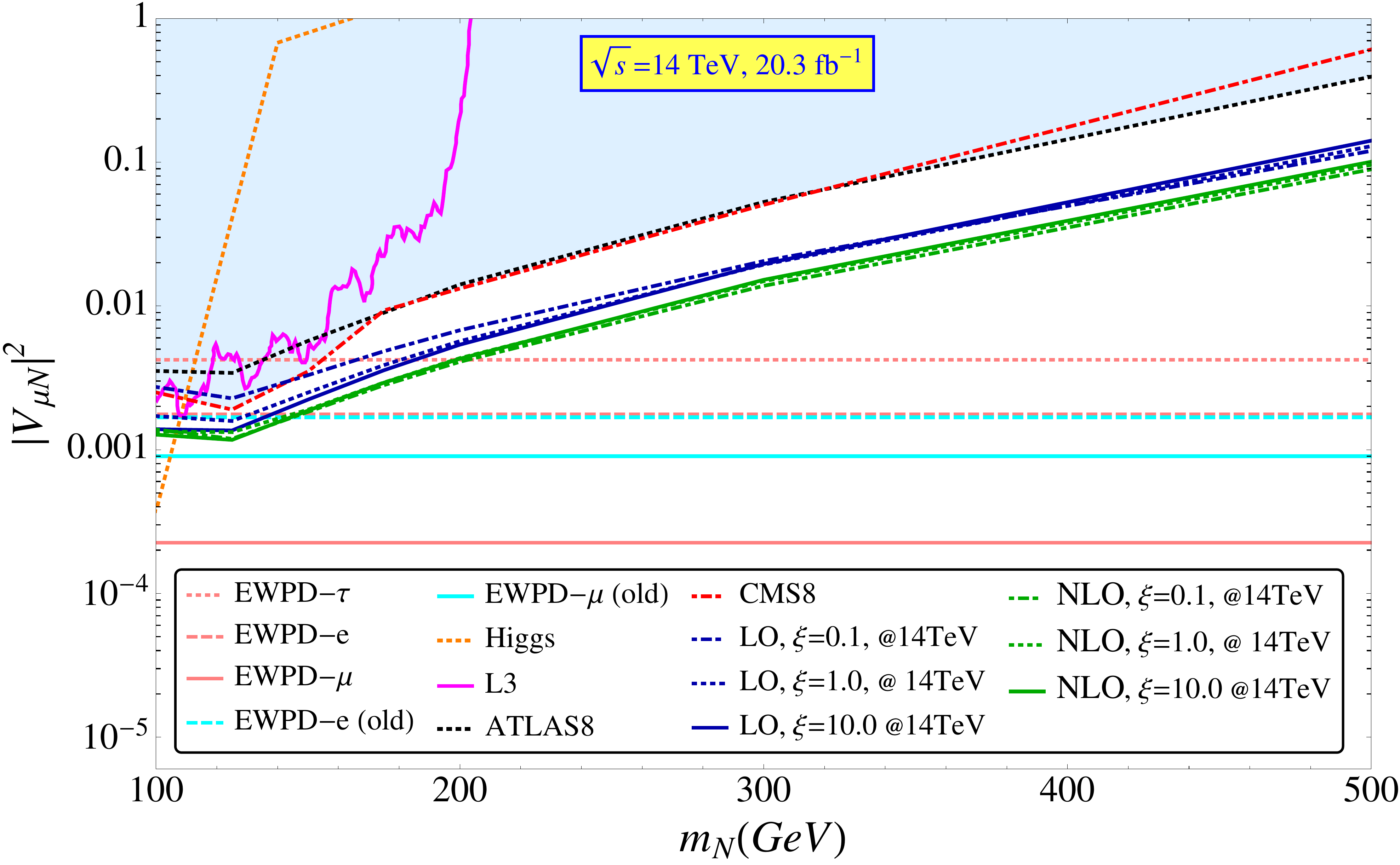}
\includegraphics[scale=0.185]{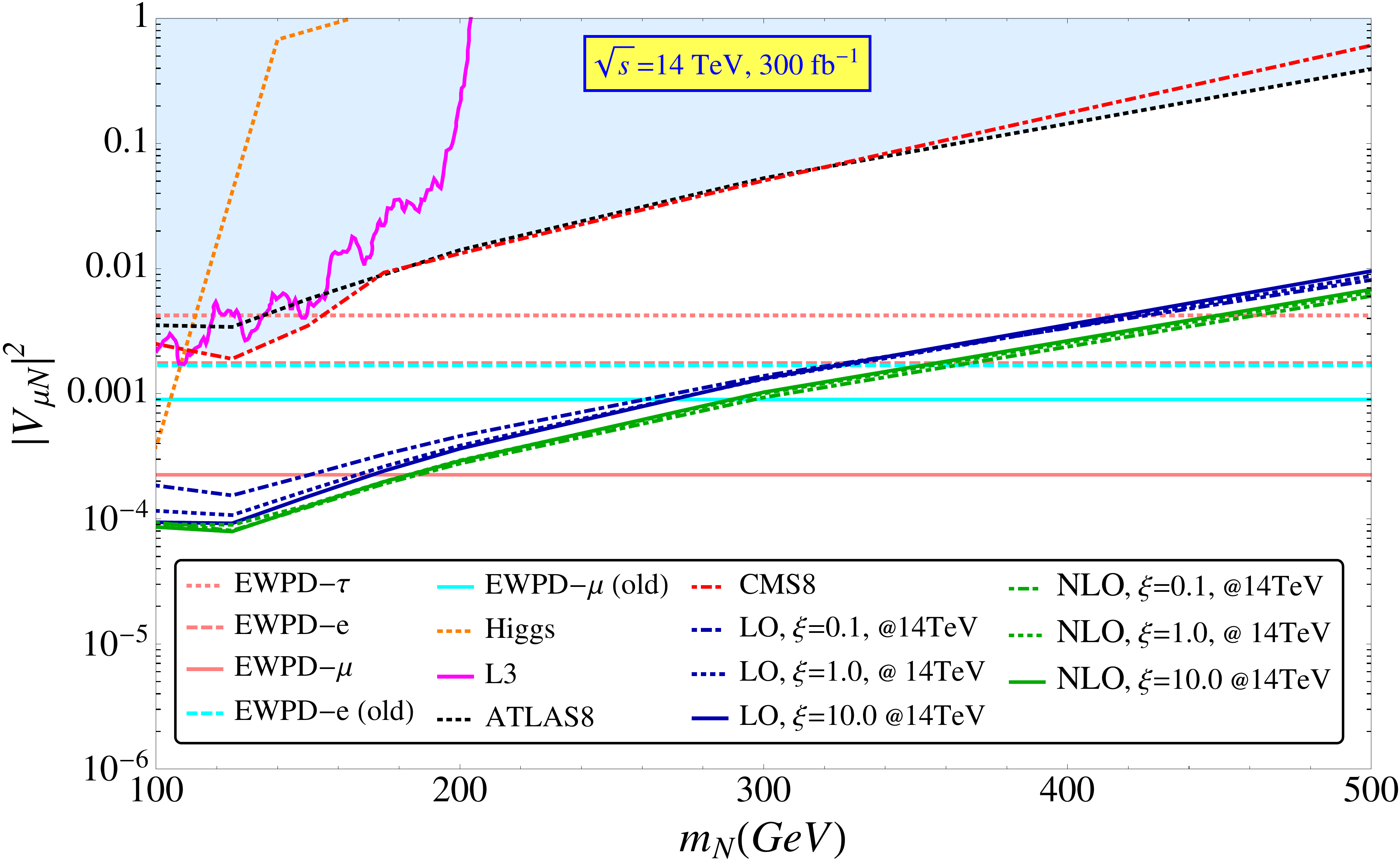}\\
\includegraphics[scale=0.185]{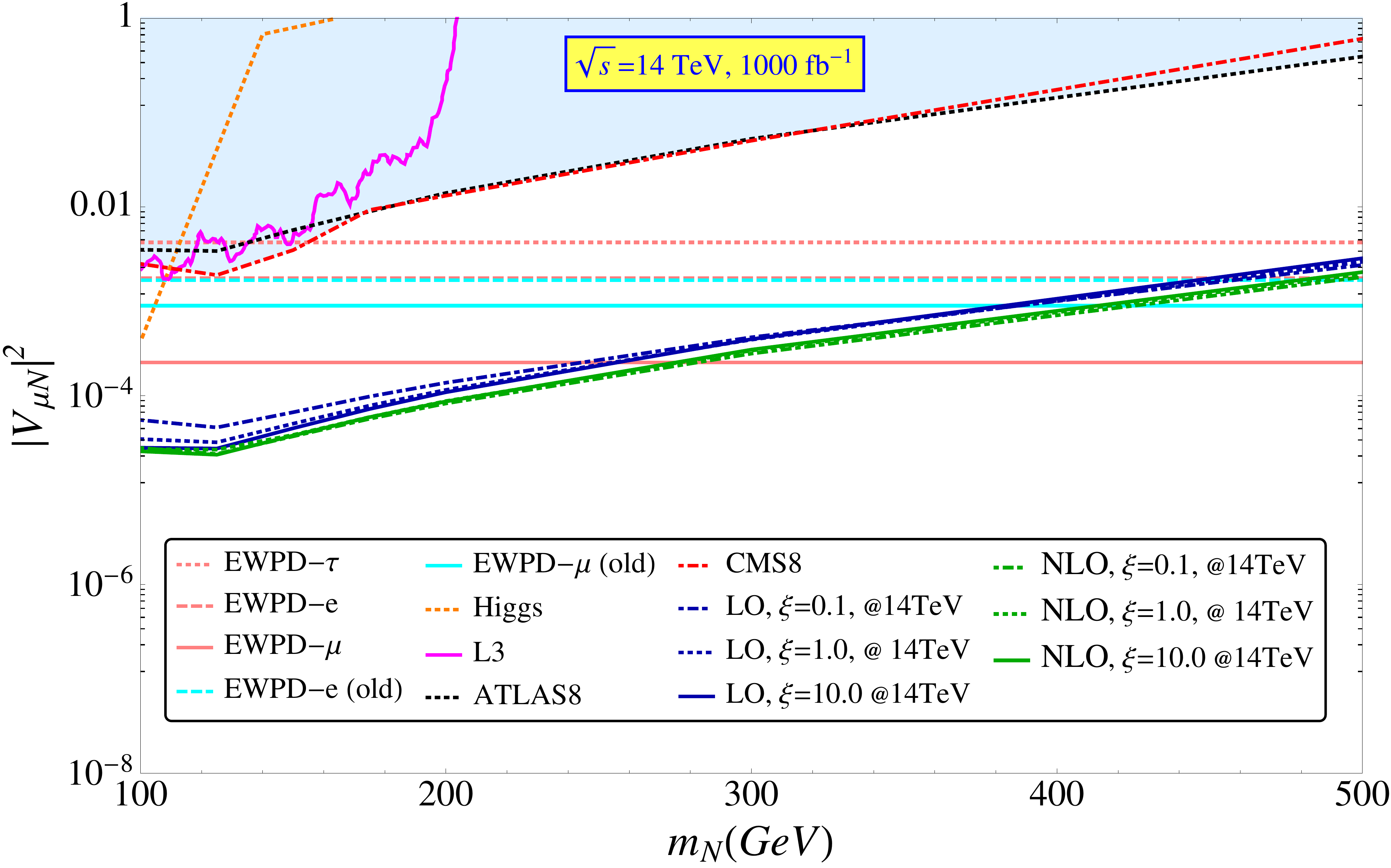}
\caption{Figure shows the prospective upper bounds of square on the mixing angles as a function of $m_{N}$ using the ATLAS data at the 8 TeV \cite{Aad:2015xaa} at $20.3$ fb$^{-1}$ luminosity for the same sign dileton plus dijet case. 
The scale dependent LO and NLO prospective upper bounds at the $14$ TeV LHC at $20.3$ fb$^{-1}$ luminosity (left panel, upper row), at $300$ fb$^{-1}$ luminosity (right panel, upper row) and
$1000$ fb$^{-1}$ (lower row) are given. These bounds are compared to (i) the $\chi^{2}$-fit to the LHC Higgs data \cite{BhupalDev:2012zg} (Higgs), (ii) from a direct search at LEP \cite{Achard:2001qv}(L3), valid only for the electron flavor, (iv) CMS limits from $\sqrt{s}=$8 TeV LHC data \cite{Khachatryan:2015gha} (CMS8) and ATLAS \cite{Aad:2015xaa} (ATLAS8), for a heavy Majorana neutrino of the muon flavor and (v) indirect limits from the global fit to the electroweak precision data (EWPD) from~\cite{deBlas:2013gla, delAguila:2008pw, Akhmedov:2013hec} for electron (cyan, EWPD-e(old)) and muon (cyan, EWPD-$\mu$(old)) flavors(new values can be found from \cite{Antusch:2015gjw} , for tau (dotted, EWPD- $\tau$) electron (solid, EWPD- $e$) and muon (dashed, EWPD- $\mu$) flavors). The shaded region is excluded by the $8$ TeV data.}
\label{fig:Bounds_1}
\end{center}
\end{figure*}
\begin{figure*}[t]
\begin{center}
\includegraphics[scale=0.175]{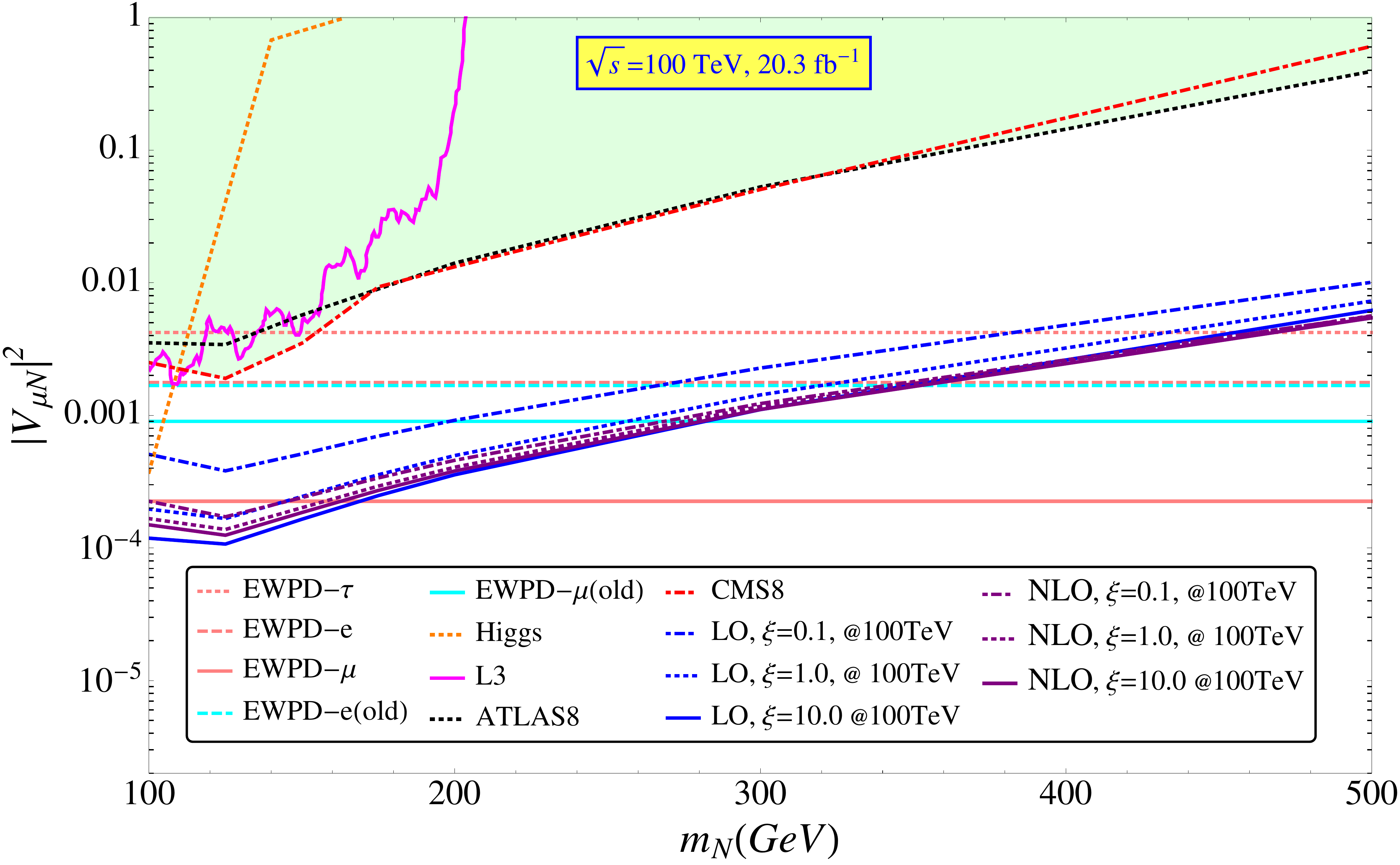}
\includegraphics[scale=0.175]{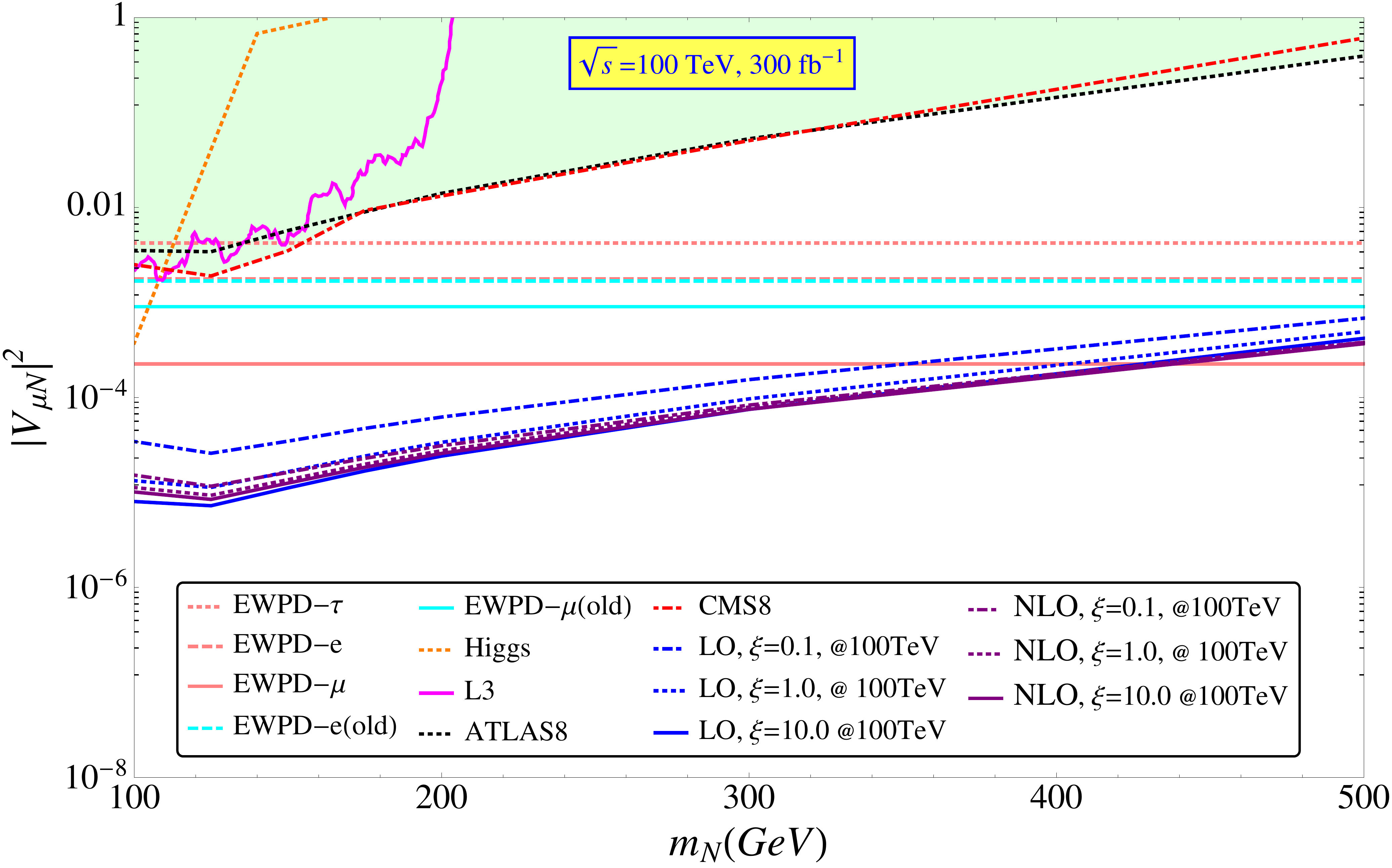}\\
\includegraphics[scale=0.175]{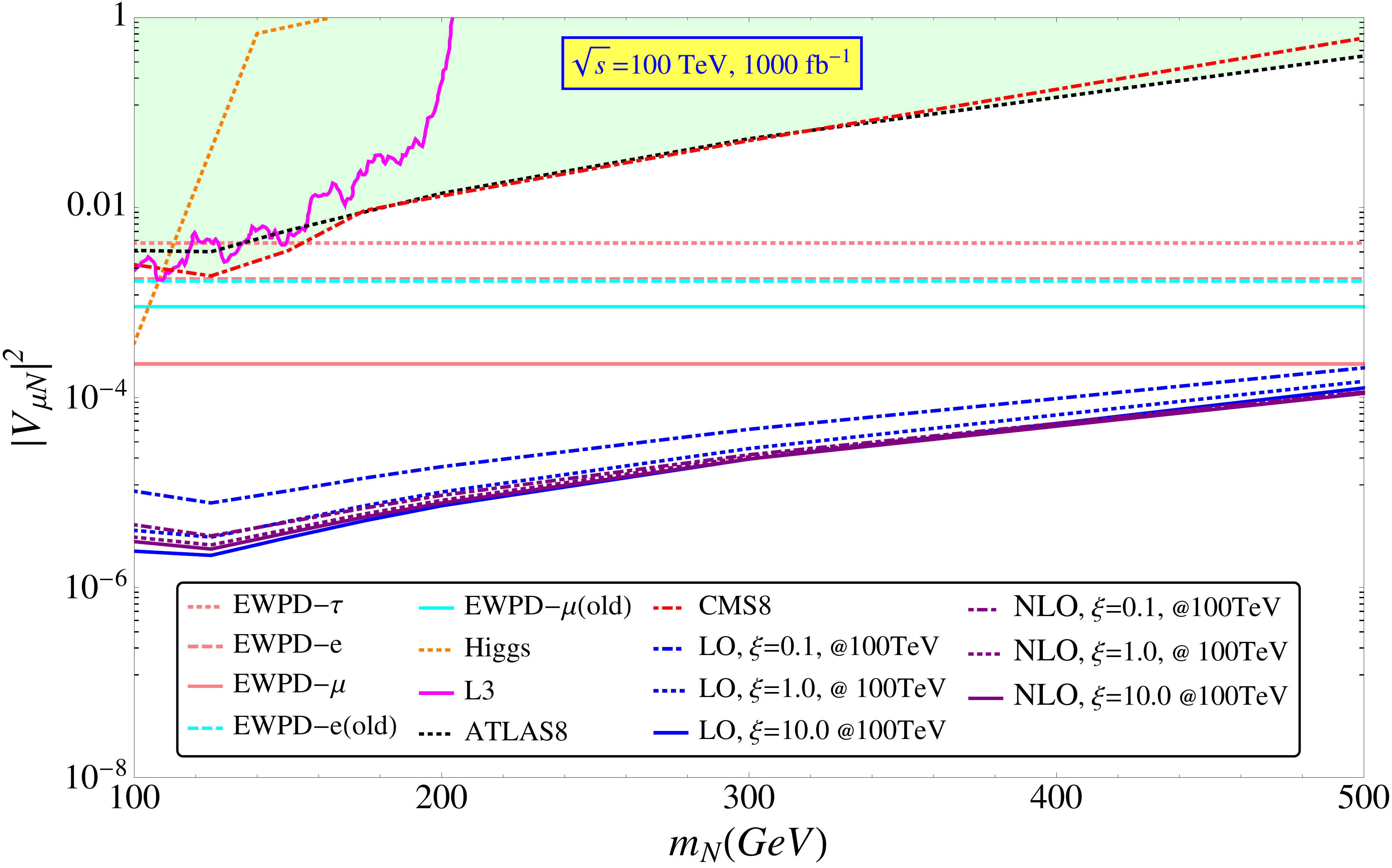}
\caption{Figure shows the prospective upper bounds of square on the mixing angles as a function of $m_{N}$ using the ATLAS data at the 8 TeV \cite{Aad:2015xaa} at $20.3$ fb$^{-1}$ luminosity for the same sign dileton plus dijet case. 
The scale dependent LO and NLO prospective upper bounds at the $100$ TeV LHC at $20.3$ fb$^{-1}$ luminosity (left panel, upper row), at $300$ fb$^{-1}$ luminosity (right panel, upper row) and
$1000$ fb$^{-1}$ (lower row) are given. These bounds are compared to (i) the $\chi^{2}$-fit to the LHC Higgs data \cite{BhupalDev:2012zg} (Higgs), (ii) from a direct search at LEP \cite{Achard:2001qv}(L3), valid only for the electron flavor, (iv) CMS limits from $\sqrt{s}=$8 TeV LHC data \cite{Khachatryan:2015gha} (CMS8) and ATLAS \cite{Aad:2015xaa} (ATLAS8), for a heavy Majorana neutrino of the muon flavor and (v) indirect limits from the global fit to the electroweak precision data (EWPD) from~\cite{deBlas:2013gla, delAguila:2008pw, Akhmedov:2013hec} for electron (cyan, EWPD-e(old)) and muon (cyan, EWPD-$\mu$(old)) flavors(new values can be found from \cite{Antusch:2015gjw} , for tau (dotted, EWPD- $\tau$) electron (solid, EWPD- $e$) and muon (dashed, EWPD- $\mu$) flavors). The shaded region is excluded by the $8$ TeV data.}
\label{fig:Bounds_2}
\end{center}
\end{figure*}

\begin{figure*}[t]
\begin{center}
\includegraphics[scale=0.185]{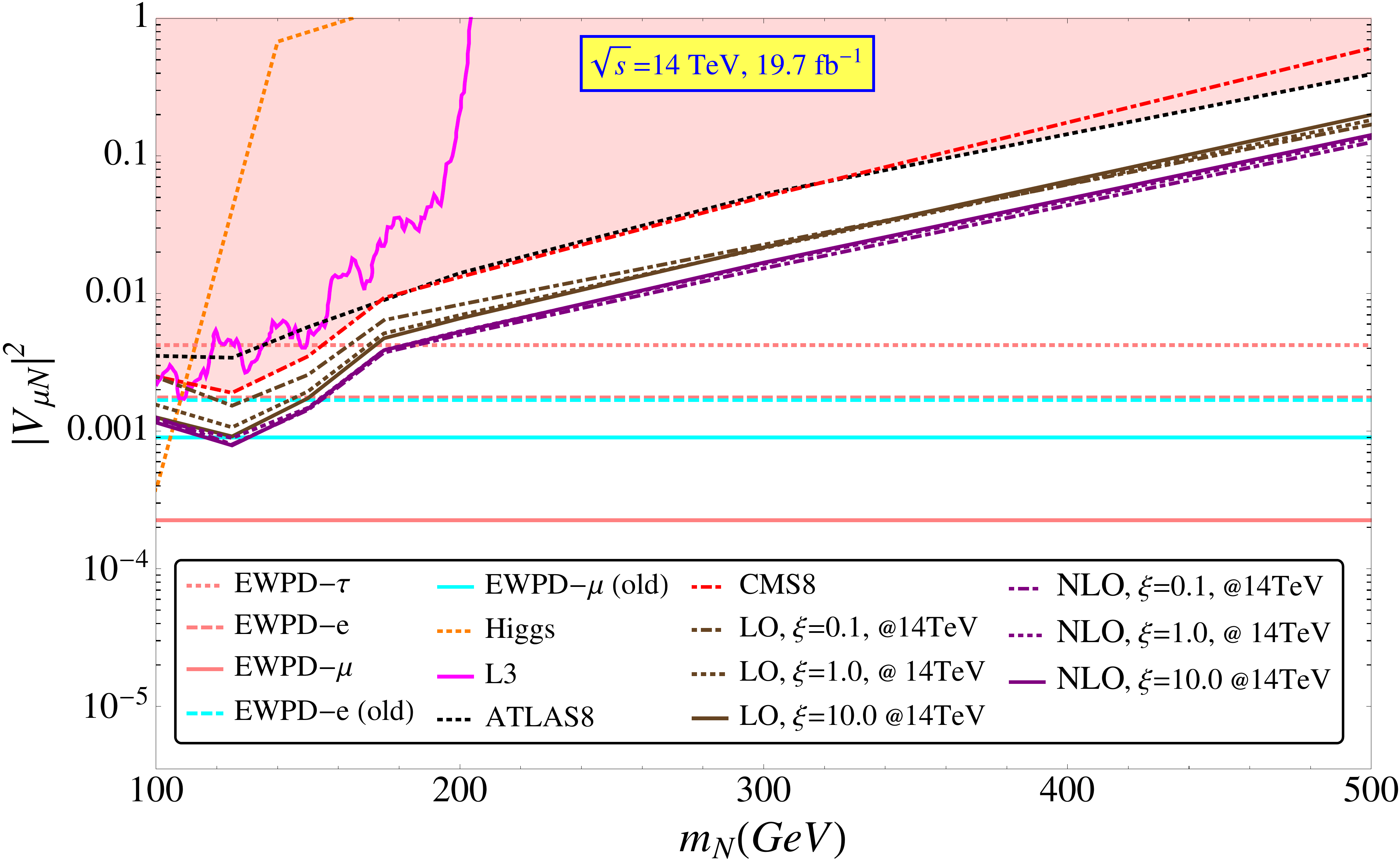}
\includegraphics[scale=0.185]{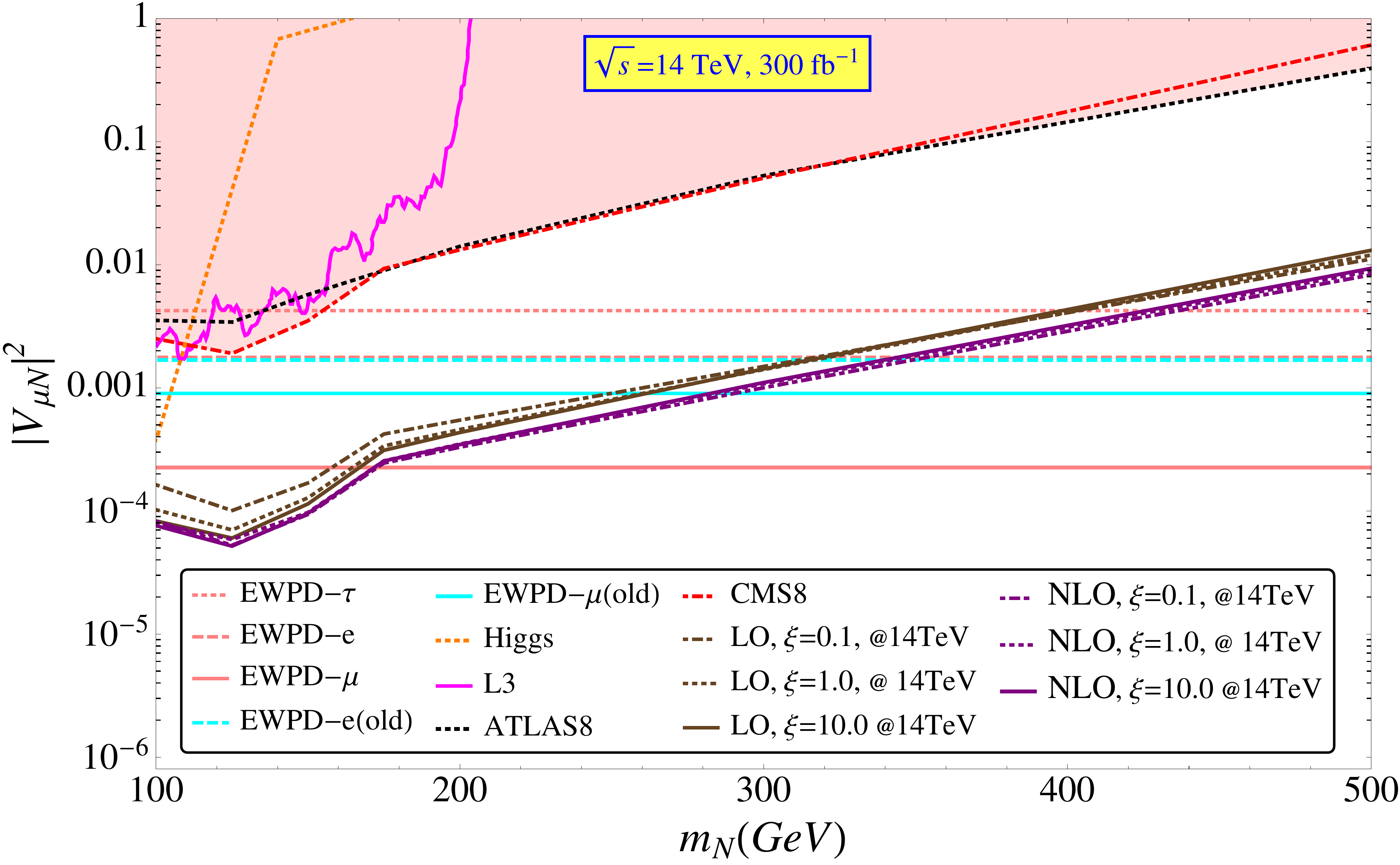}\\
\includegraphics[scale=0.185]{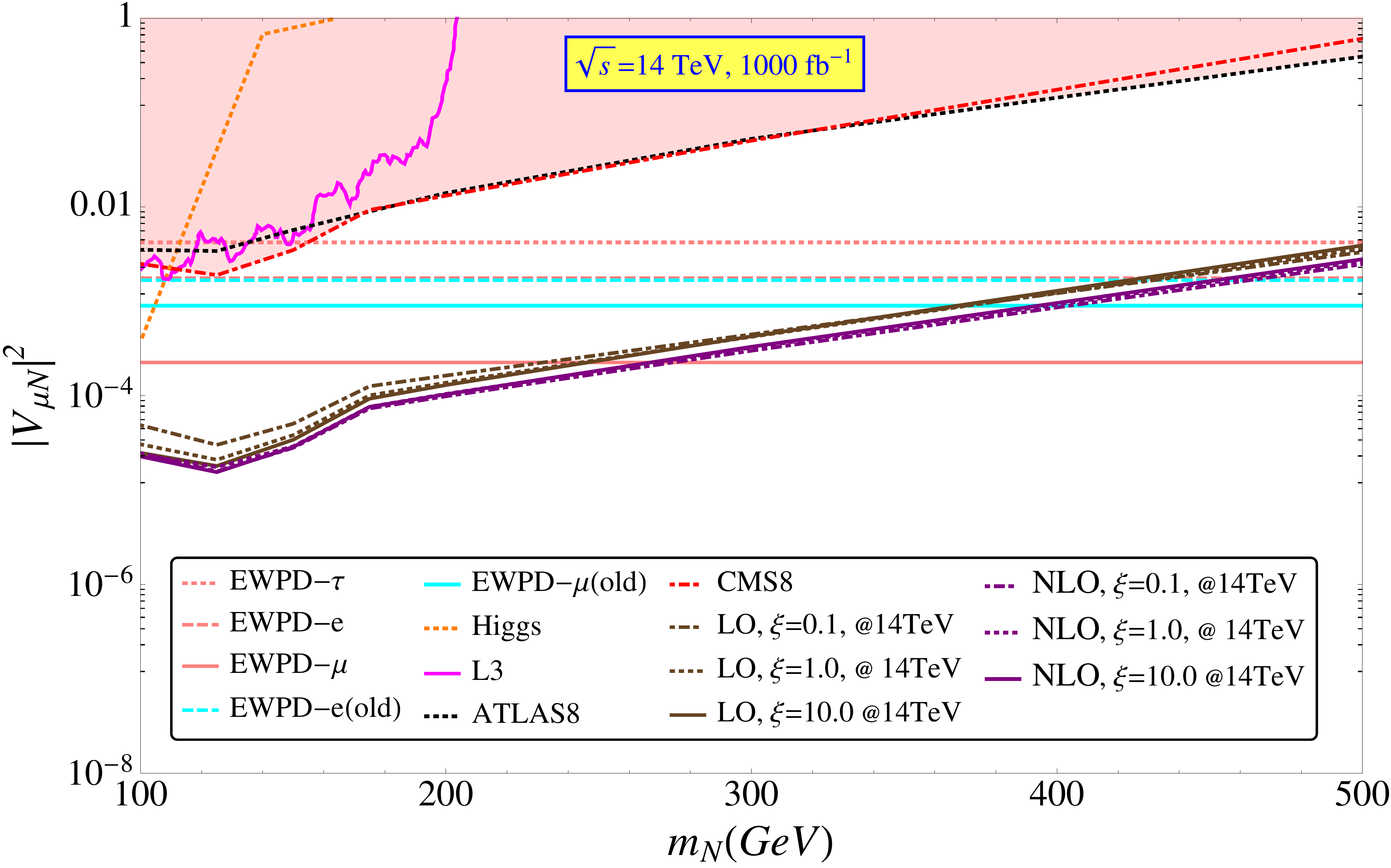}\\
\caption{Figure shows the upper bounds on square of the mixing angles as a function of $m_{N}$ using the CMS data at the 8 TeV \cite{Khachatryan:2015gha} at $19.7$ fb$^{-1}$ luminosity for the same sign dileton plus dijet case. 
The scale dependent LO and NLO prospective upper bounds at the $14$ TeV LHC at $20.3$ fb$^{-1}$ luminosity (left panel, upper row), at $300$ fb$^{-1}$ luminosity (right panel, upper row)
and $1000$ fb$^{-1}$ (lower row) are given.
The bounds are compared to (i) the $\chi^{2}$-fit to the LHC Higgs data \cite{BhupalDev:2012zg} (Higgs), (ii) from a direct search at LEP \cite{Achard:2001qv}(L3), valid only for the electron flavor, (iv) ATLAS limits from $\sqrt{s}=$8 TeV LHC data \cite{Aad:2015xaa} (ATLAS 8) and CMS \cite{Khachatryan:2015gha}, for a heavy Majorana neutrino of the muon flavor and (v) indirect limits from the global fit to the electroweak precision data (EWPD) from~\cite{deBlas:2013gla, delAguila:2008pw, Akhmedov:2013hec} for electron (cyan, EWPD-e(old)) and muon (cyan, EWPD-$\mu$(old)) flavors(new values can be found from \cite{Antusch:2015gjw} , for tau (dotted, EWPD- $\tau$) electron (solid, EWPD- $e$) and muon (dashed, EWPD- $\mu$) flavors). The shaded region is excluded by the $8$ TeV data.
}
\label{fig:Bounds_3}
\end{center}
\end{figure*}
\begin{figure*}[t]
\begin{center}
\includegraphics[scale=0.175]{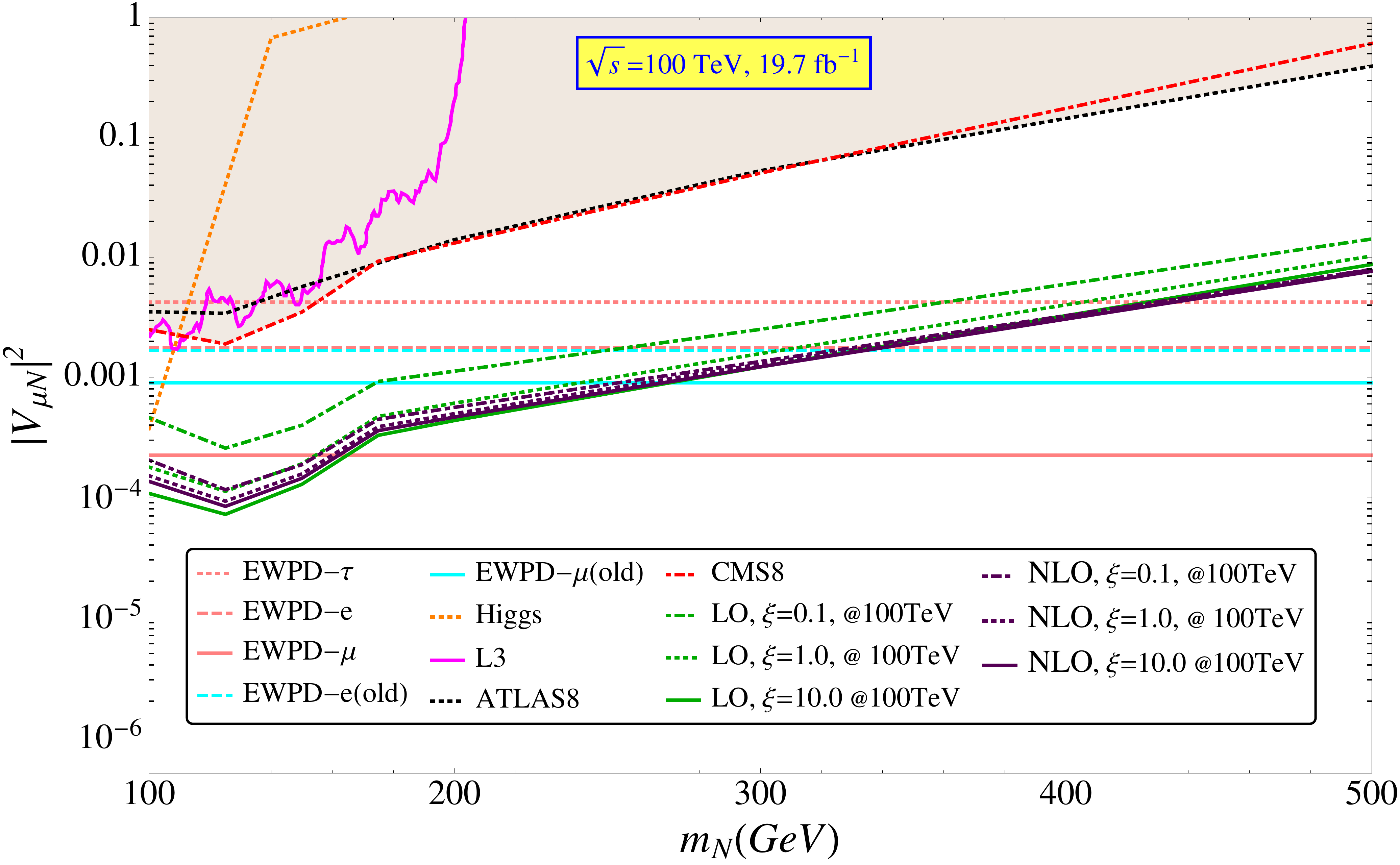}
\includegraphics[scale=0.175]{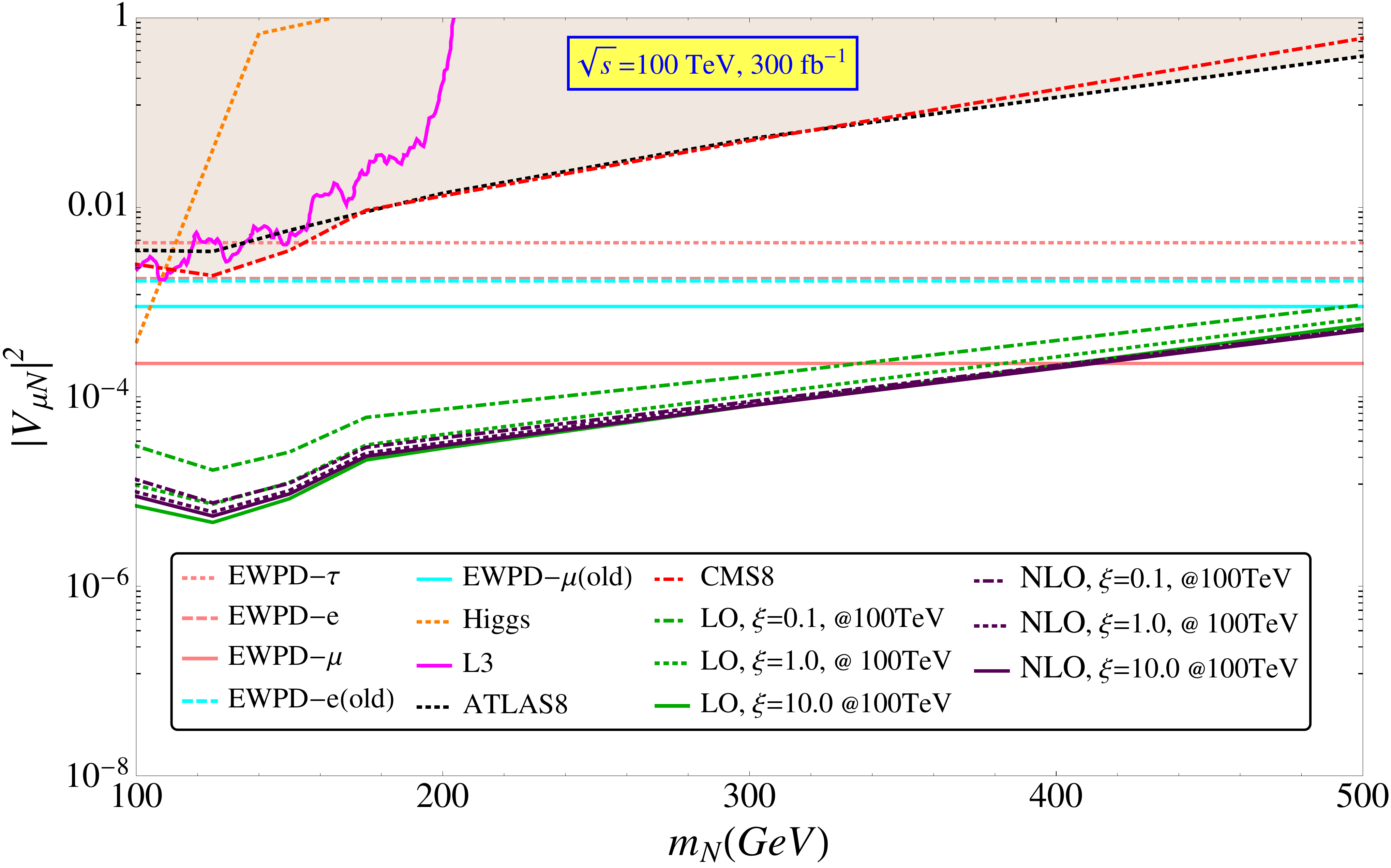}\\
\includegraphics[scale=0.175]{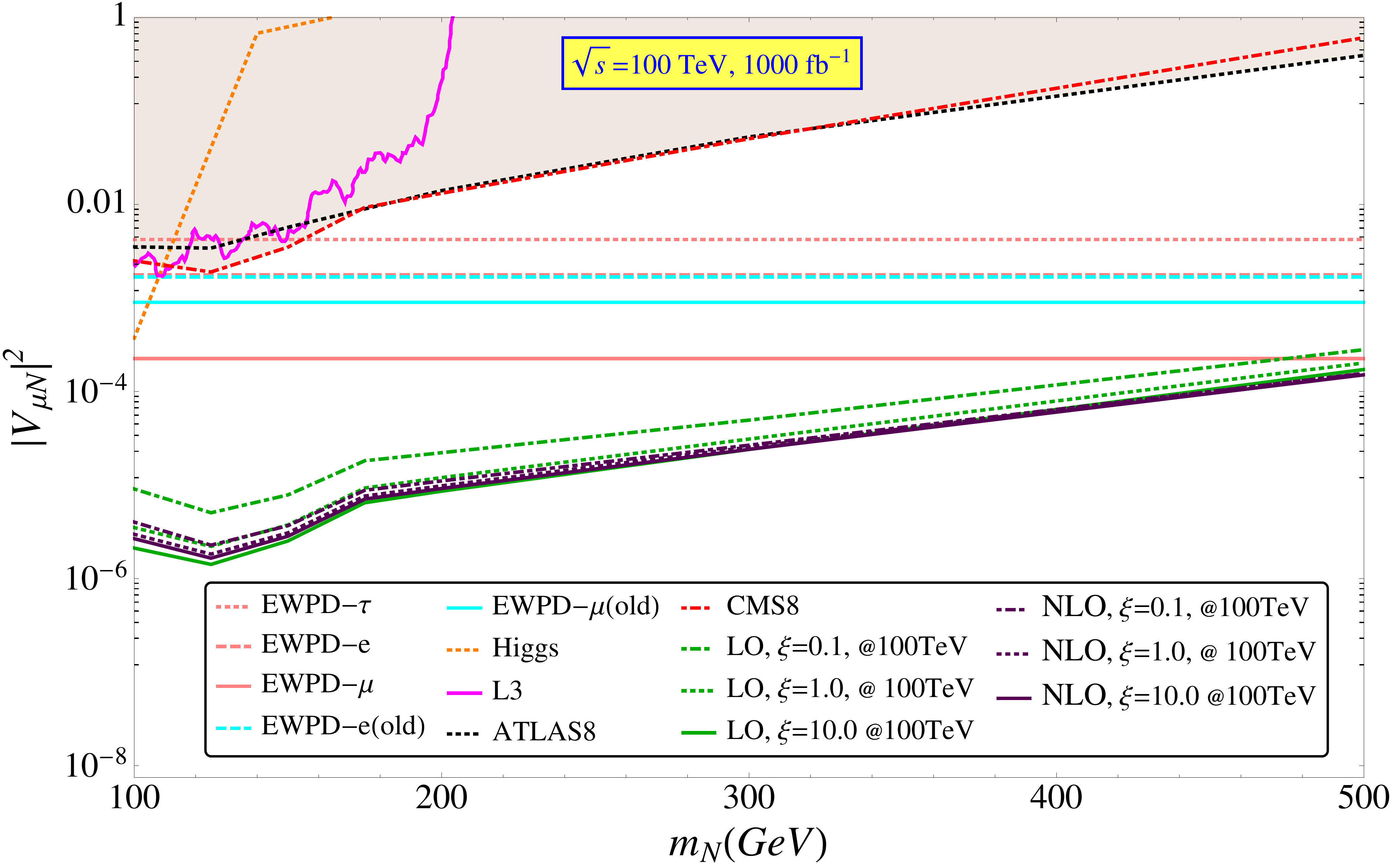}
\caption{Figure shows the upper bounds on square of the mixing angles as a function of $m_{N}$ using the CMS data at the 8 TeV \cite{Khachatryan:2015gha} at $19.7$ fb$^{-1}$ luminosity for the same sign dileton plus dijet case. 
The scale dependent LO and NLO prospective upper bounds at the $100$ TeV LHC at $20.3$ fb$^{-1}$ luminosity (left panel, upper row), at $300$ fb$^{-1}$ luminosity (right panel, upper row)
and $1000$ fb$^{-1}$ (lower row) are given.
The bounds are compared to (i) the $\chi^{2}$-fit to the LHC Higgs data \cite{BhupalDev:2012zg} (Higgs), (ii) from a direct search at LEP \cite{Achard:2001qv}(L3), valid only for the electron flavor, (iv) ATLAS limits from $\sqrt{s}=$8 TeV LHC data \cite{Aad:2015xaa} (ATLAS 8) and CMS \cite{Khachatryan:2015gha}, for a heavy Majorana neutrino of the muon flavor and (v) indirect limits from the global fit to the electroweak precision data (EWPD) from~\cite{deBlas:2013gla, delAguila:2008pw, Akhmedov:2013hec} for electron (cyan, EWPD-e(old)) and muon (cyan, EWPD-$\mu$(old)) flavors(new values can be found from \cite{Antusch:2015gjw} , for tau (dotted, EWPD- $\tau$) electron (solid, EWPD- $e$) and muon (dashed, EWPD- $\mu$) flavors). The shaded region is excluded by the $8$ TeV data.}
\label{fig:Bounds_4}
\end{center}
\end{figure*}

At the $100$ TeV we use Eqs.~(\ref{mixLO14xi}) and (\ref{mixNLO14xi}) replacing the cross-sections at the $14$ TeV LHC with those at the $100$ TeV collider. The calculated prospective upper bounds on the mixing angles are shown in Fig.~\ref{fig:Bounds_1}
along with the bounds from
  ATLAS \cite{Aad:2015xaa}, CMS \cite{Khachatryan:2015gha}, LEP (L3) \cite{Achard:2001qv}, electroweak precision data for tau (EWPD-$\tau$), electron (EWPD-e)
  and muon (EWPD-$\mu$)~\cite{Antusch:2014woa, Basso:2013jka} (see, \cite{deBlas:2013gla, delAguila:2008pw, Akhmedov:2013hec} for previous analysis), and finally LHC Higgs data (Higgs) \cite{BhupalDev:2012zg}
   (see, \cite{Antusch:2015gjw, Antusch:2015rma, Antusch:2015mia} for some updated analysis).
 Comparing our results with the $8$ TeV results given by ATLAS \cite{Aad:2015xaa} we give a prospective upper bound on the mixing angle at $14$ TeV LHC with $20.3$ fb$^{-1}$ luminosity
 for different values of $\xi=0.1$, $1.0$, $10.0$ at the LO and NLO. We notice that the scale dependence at LO is very clear for $m_{N}\lesssim300$ GeV in comparison to the NLO case at the $14$ TeV LHC.
 The LO and NLO results could be comparable to the EWPD at the $14$ TeV. Which we can easily verify using the LHC results in Run$-2$ at the $14$ TeV.
  We have also studied the prospective upper bounds on the mixing angles at the $100$ TeV collider for the LO and NLO cases at $20.3$ fb$^{-1}$ luminosity. We have noticed that for $m_{N}\lesssim 250$ GeV, the mixing angle could be a factor better than those given by EWPD. 
An improved prospective search reach by an order of magnitude (more) for $300$ fb$^{-1}$($1000$ fb$^-1$) luminosity is also given in Fig.~\ref{fig:Bounds_1} for the $14$ TeV LHC. We have also calculated a prospective search reach for the $100$ TeV Collider  at $20.3$ fb$^{-1}$, $300$ fb$^{-1}$ and $1000$
fb$^{-1}$ luminosities in Fig.~\ref{fig:Bounds_2}. The improvement in search reach of the mixing angle from a factor to an order of magnitude with respect to the EWPD can be obtained at the $100$ TeV.

Recently the CMS has performed the same-sign dilepton plus dijet search \cite{Khachatryan:2015gha}.
Using this result and adopting the same procedure for the ATLAS result we calculate 
  the prospective upper bound on the mixing angles at the 14 TeV LHC for the LO and NLO cases at $19.7$ fb$^{-1}$.
The results are shown in Fig.~\ref{fig:Bounds_2}.  A clear scale dependence is observed for $m_{N}\lesssim 300$ GeV for the LO case whereas the mixing angle around 120 GeV
is comparable to the EWPD. The scale dependence is not  very high in the NLO case in comparison to the LO case. We can easily verify using the LHC results in Run$-2$ at the $14$ TeV.
An improved prospective search reach by an order of magnitude (more) for $300$ fb$^{-1}$ ($1000$ fb$^{-1}$) luminosity is also given in Fig.~\ref{fig:Bounds_3} for the $14$ TeV LHC.
We have also depicted the prospective search reach in Fig.~\ref{fig:Bounds_4} for the $100$ TeV Collider at $19.7$ fb$^{-1}$, $300$ fb$^{-1}$ and $1000$fb$^{-1}$ luminosities where 
we can improve the upper bound on the mixing angle from a factor up to an order of magnitude with respect to the EWPD from low to high luminosities.

\subsection{Trilepton associated with missing transverse energy signal}
We consider two cases in this analysis. 
One is the Flavor Diagonal case (FD), where
  we employ three generations of the degenerate heavy neutrinos 
  and each generation couples with the single, corresponding lepton flavor.
The other one is the Single Flavor case (SF) where only one of the heavy neutrinos  
is accessible to the LHC and having mass in the Electroweak scale being coupled to 
only the first or second generation of the lepton flavor.
In this analysis we use the {\tt CTEQ6M PDF} \cite{CTEQ6} for generating the 
NLO$(\mu_{F}= \mu_{R})$ processes to compute the scale dependent trilepton plus missing energy
events with $\xi=0.1$, $1.0$ and $10.0$ at $\sqrt{s}=14$ TeV LHC using {\tt MadGreph-aMC@NLO} bundled with {\tt PYTHIA6Q} using {\tt anti-$k_{T}$} algorithm for jet clustering in {\tt FastJet}.
We use the hadronized events in {\tt Delphes} \cite{Delphes} to produce events after the detector simulation.
 The trilepton plus missing energy mode
is given in Eq.~\ref{trilep}. After the detector simulation we have considered the events with $3\ell+ E_{T}^{miss}+ n-$jets
where $n=0$,$1$ and $2$. 

Recently the CMS has studied the anomalous multilepton plus missing energy final state at the $8$ TeV \cite{Chatrchyan:2014aea}.
We adopt their search result for out trilepton analysis and compare our trilepton plus missing energy final state after the detector simulation to put a prospective upper limit on the mixing angle 
at the $14$ TeV. The cuts we used for this analysis according to \cite{Chatrchyan:2014aea} are itemized below: 
\begin{itemize}
\item [(i)] The transverse momentum of each lepton: $p^\ell_T > 10$ GeV.
\item [(ii)] The transverse momentum of at least one lepton: $p^{\ell,{\rm leading}}_{T} > 20$ GeV.
\item [(iii)]  The jet transverse momentum: $p_T^j > 30$ GeV. 
\item [(iv)] The pseudo-rapidity of leptons: $|\eta^\ell| < 2.4$ and of jets: $|\eta^j| < 2.5$.
\item [(v)] The lepton-lepton separation: $\Delta R_{\ell \ell} > 0.1$ and the lepton-jet separation: $\Delta R_{\ell j} > 0.3$. 
\item [(vi)] The invariant mass of each OSSF (opposite-sign same flavor) lepton pair: $m_{\ell^+ \ell^-}< 75$  GeV or $> 105$ GeV to avoid the on-$Z$ region which was excluded from the CMS search. Events with $m_{\ell^+ \ell^-}< 12$ GeV are rejected to eliminate background from low-mass Drell-Yan processes and hadronic decays. 
\item [(vii)]  The scalar sum of the jet transverse momenta: $H_{T} >  200$ GeV. 
\item [(viii)] The missing transverse energy: $50$ GeV$< E_{T}^{miss} < 100$ GeV. 
\end{itemize}
To derive the limits on $|V_{\ell N}|^{2}$, we calculate the signal cross-section normalized by the 
  square of the mixing angle as a function of the heavy neutrino mass $m_{N}$
  for both SF and FD cases, by imposing the CMS selection criteria listed above for different scale values of $\xi$
  at the NLO process.\footnote{It should be mentioned clearly that omitting the Z$-$pole we are excluding the effects of the other trilepton channels like $pp\to N\ell$, $N \to Z \nu$ followed by $Z\to \ell^{+}\ell^{-}$ \cite{Haba:2009sd, Matsumoto:2010zg} exclusively form the present analysis. There is another channel $pp\rightarrow N\overline{N}$ which will be suppressed by $ |V_{\ell N}|^{4}$ compared to the $W$ mediated channels in Fig.~\ref{fig:Nl}. However, see \cite{Kang:2015uoc, Abdallah:2015hma, Abdallah:2015uba, Khalil:2015naa, Elsayed:2011de, Abbas:2015zna, Khalil:2015wua, Huitu:2008gf} for some recent analyses on $N\overline{N}$ production from the B$-$L model due to its rich phenomenology.}
  Passing the generated detector events through all the cuts 
we compare them with the observed number of events at the 19.5 fb$^{-1}$ luminosity \cite{Chatrchyan:2014aea}.
For the selection criteria listed above, the CMS experiment observed:
\begin{itemize}
\item[(a)] 10 events with the SM background expectation of 11$\pm$3.8 events for $m_{\ell^{+}\ell^{-}} <$ 75 GeV.
\item[(b)] 4 events with the SM background expectation of  5.0$\pm$1.6 events for $m_{\ell^{+}\ell^{-}} >$ 105 GeV. 
\end{itemize}
In case (a) we have an upper limit of 2.8 signal events, while in 
  case (b) leads to an upper limit of 0.6 signal events.

Using these limits, we can set an upper bound on $|V_{\ell N}|^{2}$ for a given value of $m_{N}$ for the scale dependent NLO case. 
\begin{figure*}[t]
\begin{center}
\includegraphics[scale=0.170]{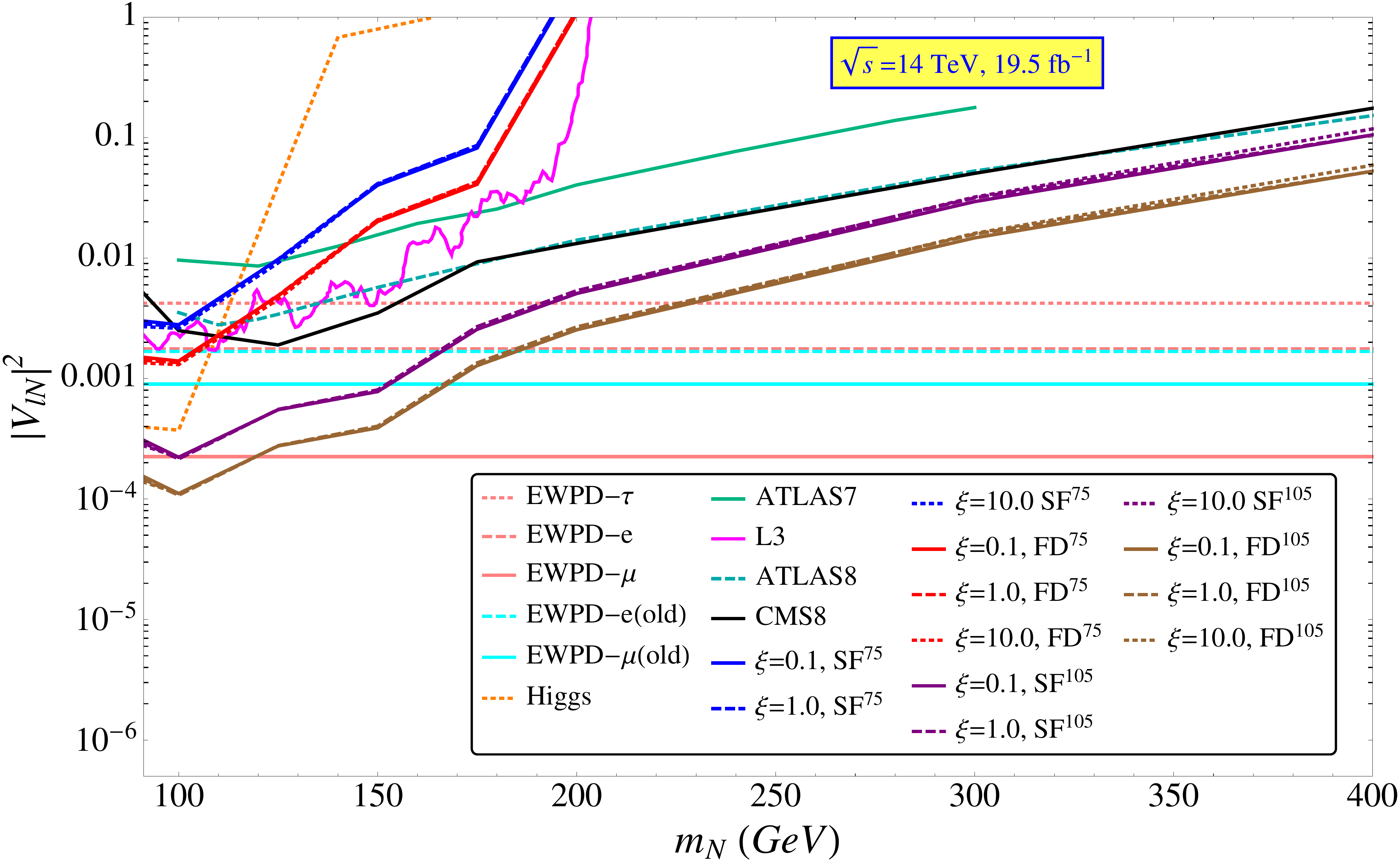}
\includegraphics[scale=0.149]{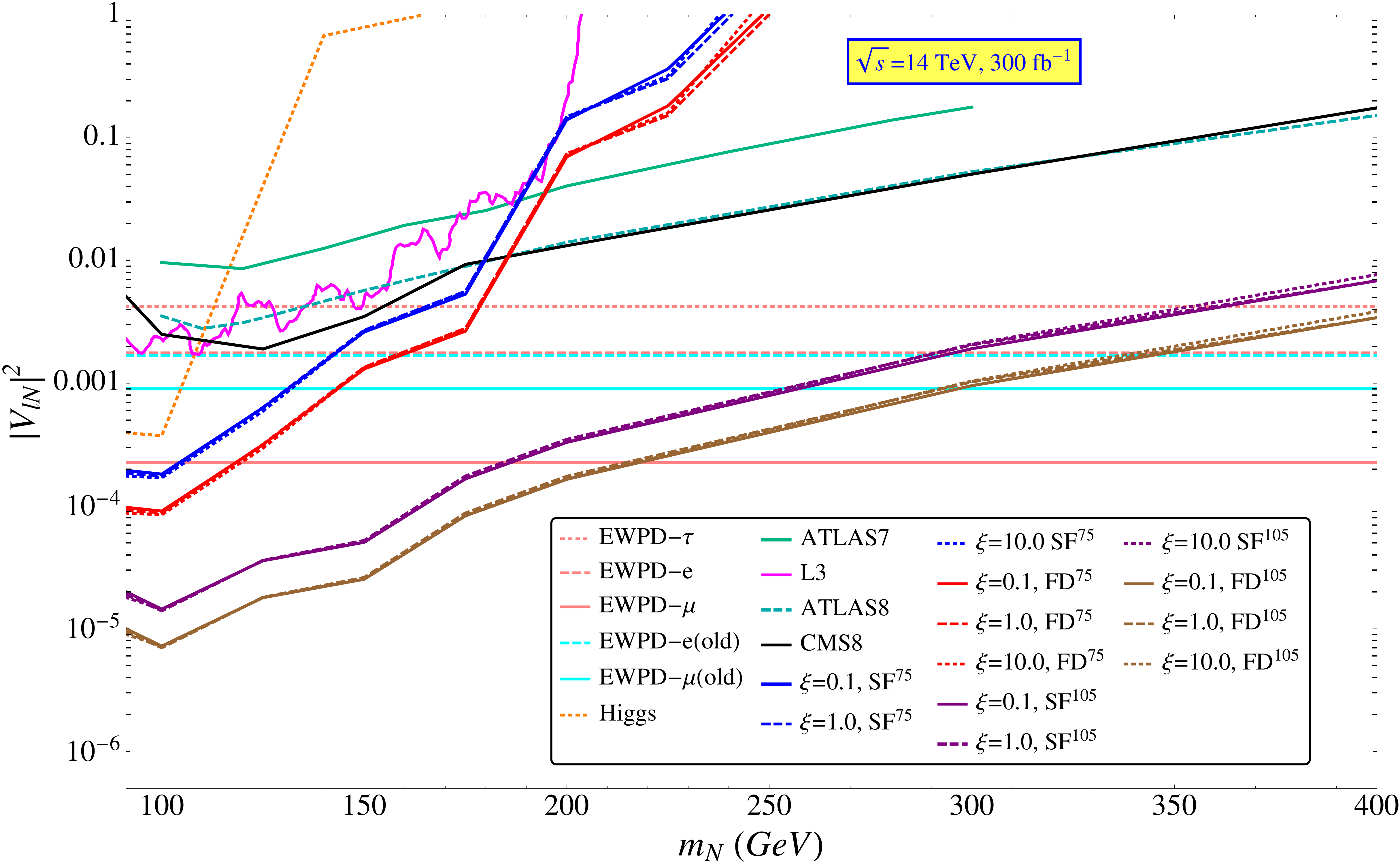}
\caption{The prospective  upper bounds on the light-heavy neutrino mixing angles $|V_{\ell N}|^2$ as a function of the heavy pseudo-Dirac neutrino mass $m_{N}$ at $14$ TeV LHC with $19.5$ fb$^{-1}$(left panel) and $300$ fb$^{-1}$(right panel) luminosities, derived from the CMS trilepton data at $\sqrt{s}=$8 TeV LHC for 19.5 fb$^{-1}$ luminosity\cite{Chatrchyan:2014aea} at 95 $\%$ CL.
We have considered the scale dependent NLO case$\left(\xi=0.1, 1.0, 10.0\right)$ for the trilepton plus missing energy final state.
 Some relevant existing upper limits (all at 95$\%$ CL) are also shown for comparison: (i) from a $\chi^{2}$-fit to the LHC Higgs data \cite{BhupalDev:2012zg} (Higgs), (ii) from a direct search at LEP \cite{Achard:2001qv}(L3), valid only for the electron flavor, (iii) ATLAS limits from $\sqrt{s}=7$ TeV LHC data \cite{Chatrchyan:2012fla} (ATLAS7) and $\sqrt{s}=$8 TeV LHC data \cite{Aad:2015xaa} (ATLAS8), valid for a heavy Majorana neutrino of the muon flavor, (iv) CMS limits from $\sqrt{s}=$8 TeV LHC data \cite{Khachatryan:2015gha} (CMS8), for a heavy Majorana neutrino of the muon flavor and (v) indirect limits from the global fit to the electroweak precision data (EWPD) from~\cite{deBlas:2013gla, delAguila:2008pw, Akhmedov:2013hec} for electron (cyan, EWPD-e(old)) and muon (cyan, EWPD-$\mu$(old)) flavors(new values can be found from \cite{Antusch:2015gjw} , for tau (dotted, EWPD- $\tau$) electron (solid, EWPD- $e$) and muon (dashed, EWPD- $\mu$) flavors). Here $SF^{75}$ and $FD^{75}$ are the single flavor and flavor diagonal cases below the $Z$-pole whereas $SF^{105}$ and $FD^{105}$ are the same above the $Z$-pole. }
\label{fig:Bounds_5}
\end{center}
\end{figure*}
In Fig.~\ref{fig:Bounds_3} we plot our results of the prospective upper bounds for the SF and FD cases for the scale dependent NLO case at the $14$ TeV. In \cite{Das:2014jxa} different $H_{T}$ and $E_{T}^{miss}$ regions are considered to calculate the upper bounds on the mixing angle which has been improved in \cite{Das:2015toa} for the LO processes. In this work we have considered a different region for $H_{T}$ and $E_{T}^{miss}$ to evaluate the upper bounds on the mixing angles which has not been tested before. In this new region we can put the prospective upper bounds on the mixing angle for the $14$ TeV LHC for $m_{N}=91.2-400$ GeV for the scale dependent NLO processes. We notice that for the trilepton case for the NLO processes at $14$ TeV the scale dependent prospective bounds on mixing angle well coincide with each other. An estimation at the $100$ TeV collider for the same study can make a legitimate improvement by approximately one order of magnitude or more and this will be tested in future. A prospective search reach for the $300$ fb$^{-1}$ luminosity at $14$ TeV LHC is also given in Fig.~\ref{fig:Bounds_5} for which we can get up to order one improvement in the upper bounds of the mixing angles. 
\section{Conclusion}
\label{sec:conc}
In this paper we have discussed the generation of the SM neutrino mass through the type-I and inverse seesaw mechanisms which involve the Majorana and the pseudo-Dirac heavy neutrinos respectively. Such heavy neutrinos, residing in the eleactroweak scale, can be produced at LHC and proposed $100$ TeV hadron collider
through a large mixing angle with the SM light neutrinos. To produce such heavy neutrinos at such high energy hadronic colliders it is important to discuss the scale dependent production cross-sections and distributions at the LO and also at the NLO QCD.

We have studied two different channels for the heavy neutrino production; one is the $W$ mediated for the associated production of lepton and the other one is the $Z$ mediated process with associated light neutrino. We have demonstrated that the heavy neutrino production cross-sections at the next-to-leading order QCD accuracy are quite stable against the scale variations, where as leading order estimated can change substantially. We also exhibit the scale dependance in different differential distributions related with the leptons and correlations between them.

We have obtained the prospective scale dependent search reach at the $14$ TeV LHC and as well as at the $100$ TeV collider for the Majorana heavy neutrino through the same sign dilepton plus dijet final state. Using the pseudo-Dirac heavy neutrinos we have studied the trilepton plus missing energy final state with jets and obtained the prospective search reach at the $14$ TeV. A collider with a higher energy can probe the mixing angle more  precisely improving the $14$ TeV result up to an order of magnitude or  more. 

\bigskip
\acknowledgments
\label{sec:acknowledgments}
We thank Oliver Mattelaer and Valentin Hirshi for useful discussions and fixing bugs on {\tt Madgraph5\_aMC@NLO} when we were implementing the SM singlet heavy neutrinos in the MadGraph.
AD would like to thank UC Davis and the organizers of Pre-SUSY 2015 and SUSY 2015 where a part of the work had been continued. Authors also thank V. Ravindran for useful discussions. 

\end{document}